\documentclass[aps,nofootinbib,amsmath,prd,onecolumn,notitlepage,showpacs,superscriptaddress,11pt]{revtex4-1}

\usepackage[american]{babel}
\usepackage{amsfonts}
\usepackage{amsmath}
\usepackage{amssymb}
\usepackage{epsfig}
\usepackage{latexsym}
\usepackage{dsfont}

\usepackage{fancyhdr}
\usepackage{graphicx}

\usepackage{hyperref}
\usepackage{etex}
\allowdisplaybreaks
\usepackage{float}
\usepackage{slashed}
\usepackage{xcolor}

\usepackage{mathtools}

\usepackage{dcolumn} 
\usepackage{bm}
\usepackage{splitidx}

\usepackage{slashed}

\fancypagestyle{plain}{
	\lhead{}
	\fancyhead[R]{\thepage}
	\fancyhead[L]{}
	
	\fancyfoot{}
}

\newcommand{\id}{\hat{\mathds{1}}}

\begin{document}

	\title{Heat kernel of non-minimal second-order operators}

	\author{Dario Sauro}
	\email{dario.sauro@uni-jena.de}
	\affiliation{
		Theoretisch-Physikalisches Institut, Friedrich-Schiller-Universität Jena,\\
		Fröbelstieg 1, 07743 Jena, Germany
	}

	\begin{abstract}
		We analyze the spectra of general non-minimal second-order operators. To do this, we derive the local part of the trace of the second Seeley-DeWitt heat kernel coefficient for such operators in a completely model-independent way. Afterwards, we provide three examples to show how our result can be applied in practical scenarios. In particular, we emphasize this discussion when dealing with a toy model of dynamical torsion, which is viewed as a simple instance of higher-spin fields. All our results are compatible with the literature, and we provide a \texttt{Mathematica} notebook with the model-independent results that are written in the paper. 
	\end{abstract}

	\maketitle

	
	\newcommand{\hD}{\hat{D}}
	\newcommand{\hF}{\hat{F}}
	\newcommand{\hnabla}{\hat{\nabla}}
	\newcommand{\hG}{\hat{G}}
	\newcommand{\hN}{\hat{N}}
	\newcommand{\ha}{\hat{a}}
	\newcommand{\hOmega}{\hat{\Omega}}
	\newcommand{\hM}{\hat{M}}
	\newcommand{\hK}{\hat{K}}
	\newcommand{\hY}{\hat{Y}}
	\newcommand{\hE}{\hat{E}}
	\newcommand{\hV}{\hat{V}}
	\newcommand{\calR}{\mathcal{R}}
	\newcommand{\hcalR}{\hat{\mathcal{R}}}
	\newcommand{\hid}{\hat{\mathds{1}}}
	\newcommand{\hsquare}{\hat{\square}}
	\renewcommand{\thefootnote}{\arabic{footnote}}
	\newcommand{\barpsi}{\overline{\psi}}
	\newcommand{\barlambda}{\overline{\lambda}}
	\setcounter{footnote}{0}

	\section{Introduction}\label{sect:intro}

	The heat kernel formalism provides a powerful method for computing the quantum corrections to the effective action \cite{Vassilevich:2003xt}. It has been developed by physicists for dealing with quantum effects on top of classical backgrounds \cite{Schwinger:1951nm,DeWitt:1964mxt,Abbott:1980hw}, while mathematicians later put it on stable grounds and connected it to the index theorem in topology \cite{Seeley:1969,Atiyah:1973ad,Gilkey:1975iq,Avramidi:2000bm}.
	
	The heat kernel method provides an elegant computational tool for deriving perturbative quantum properties in the presence of background fields in a manifestly covariant fashion. For this reason, it has been widely used in the context of gravitational theories (see, e.g.,\cite{Barvinsky:1985an,Vassilevich:2003xt,Percacci:2017fkn}). Among the many phenomenological applications we cite the computation of beta-functions \cite{Buchbinder:2021wzv,Daas:2020dyo}, working out the quantum anomalies \cite{Vassilevich:2005vk,Barvinsky:2023exr}, deriving the quantum corrections to the black hole entropy \cite{Fursaev:1994ea} and analyzing gravitational catalysis \cite{Gies:2021upb}. Furthermore, even though this formalism is mostly applied in the $1$-loop approximation, it can also be extended to take into account multi-loop diagrams \cite{Jack:1982hf,Ivanov:2020yrc,Ivanov:2024lbs,Carneiro:2024slt,Fuentes-Martin:2024agf,Carneiro:2025xny}. 
	
	Most of the applications of the heat kernel method have focused on minimal second-order operators, i.e., those whose principal part is given by a Laplace-type operator. This is because in this case one can prove that there exists an asymptotic expansion of the heat kernel for small distances \cite{Gilkey:1975iq}, which in turn suggests the introduction of the so-called Seeley-DeWitt bi-tensors. Such an expansion was found in \cite{DeWitt:1964mxt} by comparison with the known flat-space result of a diffusion process. Then, the coincidence limits of such coefficients can be found in a recursive way \cite{DeWitt:1964mxt}. Of particular interest are the first three coefficients $\ha_0$, $\ha_1$ and $\ha_2$, since these dictate the form of the beta-functions of marginal and relevant operators in $d=4$ in the $1$-loop approximation. In practice, the knowledge of the trace of $\ha_2$ suffices to yield the beta-functions of dimensionless coupling constant; for this reason the focus of this paper will be on the derivation of such a trace for a wide class of differential operators. 
	
	The most general minimal second-order operators are those that also comprise a derivative interaction. These have been dealt with in in \cite{Obukhov:1983mm,Gusynin:1988zt}, where the first Seeley-DeWitt coefficients have been derived using a squaring procedure. Furthermore, fourth-order minimal operators have been taken into account in \cite{Barvinsky:1985an,Gusynin:1990ek}, and this procedure can be applied to higher order operators.
	
	On the other hand, if one wants to check the gauge invariance of a given calculation or considers more complicated fields such as torsion, non-metricity or higher-spins, he will have to deal with non-minimal operators. To derive the heat kernel coefficients of these operators a completely different approach is needed, since there exists no asymptotic expansion when the principal part is not of Laplace type. However, one can rely on a different perturbative expansion as in \cite{Barvinsky:1985an}, or he can adopt the so-called non-local heat kernel \cite{Groh:2011dw}. Recently, such a problem has been reanalyzed in some special examples \cite{Grasso:2023qye,Farolfi:2025knq,Melichev:2025hcg,Sangy:2025jxk}. Of particular interest for the present discussion is the approach of \cite{Melichev:2025hcg}, in which a slight modification of the trick introduced in \cite{Barvinsky:1985an} was employed. Indeed, the approach of \cite{Melichev:2025hcg} significantly improves the computational time needed for the derivation of the trace of the $\ha_2$ coefficient. A radically different approach towards non-minimal operator is given by employing spin-parity decompositions, see, e.g. \cite{Groh:2011dw}. Here the idea is to trade non-minimality for higher order but minimal operators. Such an approach was applied to a dynamical torsion theory in \cite{Martini:2023apm}, but it will not be pursued here. Finally, non-minimal operators arise also in Proca-like theories, where the principal part is degenerate. In such cases one usually singles out the non-minimal part and derives the heat kernel coefficients by deriving the algebra of the pseudo-differential operators that appear in the Hessian \cite{Endo:1984sz,Endo:1994yj,Barvinsky:2025jbw}.
	
	Thus, the aim of this work is to derive local part of the trace of the second Seeley-DeWitt coefficient for general non-minimal operators in a model-independent way, and to provide some examples to show how the established recipe can be applied in practice. To this end the trick of \cite{Barvinsky:1985an} is applied in the new formulation of \cite{Melichev:2025hcg}, and vector bundles are employed to stand for an unspecified target space parametrizing the quantum fields of the theory. On a practical viewpoint the \texttt{xAct} packages of \texttt{Mathematica} \cite{Martin-Garcia:xAct,Martin-Garcia:xTensor,Martin-Garcia:2008ysv,Nutma:2013zea} have been heavily exploited, out of which \texttt{FieldsX} has played a crucial role \cite{Frob:2020gdh}. Furthermore, the complete form of the model-independent result derived in this paper is also attached in a \texttt{Mathematica} notebook to provide a handy use of the present findings \cite{Sauro:2025nb}.
	
	The structure of the paper is as follows. First of all, in Sect.\ \ref{sect:schwinger-dewitt} the heat kernel is introduced, and it is related to the trace of the logarithm of a second-order Laplace operator. Next, in Section \ref{sect:non-min-op} non-minimal second order operators are considered in a formal way, singling out the trace-log of the principal part in Subsections \ref{subsect:tr-log-pric-part-simple} and \ref{subsect:tr-log-pric-part-general} and treating perturbatively the rest in \ref{subsect:tr-log-Y-part}. To turn formal expressions into concrete ones one has to evaluate functional traces, thus in Sect.\ \ref{sect:funct-tr} a method to perform such calculations is presented. In Sect.\ \ref{sect:final-result} the final result for the trace-log of a general non-minimal second-order operator is written down. To this end, some computer-based computational tools and the associated conventions for representing field-space indices are briefly discussed in Subsect.\ \ref{subsect:result-computer}. Such conventions are employed in Subsect.s\ \ref{subsect:result-integral} and \ref{subsect:result-non-integral} to present the resulting modifications to the trace of the second Seeley-DeWitt coefficient $\ha_2$. Section \ref{sect:applications} provides three examples of applications of the general findings, dealing with a vector field \ref{subsect:appl-vector}, a Kalb-Ramond field \ref{subsect:appl-Kalb} and a toy model of the torsion \ref{subsect:appl-torsion}. The results are summarized and some possible future developments are discussed in Sect.\ \ref{sect:outro}. Finally, the Appendices contain useful commutator identities \ref{sect:app:comm-ids}, they provide all coincidence limits up to dimension-four terms \ref{sect:app:coinc-limits}, and they collect the tensor structures that concur to ${\rm tr} \, \ha_2$ that were omitted in Sect.\ \ref{sect:final-result} to avoid overburdening the flow of the main text \ref{sect:app:final-result-remaining-terms}. Finally, the mostly-plus Minkowski signature will be used throughout the paper.

	\section{The Schwinger-DeWitt technique}\label{sect:schwinger-dewitt}

	Let us consider a minimal second-order differential operator, i.e.,
	\begin{equation}\label{eq:min-op}
		\hF(\hnabla) = \hsquare + \hE \, ,
	\end{equation}
	where $\hE$ is an endomorphism. The covariant derivative $\hnabla_\mu$ is assumed to be compatible with the metric, and such that the following commutation relations hold
	\begin{equation}
		[\hnabla_\mu,\hnabla_\nu] \hnabla_\rho \hat{\varphi} = - R^\lambda{}_{\rho\mu\nu} \hnabla_\lambda \hat{\varphi} + \hcalR_{\mu\nu} \hnabla_\rho \hat{\varphi} \, ,
	\end{equation}
	where $\hcalR_{\mu\nu}$ is the generalized curvature acting on field space, and hats stand for field-valued tensors. In practice, an operator like Eq.\ \eqref{eq:min-op} is usually associated with the Hessian $\hat{\cal{H}}$ of a bosonic field theory, or to the square of a fermionic Hessian, through
	\begin{equation}
		\hat{\cal{H}} = \hF(\nabla) - m^2 \id \, ,
	\end{equation}
	where $\id$ is the identity in field space and $m$ is some mass parameter. Then, the Green's function is defined by
	\begin{equation}
		\left( \hF(\nabla) - m^2 \right) \hG(m^2) = - \id \, .
	\end{equation}
	By giving an infinitesimal negative imaginary part to the mass, we obtain Feynman's prescription for the Green's function, which allows to write the latter through the following integral representation over a ``proper time" $s$ \cite{DeWitt:1964mxt}
	\begin{equation}
		\hat{G} = - \frac{\id}{\hF(\nabla) - m^2 + i \epsilon } = i \int_0^\infty ds \, e^{\, i \, s\,(\hF(\nabla) -m^2) } \, .
	\end{equation}
	Taking the matrix elements of this equation yields
	\begin{equation}
		\hG (x,y) = i \int_0^\infty ds \, \langle x, s | y , 0 \rangle \, e^{-i \, m^2 s} \, ,
	\end{equation}
	where the ``transition amplitude" is defined by
	\begin{equation}
		\langle x, s | y , 0 \rangle \equiv \langle  x| e^{i \, \hF(\nabla) s} | y \rangle \, ,
	\end{equation}
	and it satisfies a Cauchy problem described by the Schr\"odinger-like equation
	\begin{equation}\label{eq:schroedinger-cauchy-problem}
		i \partial_s \langle x, s | y , 0 \rangle + \langle x ,s| \hF(\nabla) |y, 0 \rangle  = 0 \, , \qquad \langle x, 0 | y, 0 \rangle = \delta(x,y) \, .
	\end{equation}
	For notational convenience, we shall denote the ``transition amplitude" by
	\begin{equation}\label{eq:def-heat-kernel}
		 \hat{\cal{K}} (x,y;s) \equiv \langle x, s | y , 0 \rangle = \langle  x| e^{i \, \hF(\nabla) s} | y  \rangle \, .
	\end{equation}
	In Euclidean signature the differential equation in \eqref{eq:schroedinger-cauchy-problem} becomes a heat-like equation, whence $\hat{\cal{K}}$ is referred to as the heat kernel. The Cauchy problem \eqref{eq:schroedinger-cauchy-problem} is solved by the following ansatz \cite{DeWitt:1964mxt}
	\begin{equation}\label{eq:HK-asympt-min}
		\hat{ \cal{K} } (x,y;s) = \frac{ i \mathcal{D}^{\frac{1}{2}} (x,y) }{(4 \pi i s)^{\omega}} {\rm e}^{i \frac{\sigma(x,y)}{2 \, s}} \hat{{\Lambda}}(x,y;s) \, , \qquad  \lim_{x\rightarrow y} \hat{{\Lambda}}(x,y;0) = \id \, ,
	\end{equation}
	which has to be thought of as a modification to the known flat-space solution in absence of background fields \cite{Schwinger:1951nm}. Here and in the following $2 \omega=d$, where $d$ is the spacetime dimension. Moreover, $\sigma(x,y)$ is the Synge world function, which is equal to half the square of the geodesic distance between $x$ and $y$. Thus, it generalizes the flat-space modulus of the difference of the $x$ and $y$ coordinates, and it obeys
	\begin{eqnarray}\label{eq:world-function}
		\sigma = \frac{1}{2} \sigma_\mu \sigma^\mu \, , && \qquad \sigma_\mu \equiv \nabla_\mu \sigma \, .
	\end{eqnarray}
	Furthermore, $\mathcal{D} (x,y)$ is the Van Vleck-Morette determinant
	\begin{equation}\label{eq:van-vleck}
		\mathcal{D} (x,y) = - {\rm det} \left( - \frac{\partial^2 \sigma}{\partial x^\mu \partial y^\nu} \right) \, ,
	\end{equation}
	that describes the divergence or convergence of geodesics emanating from the point $x$, and it becomes singular whenever we encounter a caustic surface. Such a bi-scalar density can be rewritten to single out the densities as
	\begin{equation}\label{eq:van-vleck-reduced}
		\mathcal{D} (x,x^\prime) = \sqrt{-g(x)} \sqrt{-g(x^\prime)} \Delta (x,x^\prime) \, ,
	\end{equation}
	where $\Delta (x,x^\prime)$ is a bi-scalar which is commonly called the de-densitized Van Vleck determinant. The latter satisfies
	\begin{equation}\label{eq:van-vleck-world-function}
		\Delta^{-1} \nabla_\mu \left( \sigma^\mu \Delta \right) = d \, .
	\end{equation}
	Local information is extracted from the previous equations by first differentiating and then taking the coincidence limit. This enables to approximate these bi-tensors by expanding them in a covariant Taylor series \cite{Barvinsky:1985an}. The coincidence limits up to mass dimension four are given in the Appendix \ref{sect:app:coinc-limits}.
	
	While the Synge world function and the Van Vleck determinant pertain only to the underlying pseudo-Riemannian structure of spacetime, all the information about the operator $\hat{F}$ is encoded in $\hat{\Lambda}$. It is convenient to expand the latter as an infinite series, whose coefficients are the Seeley-DeWitt bi-tensors $\hat{a}_n$
	\begin{equation}\label{eq:seeley-dewitt-min}
		\hat{{\Lambda}}(x,x';s) = \sum_{n=0}^{\infty} (is)^n \hat{a}_n (x,x') \, .
	\end{equation}
	By plugging Eq.\ \eqref{eq:HK-asympt-min} into the defining equation \eqref{eq:def-heat-kernel} we find that $\hat{{\Lambda}}$ satisfies the following differential equation \cite{DeWitt:1964mxt}
	\begin{equation}\label{eq:diff-eq-lambda}
		-i \partial_s \hat{{\Lambda}} - \frac{i}{s} \sigma^\mu \nabla_\mu \hat{{\Lambda}} = \Delta^{-1/2} \hat{F} \left( \Delta^{1/2} \hat{{\Lambda}} \right) \, .
	\end{equation}
	Such a first-order differential equation is solved by employing the expansion \eqref{eq:seeley-dewitt-min} and iteratively deriving local expressions for the Seeley-DeWitt coefficients from the following recurrence relation 
	\begin{align}\label{eq:rec-rel-min}
		\begin{split}
			& n \hat{a}_n + \sigma^\mu \nabla_\mu \hat{a}_n = \Delta^{-1/2} \hat{F} \left(\Delta^{1/2} \hat{a}_{n-1} \right) \, ;\\
			& \sigma^\mu \nabla_\mu \hat{a}_0 = 0 \, , \qquad a_0(x,x) = \hat{\mathds{1}} \, .
		\end{split}
	\end{align}
	Thus, once again local information is extracted by repeatedly differentiating it and taking the coincidence limit afterwards. Such limits are given in App.\ \ref{sect:app:coinc-limits} for $\hF(\hnabla)=\hsquare +\hE$.
	
	Let us now connect the heat kernel of the differential operator $\hF(\hnabla)$ to the trace of the logarithm of this operator following Schwinger's presentation \cite{Schwinger:1951nm}. Let $\hF(\hnabla)$ be a differential operator, and $\hG(\hnabla)$ its Green's function. Then, for a general functional variation we have
	\begin{align}
		\delta {\rm Tr} \log \hF = & - {\rm Tr} \left(  \hG \delta \hF \right) =  - i \,{\rm Tr} \int_0^\infty ds \, \delta \hF \, e^{is \hF} \\\nonumber
		= & - {\rm Tr} \int_0^\infty \frac{ds}{s}   \delta e^{is\hF} = - \delta \left( {\rm Tr} \int_0^\infty \frac{ds}{s} e^{is \hF} \right) \, .
	\end{align}
	Thus, up to an integration constant we have that the trace of the logarithm of $\hF$ can be written in terms of the trace of the heat kernel as
	\begin{equation}\label{eq:tr-log-to-HK}
		{\rm Tr} \log (\hF(\hnabla)) = - \int_0^\infty \frac{ds}{s} {\rm Tr}\, e^{is\hF(\hnabla)} = - \int_0^\infty \frac{ds}{s} \int \sqrt{-g} \,\, {\rm tr} \, \hat{ \cal{K}} (x,x;s)   \, .
	\end{equation}
	The last trace is not capitalized since it is to be taken on internal and spacetime indices only. While this equation is valid in general, the asymptotic expansion of the heat kernel given by \eqref{eq:HK-asympt-min} and \eqref{eq:seeley-dewitt-min}	is valid only for Laplace-type second-order operators. Thus, when the principal part is not given by $\hsquare$ we have to resort to different methods for evaluating the trace log and singling out its divergent parts.
	
	Before focusing on the evaluation of the divergent parts of complicated functional traces, let us make a brief remark. When we are not dealing with scalar fields the Hessian ${\cal{H}}_{ab}$ of a given Lagrange density is not a differential operator in a strict sense, since it does not provide a map to tensors or spinors of the same type. Thus, to make sense out of the tracing procedure in Eq.\ \eqref{eq:tr-log-to-HK}, we need to factor out a ``configuration metric", yielding a true differential operator
	\begin{equation}\label{eq:operator-from-hessian}
		{\cal F}_a{}^b (\hnabla) = {\cal G}^{bc} {\cal H}_{ac} (\hnabla) \, .
	\end{equation}
	In practice, such a procedure becomes non-trivial only when the tensor fields that we are integrating out are $O(d)$-reducible, like a symmetric $2$-tensor. We shall comment on another type of such a field in the last example in Subsect.\ \ref{subsect:appl-torsion}.
	
	\section{Non-minimal operators}\label{sect:non-min-op}

	In this section we address the main concern of the present work, i.e., a general derivation of the trace of the second Seeley-DeWitt coefficient $\ha_2$ for non-minimal second-order operators. First of all, let us point out that the adjective \emph{non-minimal} has had different interpretation in the literature. For example, in \cite{Gusynin:1988zt} operators involving a derivative interaction are dubbed non-minimal, whereas here we follow the nomenclature of \cite{Barvinsky:1985an}, and we refer to non-minimality whenever the principal part of the operator is not just given by $\hsquare$. Therefore, our task is work out the correction to ${\rm tr}\,\ha_2$ in the presence of these non-minimal terms.
	
	The class of operators on which we are focusing is quite broad, and we must make the two following assumptions. First of all, the non-minimality cannot be completely general, in that we require the principal part of the operator to be non-degenerate. Let this operator be written as 
	\begin{equation}\label{eq:non-min-op}
		\hF(\hnabla) = \hD (\hnabla) + \hY (\hnabla) = \square + \hN (\hnabla) + \hY (\hnabla) \, ,
	\end{equation}
	where $\hD(\hnabla)$ is the principal part, whereas $\hY$ comprises the derivative interactions and endomorphisms. Thus, the non-minimality is parametrized solely by $\hN$. Let us define the principal symbol of the operator by replacing the covariant derivative $\hnabla_\mu$ with an arbitrary vector $n_\mu$ in $\hD$, i.e.,
	\begin{equation}
		\hD(n) = \hD^{\mu\nu} n_\mu n_\nu \, ,
	\end{equation}
	Then, non-degeneracy means that
	\begin{equation}\label{eq:non-deg-princ-symb}
		\det \hD^{\mu\nu} n_\mu n_\nu \neq 0 \, .
	\end{equation}
	Furthermore, we also restrict on principal symbols that are parametrized by metric-compatible tensors, i.e.,
	\begin{equation}
		\hnabla_\rho \hD^{\mu\nu} = 0 \, .
	\end{equation}
	When the background geometry is curved the last assumption merely asserts that the quantum fields that we are integrating out must bear their canonical Weyl weights.
	
	The existence of the asymptotic expansion of the heat kernel Eq.\ \eqref{eq:seeley-dewitt-min} rests upon the hypothesis that the operator is of Laplace type. Clearly, this is not the case when we introduce a non-minimal term $\hN(\hnabla)$ in the principal part. Nevertheless, in this situation we can derive the divergent parts of the effective action employing a trick first introduced by Barvinksy and Vilkovisky \cite{Barvinsky:1985an}, and recently slightly modified by Melichev in \cite{Melichev:2025hcg}. The main observation of \cite{Melichev:2025hcg} that we will employ here is that in absence of multiplicative anomalies \cite{Evans:1998pd}, our calculation can actually be split in two parts as
	\begin{equation}\label{eq:tr-log-F-formal}
		{\rm Tr} \log \hF(\hnabla) = {\rm Tr} \log \hD(\hnabla) + {\rm Tr} \log \left( 1 + \hD^{-1} \hY \right) \, .
	\end{equation}
	The trick introduced in \cite{Barvinsky:1985an} is to parameterize the principal $\hD(\hnabla)$ as a one parameter family of operators such that
	\begin{equation}\label{eq:1-parameter}
		\hD(\hnabla;\zeta)\Big|_{\zeta=1} = \hD(\hnabla) \, , \qquad \hD(\hnabla;\zeta)\Big|_{\zeta=0} = \hsquare \, ,
	\end{equation}
	which allows to perform a perturbative expansion of the first term on the r.h.s.\ of Eq.\ \eqref{eq:tr-log-F-formal}. If the dependence of $\hN$ on $\zeta$ is linear we have some important simplifications, which are discussed in the next subsection \ref{subsect:tr-log-pric-part-simple}. However, such an assumption cannot always made in practice, as we will show in one of our applications in \ref{sect:applications}. Thus, the subsequent subsection \ref{subsect:tr-log-pric-part-general} deals with those scenarios in which the dependence of $\zeta$ is the most general one. Finally, in \ref{subsect:tr-log-Y-part} we study the second term on the r.h.s. of Eq.\ \eqref{eq:tr-log-F-formal}. Importantly, the latter does not require the introduction of any auxiliary parameter.
	
	\subsection{Trace log of the principal part: The simple case}\label{subsect:tr-log-pric-part-simple}
	
	The trick employed in \cite{Barvinsky:1985an} to evaluate the divergent parts of the trace log of non-minimal operator relies on the introduction of an auxiliary parameter $\zeta$ such as Eq.\ \eqref{eq:1-parameter}. In the simplest case this parameter appears linearly in $\hD(\hnabla)$, i.e.,
	\begin{equation}
		\hD(\hnabla) = \left( \hsquare + \zeta \hN \right)\Big|_{\zeta=1} = \hD(\hnabla;\zeta)\Big|_{\zeta=1} \, .
	\end{equation}
	Obviously, for $\zeta=0$ the operator is minimal, and its heat kernel coefficients can be found by means of the standard asymptotic expansion. We consider the inverse of the principal symbol
	\begin{equation}\label{eq:def-G}
		\hD (\hnabla;\zeta)  \, \hG_0 (\hnabla;\zeta) = \hid \, ,
	\end{equation}
	which exists thanks to Eq.\ \eqref{eq:non-deg-princ-symb}. Thus, using the fact that $\hD(\hnabla;\zeta=0)= \hsquare$, we can write the trace of the logarithm of $\hD(\hnabla)$ as
	\begin{align}\label{eq:trace-log-princ-part}
		{\rm Tr} \log \hD(\hnabla) = & \, {\rm Tr} \log \hD(\hnabla;\zeta)\Big|_{\zeta=1} = {\rm Tr} \log \hsquare + \int_0^1 d \zeta \, \frac{d }{d\zeta} {\rm Tr} \log \hD(\hnabla;\zeta) \\\nonumber
		= & \, {\rm Tr} \log \hsquare + \int_0^1 d \zeta {\rm Tr} \left( \hN \hG_0(\zeta) \right) = {\rm Tr} \log \hsquare + \int_0^1 \frac{d \zeta}{\zeta} \, {\rm Tr} \left[ \left( \hD(\hnabla;\zeta) - \hsquare\right) \hG_0(\hnabla;\zeta) \right] \\\nonumber
		= & \, {\rm Tr} \log \hsquare - \int_0^1 {d \log\zeta} \, {\rm Tr} \left( \hG_0(\hnabla;\zeta) \, \hsquare \right) \, .
	\end{align}
	Thus, our new task is to work out a perturbative expansion of $\hG_0 (\hnabla;\zeta)$, which will be used to evaluate the trace of the logarithm of $\hD (\hnabla;\zeta)$ through the previous equation. The first step in this direction is provided by inverting the principal symbol in momentum space, i.e., solving the following equation for the matrix $\hK(n;\zeta)$
	\begin{equation}\label{eq:def-K(n)}
		\hD (n;\zeta) \, \hK (n;\zeta) = n^4 \id \, .
	\end{equation}
	Then, promoting the vectors $n_\mu$ to covariant derivatives defines $\hK (\hnabla;\zeta)$ by fixing the arbitrary ordering of covariant derivatives. Next, due to the previous equation, we automatically find that
	\begin{equation}\label{eq:def-M}
		\hD (\hnabla;\zeta) \, \hK (\hnabla;\zeta) = \hsquare^2 + \hM (\hnabla;\zeta) \, ,
	\end{equation}
	where $\hM$ is a differential operator of order two. Then, we compare Eqs.\ \eqref{eq:def-G} and \eqref{eq:def-M} to find a perturbative expression of $\hG_0(\hnabla;\zeta)$, which reads
	\begin{align}\label{eq:pert-exp-G}
		\hG_0 (\hnabla;\zeta) = \hK \frac{1}{\hsquare^2} - \hK \hM \frac{1}{\hsquare^4} + 2 \hK [\hsquare,\hM] \frac{1}{\hsquare^5} + \hK \hM \hM \frac{1}{\hsquare^6} + \dots \, ,
	\end{align}
	where we use ellipsis to stand for higher-dimensional operators that we are not interested in. To derive this equation we have commuted the inverse box operators to the right using Eqs.\ \eqref{eq:comm-box^(-1)-M} and \eqref{eq:comm-box^(-2)-M}. Finally, we can insert Eq.\ \eqref{eq:pert-exp-G} into Eq. \eqref{eq:trace-log-princ-part}, obtaining the desired relation for the trace-log of the principal part
	\begin{align}\label{eq:integral-part-general-simple-case}
		{\rm Tr} \log \hD(\hnabla) & = {\rm Tr} \log \hsquare - \int_0^1 {d \log\zeta} \, {\rm Tr} \left( \hG_0(\zeta) \, \hsquare \right)\\\nonumber
		& = {\rm Tr} \log \hsquare - \int_0^1 {d \log\zeta} \, {\rm Tr}\left[ \hK \frac{1}{\hsquare} - \hK \hM \frac{1}{\hsquare^3} + 2 \hK [\hsquare,\hM] \frac{1}{\hsquare^4} + \hK \hM \hM \frac{1}{\hsquare^5} \right] \, ,
	\end{align}
	where the dependence on $\zeta$ is understood.
	
	\subsection{Trace log of the principal part: The general case}\label{subsect:tr-log-pric-part-general}
	
	Now we want to address the problem of finding an expression for ${\rm Tr} \log (\hD(\hnabla))$ in the most general case. As in the previous simpler case, our starting point is Eq.\ \eqref{eq:1-parameter}, while the difference is that
	\begin{equation}
		\frac{d \hD(\hnabla;\zeta)}{d\zeta} \neq \hD(\hnabla;\zeta=1) - \hsquare \, .
	\end{equation}
	Therefore, using the trick of \cite{Barvinsky:1985an} we find
	\begin{align}\label{eq:trace-log-princ-part-general-case}
		{\rm Tr} \log \hD(\hnabla) = & \, {\rm Tr} \log \hD(\hnabla;\zeta)\Big|_{\zeta=1} = {\rm Tr} \log \hsquare + \int_0^1 d \zeta \, \frac{d }{d\zeta} {\rm Tr} \log \hD(\hnabla;\zeta) \\\nonumber
		= & \, {\rm Tr} \log \hsquare + \int_0^1 d \zeta {\rm Tr} \left( \frac{d\hN(\zeta)}{d \zeta} \hG_0(\zeta) \right) ,
	\end{align}
	where we have used the definition of the principal part. The expansion of the inverse $\hG_0$ of $\hD$ Eq.\ \eqref{eq:pert-exp-G} can still be used. Thus, we can rewrite the previous equation as
	\begin{align}\label{eq:integral-part-general-complicated-case}
		 {\rm Tr} \log \hD(\hnabla) = \, {\rm Tr} \log \hsquare + & \int_0^1 d \zeta {\rm Tr} \left[ \frac{d\hN(\zeta)}{d \zeta} \hK \frac{1}{\hsquare^2} - \frac{d\hN(\zeta)}{d \zeta} \hK \hM \frac{1}{\hsquare^4} \right.\\\nonumber
		 & \left. + \frac{d\hN(\zeta)}{d \zeta} 2 \hK [\hsquare,\hM] \frac{1}{\hsquare^5} + \frac{d\hN(\zeta)}{d \zeta} \hK \hM \hM \frac{1}{\hsquare^6} + \dots \right] \, .
	\end{align}

	\subsection{Trace-log of the rest}\label{subsect:tr-log-Y-part}
	
	Let us now come back to the starting point, which is to derive the divergent parts of the trace log of the operator Eq.\ \eqref{eq:non-min-op}. Assuming the absence of multiplicative anomalies \cite{Evans:1998pd} and using the definition of $\hG(\hnabla)$ \eqref{eq:def-G} we can write
	\begin{equation}\label{eq:tr-log-F-formal22}
		{\rm Tr} \log \hF(\hnabla) = {\rm Tr} \log \hD(\hnabla) + {\rm Tr} \log \left( 1 + \hG_0 \hY \right) \, .
	\end{equation}
	The first term on the right-hand side is given by Eq. \eqref{eq:integral-part-general-simple-case} (or by Eq.\ \eqref{eq:integral-part-general-complicated-case} in the most general case), therefore we now focus on the second one. By using the known Taylor expansion of the logarithm and commuting the $\hsquare^{-2}$ operators to the right we find
	\begin{align}\label{eq:non-integral-part-general}
		{\rm Tr} \log \left( 1 + \hG_0 \hY \right) = & {\rm Tr} \left[ \hG_0 \hY - \frac{1}{2} (\hG_0 \hY)^2 + \frac{1}{3} (\hG_0 \hY)^3 - \frac{1}{4} (\hG_0 \hY)^4 + \dots \right] \\\nonumber
		= & {\rm Tr} \left[ \hY \hK \frac{\hid}{\,\,\hsquare^2} - \hY \hK \hM \frac{\hid}{\,\,\hsquare^4} + 2 \hY \hK [\hsquare,\hM] \frac{\hid}{\,\,\hsquare^5} - \frac{1}{2} \hY \hK \hY \hK \frac{\hid}{\,\,\hsquare^4} \right.\\\nonumber
		& \,\,\, \left. + \hY \hK [\hsquare,\hY] \hK \frac{\hid}{\,\,\hsquare^5} + \hY \hK \hY [\hsquare,\hK] \frac{\hid}{\,\,\hsquare^5}  - \frac{3}{2} \hY \hK [\hsquare,[\hsquare,\hY]] \hK \frac{\hid}{\,\,\hsquare^6} + \hY \hK \hM \hY \hK \frac{\hid}{\,\,\hsquare^6} \right.\\\nonumber
		& \,\,\, \left. - 2 \hY \hK \hY \hK [\hsquare,\hY] \hK \frac{\hid}{\,\,\hsquare^7} + \frac{1}{3} (\hY \hK)^3 \frac{\hid}{\,\,\hsquare^6} - \frac{1}{4} (\hY \hK)^4 \frac{\hid}{\,\,\hsquare^8} + \dots \right] \, .
	\end{align}
	Again, our perturbative expansion stops at dimension four operators, and we also exclude total derivatives. Furthermore, we split the lower derivative part of the operator $\hY(\hnabla)$ into a derivative interaction and an endomorphism as
	\begin{equation}\label{eq:YtoV-E}
		\hY = \hV + \hE = \hV^\mu \hnabla_\mu + \hE \, .
	\end{equation}
	Thus, the contributions encoded in Eq. \eqref{eq:non-integral-part-general} can be expressed as a series expansion in $\hV$. The affine part of the latter is given by
	\begin{align}\label{eq:endomorphism-part-general}
		{\rm Tr} \log \left( 1 + \hG_0 \hY \right)\Big|_{V^0} = {\rm Tr} \left[ \hE \hK \frac{\hid}{\,\,\hsquare^2} - \hE \hK \hM \frac{\hid}{\,\,\hsquare^4}  - \frac{1}{2} \hE \hK \hE \hK \frac{\hid}{\,\,\hsquare^4} + \dots \right] \, .
	\end{align}
	Next, we consider the terms linear in $\hV$, whose contribution is
	\begin{align}\label{eq:V1-part-general}
		{\rm Tr} \log \left( 1 + \hG_0 \hY \right)\Big|_{V^1} = {\rm Tr} & \left[ \hV \hK \frac{\hid}{\,\,\hsquare^2} - \hV \hK \hM \frac{\hid}{\,\,\hsquare^4} + 2 \hV \hK [\hsquare,\hM] \frac{\hid}{\,\,\hsquare^5} - \frac{1}{2} \hV \hK \hE \hK \frac{\hid}{\,\,\hsquare^4} \right.\\\nonumber
		& \left. \,\,\,\, - \frac{1}{2} \hE \hK \hV \hK \frac{\hid}{\,\,\hsquare^4} + \hE \hK [\hsquare,\hV] \hK \frac{\hid}{\,\,\hsquare^5} + \hV \hK [\hsquare,\hE] \hK \frac{\hid}{\,\,\hsquare^5}  + \dots \right] \, .
	\end{align}
	On the other hand, taking into account terms quadratic in $\hV$ yields
	\begin{align}\label{eq:V2-part-general}
		{\rm Tr} \log \left( 1 + \hG_0 \hY \right)\Big|_{V^2} = {\rm Tr} & \left[ - \frac{1}{2} \hV\hK \hV \hK \frac{\hid}{\,\,\hsquare^4} + \hV \hK [\hsquare,\hV] \hK \frac{\hid}{\,\,\hsquare^5} - \frac{3}{2} \hV \hK [ \hsquare, [\hsquare,\hV] ] \hK \frac{\hid}{\,\,\hsquare^6} \right.\\\nonumber
		& \left. \,\,\,\, + \hV \hK \hM \hV \hK \frac{\hid}{\,\,\hsquare^6} + \hE \hK \hV \hK \hV \hK \frac{\hid}{\,\,\hsquare^6} + \dots \right] \, .
	\end{align}
	We have listed all terms that depend on the $\hM$ operator and on the endomorphism $\hE$. Cubic terms in this expansion have the following form
	\begin{align}\label{eq:V3-part-general}
		{\rm Tr} \log \left( 1 + \hG_0 \hY \right)\Big|_{V^3} = {\rm Tr} \left[ \frac{1}{3} \left(\hV \hK \right)^3 \frac{\hid}{\,\,\hsquare^6} - 2 \left(\hV \hK \right)^2 [\hsquare,\hV] \hK \frac{\hid}{\,\,\hsquare^7} + \dots \right] \, ,
	\end{align}
	whereas quartic ones are given by
	\begin{align}\label{eq:V4-part-general}
		{\rm Tr} \log \left( 1 + \hG_0 \hY \right)\Big|_{V^4} = {\rm Tr} \left[ - \frac{1}{4} \left(\hV \hK \right)^4 \frac{\hid}{\,\,\hsquare^8} + \dots \right] \, .
	\end{align}
	
	After having derived these formal expressions, let us briefly discuss the advantages of following \cite{Melichev:2025hcg} instead of \cite{Barvinsky:1985an}. First of all, in practical applications the integration over $\zeta$ is usually computationally expensive, thus, reducing the number of terms to integrate over is more efficient. The second reason is explained by the subsequent observation. Both methods rely on the evaluation of functional traces of the form
	\begin{equation}
		{\rm Tr} \left( \hat{O} \frac{\id}{\,\,\hsquare^n} \right) = {\rm Tr} \left( \hat{O}^{\mu_1 \dots \mu_k} \hnabla_{\mu_1} \dots \hnabla_{\mu_k} \frac{\id}{\,\,\hsquare^n} \right) \, .
	\end{equation}
	When the starting operator involves a derivative interaction $\hV$, the method of \cite{Barvinsky:1985an} requires the evaluation of expressions with up to $16$ covariant derivatives acting on $\hsquare^{-10}$. On the other hand, in the present case we must handle expressions with up to $12$ derivatives acting on $\hsquare^{-8}$. As we will show in the next section working out the precise form of these functional traces is not particularly involved. Instead, the computational cost increases with the number of derivatives, regardless the overall mass dimension of a given functional trace. This is due to the appearance of a number of symmetrized metric tensors which is at most equal to half the number of covariant derivatives. In general, handling these kinds of expressions embodies a very expensive part of any such computation. Therefore, the present method significantly improves the computational efficiency of the calculation.

	\section{Functional traces}\label{sect:funct-tr}
	
	In this section we shall deal with the universal functional traces, i.e., objects like
	\begin{equation}\label{eq:functional-trace}
		\hnabla_{\mu_1} \dots \hnabla_{\mu_k} \frac{\hid}{\,\,\hsquare^n} \delta(x,x')|_{x'=x} \, ,
	\end{equation}
	which represent derivative interactions in loop diagrams with $n$ internal lines. Here we are only interested in the logarithmically divergent parts of these diagrams, which will be combined with the explicit form of the differential operators entering Eqs.\ \eqref{eq:integral-part-general-simple-case} and \eqref{eq:non-integral-part-general}.
	
	The first step in the evaluation of Eq.\ \eqref{eq:functional-trace} is provided by employing a Laplace transform to rewrite the $\hsquare^{-n}$ operator as
	\begin{equation}
		\frac{\id}{\,\,\hsquare{}^n} = \frac{(-i)^n}{(n-1)!}\int_0^\infty ds s^{n-1} {\rm e}^{i s \hsquare} \, .
	\end{equation}
	Next, we exploit the asymptotic expansion of the heat kernel Eq.\ \eqref{eq:HK-asympt-min} to find a more convenient expression for the exponential that appears in the previous equation. The expansion of $\hat{\Lambda}$ in Eq.\ \eqref{eq:HK-asympt-min} reads
	\begin{equation}
		\hat{\Lambda} (x,y;s) = \hat{\Lambda}_{\hsquare} (x,y;s) = \sum_{\scriptscriptstyle n\geq0} \left( i s \right)^n  \ha_n (x,y)
	\end{equation}
	where the $\hat{a}_n$ are the Seeley-DeWitt coefficients of the $\hsquare$ operator. At this point it is advantageous to modify our covariant derivatives to automatically account for the presence of the exponential of the Synge world function in \eqref{eq:HK-asympt-min}. This is done by defining
	\begin{equation}
		\hat{\frak{D}}_\mu = {\rm e}^{-i \frac{\sigma}{2s}} \hnabla_\mu {\rm e}^{i \frac{\sigma}{2s}} = \hnabla_\mu + \frac{i}{2s} \sigma_\mu \, \id \, .
	\end{equation}
	Finally, we are able to write a concise expression for the functional trace \eqref{eq:functional-trace} as the integral over the proper time of some coincidence limit, i.e.,
	\begin{equation}\label{eq:func-traces-general-formula}
		\hnabla_{\mu_1} \dots \hnabla_{\mu_k} \frac{\id}{\,\,\hsquare^n} \delta(x,x')|_{x'=x} = - \frac{(-i)^{n+1}}{(n-1)!} \frac{\sqrt{-g}}{(4 \pi i)^\omega} \int_0^\infty \frac{ds}{s^{\omega+1-n}} \lfloor \hat{\frak{D}}_{\mu_1} \dots \hat{\frak{D}}_{\mu_k} \left( \Delta^{\frac{1}{2}} \hat{\Lambda}_{\hsquare}(x,x';s) \right) \rfloor \, ,
	\end{equation}
	where $\lfloor \hat{O} \rfloor$ stands for the coincidence limit of the operator $\hat{O}$. The integral in $s$ depends on the spacetime dimension through the $\omega$ parameter, and we are only interested in the logarithmically divergent contributions of these diagrams. Furthermore, to obtain tensor structures with the correct mass or background dimensions we need only consider integrals of the form
	\begin{equation}
		\int_0^\infty ds s^{1-\omega} f(s,\omega) = \frac{1}{2-\omega} f(0,2) + \dots \, ,
	\end{equation}
	where the dots stand for non-divergent terms in dimensional regularization. Following \cite{Barvinsky:1985an} we shall parameterize the pole as
	\begin{equation}\label{eq:div-part-numerical-factor}
		\frac{1}{2-\omega} = \log L^2 \quad {\rm for} \quad \omega \rightarrow 2 \, .
	\end{equation}
	Nevertheless, we will factor out this divergent logarithm at the end of the calculation, thus re-writing our final result as a modification of the trace of the second Seeley-DeWitt coefficient $\ha_2$, i.e., in a regularization-independent way.

	By inspecting of Eqs.\ \eqref{eq:integral-part-general-simple-case}, \eqref{eq:integral-part-general-complicated-case}, \eqref{eq:endomorphism-part-general}, \eqref{eq:V1-part-general}, \eqref{eq:V2-part-general}, \eqref{eq:V3-part-general} and \eqref{eq:V4-part-general} we can list the functional traces that we need to evaluate
	\begin{align}\label{eq:all-functional-traces}
		& \hnabla_{\mu_1}\hnabla_{\mu_2} \frac{\id}{\,\hsquare} \delta(x,x')|^{\rm div}_{x'=x} \, , \qquad\qquad \hnabla_{\mu_1} \hnabla_{\mu_2} \frac{\id}{\,\,\hsquare^2} \delta(x,x')|^{\rm div}_{x'=x} \, ;\\\nonumber
		& \hnabla_{\mu_1}\dots\hnabla_{\mu_4} \frac{\id}{\,\hsquare^2} \delta(x,x')|^{\rm div}_{x'=x} \, , \qquad \hnabla_{\mu_1} \dots \hnabla_{\mu_3} \frac{\id}{\,\,\hsquare^2} \delta(x,x')|^{\rm div}_{x'=x} \, ; \\\nonumber
		& \hnabla_{\mu_1} \dots \hnabla_{\mu_4} \frac{\id}{\,\,\hsquare^3} \delta(x,x')|^{\rm div}_{x'=x} \, , \qquad \hnabla_{\mu_1} \dots \hnabla_{\mu_6} \frac{\id}{\,\,\hsquare^4} \delta(x,x')|^{\rm div}_{x'=x} \, ;  \\\nonumber
		& \hnabla_{\mu_1} \dots \hnabla_{\mu_4} \frac{\id}{\,\,\hsquare^4} \delta(x,x')|^{\rm div}_{x'=x} \, , \qquad \hnabla_{\mu_1} \dots \hnabla_{\mu_6} \frac{\id}{\,\,\hsquare^5} \delta(x,x')|^{\rm div}_{x'=x} \, ; \\\nonumber
		& \hnabla_{\mu_1} \dots \hnabla_{\mu_8} \frac{\id}{\,\,\hsquare^6} \delta(x,x')|^{\rm div}_{x'=x} \, , \qquad \hnabla_{\mu_1} \dots \hnabla_{\mu_{10}} \frac{\id}{\,\,\hsquare^7} \delta(x,x')|^{\rm div}_{x'=x} \, .
	\end{align}
	Instead of simply providing the specific expressions for all these quantities, we prefer to analyze more deeply some general features of functional traces up to dimension-four operators.
	
	Dimensional analysis already suggests that the structure of these objects may be written in an universal way, as we will show below. Let us start by considering $2n-4$ covariant derivatives acting on the $n$th inverse power of the box operator. By employing Eq.\ \eqref{eq:func-traces-general-formula} we find
	\begin{equation}\label{eq:functional-trace-BD0}
		\hnabla_{\mu_1} \dots \hnabla_{\mu_{2n-4}} \frac{\id}{\,\,\hsquare^n} \delta(x,x')|^{\rm div}_{x'=x} = \frac{i \log L^2}{16 \pi^2} \frac{\sqrt{-g}}{2^{n-2}(n-1)!} g^{(n-2)}_{\mu_1 \dots \mu_{2n-4}} \id \, ,
	\end{equation}
	where $g^{(n-2)}_{\mu_1 \dots \mu_{2n-4}}$ are completely symmetric tensors defined by
	\begin{align}\label{eq:def-symmetrized-metrics}
		\begin{split}
			& g^{(0)}= 1 \, , \qquad g^{(1)}_{\mu\nu} = g_{\mu\nu} \, , \qquad g^{(k)} = 0 \quad {\rm for} \quad k<0\,;\\
			& g^{(n)}_{\mu_1 \dots \mu_{2n}} = \sum_{j=2}^{2n} g_{\mu_1 \mu_j} \, g^{(n-1)}_{\mu_2 \dots \mu_{j-1} \mu_{j+1} \dots \mu_{2n}} \, .
		\end{split}
	\end{align}
	We point out that we have extended the definition of these symmetric tensors to negative arguments. This choice is convenient for the present analysis, as it will allow us to write down in a simple fashion all the functional traces that we are interested in.
	
	The tensor structure of Eq.\ \eqref{eq:functional-trace-BD0} could have been guessed by dimensional arguments. Indeed, since there are no dimension-one background tensors by applying one more covariant derivative we automatically obtain zero, i.e.,
	\begin{equation}
		\hnabla_{\mu_1} \dots \hnabla_{\mu_{2n-3}} \frac{\id}{\,\,\hsquare^n} \delta(x,x')|^{\rm div}_{x'=x} = 0 \, .
	\end{equation}
	Obviously, the situation changes when we further increase the number of covariant derivatives by one. In this case the non-trivial tensor structure turns out to be
	\begin{align}\label{eq:functional-trace-BD2}
		\hnabla_{\mu_1} \dots \hnabla_{\mu_{2n-2}} & \frac{\id}{\,\,\hsquare^n} \delta(x,x')|^{\rm div}_{x'=x} = - \frac{i \, \log L^2 }{16 \, \pi^2} \, \frac{\sqrt{-g}}{2^{n-2} \, (n-1)!} \left[ \frac{1}{12} R \,g^{(n-1)}_{\mu_1 \dots \mu_{2n-2}} \, \id \right.\\\nonumber
		& \left. - \sum_{i<j} \left( \frac{1}{6} R_{ij} \id + \frac{1}{2} \hcalR_{ij} \right) g^{(n-2)}_{\mu_1 \dots \hat{i}\dots \hat{j} \dots \mu_{2n-2}}  + \frac{1}{3} \sum_{{\scriptscriptstyle i<j<k<l}} \left( R_{kilj} + R_{likj} \right) g^{(n-3)}_{\mu_1 \dots \hat{i} \dots \hat{l} \dots \mu_{2n-2}} \id \right] \, .
	\end{align}
	Notice that we have employed our generalized definition of $g^{(k)}=0$ for any $k<0$, see Eq.\ \eqref{eq:def-symmetrized-metrics}, and we are using the shorthand $\hnabla_i=\hnabla_{\mu_i}$. By taking a further covariant derivative we end up with dimension-three terms on the right-hand side, whose precise structure is 
	\begin{align}\label{eq:functional-trace-BD3}\nonumber
		\hnabla_{\mu_1} \dots \hnabla_{\mu_{2n-1}} & \frac{\id}{\,\,\hsquare^n} \delta(x,x')|^{\rm div}_{x'=x} = - \frac{i \, \log L^2 }{16 \, \pi^2} \, \frac{\sqrt{-g}}{2^{n-2} \, (n-1)!} \left[ \frac{1}{12} \sum_i \left( \hnabla^\alpha \hcalR_{\mu_i\alpha} + \frac{\id}{2} \nabla_{\mu_i} R \right) g^{(n-1)}_{\mu_1 \dots \hat{i} \dots \mu_{2n-1}} \right.\\
		& \left. - \sum_{\scriptscriptstyle i<j<k} \left( \tfrac{2}{3} \hnabla_{(i} \hcalR_{j)k} + \tfrac{1}{4} \nabla_{(i} R_{jk)} \id  \right) g^{(n-2)}_{\mu_1 \dots \hat{i} \dots \hat{k} \dots \mu_{2n-1}} \right.\\\nonumber
		& \left. + \, \frac{1}{2} \sum_{{\scriptscriptstyle i<j<k<l<m}} \left( \nabla_i R_{(l|j|m)k} + \nabla_j R_{(l|i|m)k} + \nabla_k R_{(l|i|m)j} \right) g^{(n-3)}_{\mu_1 \dots \hat{i} \dots \hat{m} \dots \mu_{2n-1}} \id \, \right] \, .
	\end{align}
	Finally, we consider the dimension-four contributions. In such a case many non-trivial tensor structures arise, and the general expression of this type of functional traces is quite complicated. Thus, we choose to write it implicitly in terms of the coincidence limits summarized in the Appendix \ref{sect:app:coinc-limits}. Moreover, we assume that the summed indices never enter the symmetrized metrics. Using this kind of notation we have
	\begin{align}\label{eq:functional-trace-BD4}\nonumber
		& \hnabla_{\mu_1} \dots \hnabla_{\mu_{2n}} \frac{\id}{\,\,\hsquare^n} \delta(x,x')|^{\rm div}_{x'=x} = - \frac{i \, \log L^2 }{16 \, \pi^2} \, \frac{\sqrt{-g}}{2^{n-2} \, (n-1)!} \left[ \sum_{{\scriptscriptstyle i_1<\dots<i_4}} g^{(n-2)}_{\mu_1 \dots \hat{i}_1 \dots \mu_{2n}} \left( \lfloor \nabla_{i_1} \dots \nabla_{i_4} \Delta^{1/2} \rfloor \hid \right.\right.\\\nonumber
		& \left.\left. + \lfloor \hnabla_{i_1} \dots \hnabla_{i_4} \ha_0 \rfloor \right) + \sum_{{\scriptscriptstyle i_1<i_2}} \sum_{{\scriptscriptstyle j_1<j_2}} g^{(n-2)}_{\mu_1 \dots \mu_{2n}} \lfloor \nabla_{i_1} \nabla_{i_2} \Delta^{1/2} \rfloor \lfloor \hnabla_{j_1} \hnabla_{j_2} \ha_0 \rfloor + \frac{1}{2} \sum_{i<j} g^{(n-1)}_{\mu_1 \dots \mu_{2n}} \left( \lfloor \hnabla_i \hnabla_j \ha_1 \rfloor \right.\right.\\\nonumber
		& \left.\left. + \lfloor \nabla_i \nabla_j \Delta^{1/2} \rfloor \lfloor \ha_1 \rfloor \right) - \sum_{i_1 < i_2} \sum_{{\scriptscriptstyle j_1<\dots<j_4}} g^{(n-3)}_{\mu_1 \dots \mu_{2n}} \lfloor \nabla_{j_1} \dots \nabla_{j_4} \sigma \rfloor \left( \lfloor \nabla_{i_1} \nabla_{i_2} \Delta^{1/2} \rfloor \id + \lfloor \hnabla_{i_1} \hnabla_{i_2} \ha_0 \rfloor \right)  \right.\\\nonumber
		& \left. + \frac{1}{2} \sum_{{\scriptscriptstyle i_1<\dots<i_4}} g^{(n-2)}_{\mu_1 \dots \mu_{2n}} \lfloor \nabla_{i_1} \dots \nabla_{i_4} \sigma \rfloor \lfloor \ha_1 \rfloor - \sum_{{\scriptscriptstyle i_1<\dots<i_6}} g^{(n-3)}_{\mu_1\dots \mu_{2n}} \lfloor \nabla_{i_1} \dots \nabla_{i_6} \sigma \rfloor \id \right.\\
		& \left. - \sum_{{\scriptscriptstyle i_1<\dots<i_4}} \sum_{{\scriptscriptstyle j_1<\dots<j_4}} g^{(n-4)}_{\mu_1 \dots \mu_{2n}} \lfloor \nabla_{i_1} \dots \nabla_{i_4} \sigma \rfloor \lfloor \nabla_{j_1} \dots \nabla_{j_4} \sigma \rfloor \id - \frac{1}{4} g^{(n)}_{\mu_1 \dots \mu_{2n}} \lfloor \ha_2 \rfloor  \right] \, .
	\end{align}
	
	All the functional traces that we have listed in Eq. \eqref{eq:all-functional-traces} fall into the three classes that we have just taken into account. Thus, their specific contribution to the derivation of the $1$-loop divergent terms can be found by specializing these general formulae.

	\section{Final result}\label{sect:final-result}
	
	In this section we wrap up our previous calculations, and we present the modification to the trace of the second Seeley-DeWitt coefficient due to non-minimal terms. Furthermore, we discuss how known results can be recovered as liming cases. In Sect.\ \ref{sect:applications} and we will show three practical applications of these results.
	
	First of all, by using Eqs.\ \eqref{eq:tr-log-to-HK}, \eqref{eq:HK-asympt-min}, \eqref{eq:seeley-dewitt-min} and \eqref{eq:div-part-numerical-factor} we can write down the divergent part due to the box operator as
	\begin{equation}
		{\rm Tr} \log \hsquare \bigg|_{{\rm div}} = - \frac{i\, \log L^2}{16 \pi^2} \int \sqrt{-g} \, {{\rm tr} \, \ha_2(\hsquare)} \, .
	\end{equation}
	Thus, to derive the modifications to ${\rm tr} \, \ha_2$ we need to factor out the divergent pre-factor, which is the reason why we have already singled out such a numerical factor in the functional traces computed in the previous section.
	
	Before delving into the details of out result, we take on a brief detour to discuss how we have extensively exploited computer algebra to perform most of our calculations. 
	
	\subsection{On our computational tools}\label{subsect:result-computer}
	
	Combining expressions like Eq.\ \eqref{eq:V2-part-general} with functional traces like Eq.\ \eqref{eq:functional-trace-BD2} is generally very lengthy, and numerical mistakes may occur. Therefore, we have opted for performing such a calculation in \texttt{Mathematica}, employing the extension \texttt{FieldsX} \cite{Frob:2020gdh} of the \texttt{xAct} packages \cite{Martin-Garcia:xAct,Martin-Garcia:xTensor,Martin-Garcia:2008ysv,Nutma:2013zea}. 
	
	First of all, we have used inner vector bundles to parameterize field space, which is done through the commands \texttt{DefVBundleWithMetric} and \texttt{DefVBundle}. Two such bundles are needed to derive the coincidence limits of Seeley-DeWitt coefficients and their derivatives, since they are bi-local tensors and covariant derivatives must be performed before letting $y \rightarrow x$.
	
	Covariant derivatives for differentiating vector bundle tensors can be defined using \texttt{DefCovD}. This is made compatible with the underlying (psuedo-)Riemannian structure and the vector bundle metric through the options \texttt{ExtendedFrom} and \texttt{FromMetric}. A drawback of these packages is that, even though we have used the aforementioned definitions and options, covariant differentiation of objects in the vector bundles is still not directly allowed. Problems are due to the automatic canonicalization of tensorial expressions, which is prevented by explicitly specifying the metrics that we have defined as those to be used by \texttt{ToCanonical}, i.e, employing the \texttt{UseMetricOnVBundle} option of the latter.
	
	Changing the options of \texttt{ToCanonical} also reduces the ability of \texttt{CollectTensors} to recognize like scalars contractions, which is biggest drawback that we have encountered using \texttt{FieldsX}. For example, $\hcalR^\mu{}_\nu \hcalR^{\nu}{}_\mu$ and $\hcalR^{\mu\nu} \hcalR_{\mu\nu}$ may not be seen to be the proportional to each other. A possible way to circumvent this issue is to define additional tensors like $\hat{derV}_\alpha{}^\mu$ and to substitute them whenever $\hnabla_\alpha \hV^\mu$ is present. After making sure that all the covariant derivatives that are present have the form $\hnabla_\mu = \id \nabla_\mu$, we can reset the \texttt{UseMetricOnVBundle} option of \texttt{ToCanonical}, and employ \texttt{CollectTensors} afterwards. Finally, by substituting back objects like $\hnabla_\alpha \hV^\mu$ we recover a simplified expression of the starting tensor quantity.
	
	Due to the generality and complicated nature of our calculation, the final expression of the divergent part of Eq.\ \eqref{eq:tr-log-F-formal} is still very long. Thus, in order to avoid typos in the transcription, we have chosen to directly print out the resulting scalar contraction from the code using $\texttt{TexPrint}$, and to attach the code itself to the arXiv version of the paper \cite{Sauro:2025nb}. This implies that the vector bundle indices are displayed explicitly as
	\begin{equation}
		\hcalR_{\mu\nu} \, \varphi \equiv \calR_{\mu\nu a}{}^{a1} \varphi_{a1} \, .
	\end{equation}
	In an analogous way, we have that
	\begin{align}
		\hK^{\mu\nu} \equiv K^{\mu\nu}{}_a{}^{a1} \, , \qquad \hV^\mu \equiv V^\mu{}_a{}^{a1} \, , \qquad \hE \equiv E_a{}^{a1} \, .
	\end{align}
	To specialize our general result to specific tensor fields, we will then introduce soldering forms like $\theta^{\mu \nu}{}_a$ for a $2$-tensor. In this way
	\begin{equation}
		\varphi_a = \theta^{\mu\nu}{}_a \phi_{\mu\nu} \, ,
	\end{equation}
	where $\phi_{\mu\nu}$ is the quantum field whose fluctuations we are taking into account. Then, we have that
	\begin{equation}
		\theta^{\mu\nu}{}_a \theta_{\rho\sigma}{}^a = \mathds{1}^{\mu\nu}{}_{\rho\sigma} \, ,
	\end{equation}
	where $\mathds{1}^{\mu\nu}{}_{\rho\sigma}$ is the identity on the (in general reducible) representation of ${\frak{gl}}(4)$ in which $\phi_{\mu\nu}$ takes values.
	
	\subsection{First part of the result}\label{subsect:result-integral}
	
	Having set the notation, we now focus on the first part of our result, i.e., that which is given by the integral over the $\zeta$ parameter. Such a contribution to the trace of the second Seeley-DeWitt coefficient is not too complicated when the assumptions of Subsect.\ \ref{subsect:tr-log-pric-part-simple} are valid, especially when it is compared to the most general case of Subsect.\ \ref{subsect:tr-log-pric-part-general}. This is because the functional traces needed to evaluate Eq.\ \eqref{eq:integral-part-general-simple-case} have fewer tensor structures than those of Eq.\ \eqref{eq:integral-part-general-complicated-case}. For this reason, here we shall only provide the complete result in the first case, deferring the reader to the Appendix \ref{subsect:app-final-result-integrand-general-case} for the general one.
	
	First of all, let us notice that $\hM$ is a second-order differential operator which also contains first-order parts. Thus, in general we have
	\begin{equation}
		\hM = \hM_2{}^{\mu\nu}{}_a{}^{a1} \hnabla_\mu \hnabla_\nu + \hM_3{}^{\mu}{}_a{}^{a1} \hnabla_\mu + {\rm b.t.} \, , 
	\end{equation}
	where the numerical subscripts indicate the mass-dimension of the tensors. Moreover, mass-dimension-four terms are always total-derivative endomorphisms, thence we discard them. Finally, $\hM_3$ can only enter the result when considering terms linear in the derivative interaction tensor $\hV$. 
	
	The part of Eq.\ \eqref{eq:integral-part-general-simple-case} that does not explicitly depend on the Riemann tensor and its contractions is given by
	{\small
		\begin{align}\label{eq:result-int-no-Riem-terms}\nonumber
			{\rm tr} \, \ha_2 \supset & \int_{0}^{1} d \log \zeta \left[ \tfrac{1}{24} \calR_{\beta \lambda a2a} \calR^{\beta \lambda}{}_{a1}{}^{a2} K^{\alpha }{}_{\alpha }{}^{aa1} -  \tfrac{1}{12} \calR_{\alpha }{}^{\lambda a2}{}_{a} \calR_{\beta \lambda a1a2} K^{\alpha \beta aa1} -  \tfrac{1}{12} \calR_{\alpha }{}^{\lambda}{}_{a1}{}^{a2} \calR_{\beta \lambda a2a} K^{\alpha \beta aa1}\right. \\\nonumber
			& \left.-  \tfrac{1}{8} \calR_{\beta \lambda a2a} K^{\alpha \beta aa1} M_2{}_{\alpha }{}^{\lambda}{}_{a1}{}^{a2} -  \tfrac{1}{8} \calR_{\beta \lambda a2a} K^{\alpha \beta aa1} M_2{}^{\lambda}{}_{\alpha a1}{}^{a2} -  \tfrac{1}{8} \calR_{\alpha \beta a2a} K^{\alpha \beta aa1} M_2{}^{\lambda}{}_{\lambda a1}{}^{a2} \right. \\\nonumber
			& \left. -  \tfrac{1}{8} \calR_{\alpha \lambda a2a} K^{\alpha \beta aa1} M_2{}^{\lambda}{}_{\beta a1}{}^{a2} -  \tfrac{1}{8} \calR_{\alpha \lambda a2a} K^{\alpha \beta aa1} M_2{}_{\beta }{}^{\lambda}{}_{a1}{}^{a2} -  \tfrac{1}{8} \calR_{\beta \lambda a2a} K^{\alpha }{}_{\alpha }{}^{aa1} M_2{}^{\beta \lambda}{}_{a1}{}^{a2} \right. \\\nonumber		
			& \left. -  \tfrac{1}{6} \calR_{\mu}{}^{\nu}{}_{a2a} K^{\alpha \beta aa1}  M_2{}^{\lambda \mu}{}_{a1}{}^{a2} g^{(2)}{}_{\alpha \beta \lambda \nu} -  \tfrac{1}{6} \calR_{\lambda}{}^{\nu}{}_{a2a} K^{\alpha \beta aa1}  M_2{}^{\lambda \mu}{}_{a1}{}^{a2} g^{(2)}{}_{\alpha \beta \mu \nu} \right. \\
			& \left. + \tfrac{1}{192} K^{\alpha \beta aa1}  M_2{}^{\lambda \mu}{}_{a1}{}^{a2} M_2{}^{\nu \rho}{}_{a2a} g^{(3)}{}_{\alpha \beta \lambda \mu \nu \rho} \right] \, .
		\end{align}
	}
	On the other hand, the part that comprises both $\hM_2$ and the Riemann tensor (or ts contractions) takes the following form
	{\small
		\begin{align}\label{eq:result-int-no-FRiem-terms}
			{\rm tr} \, \ha_2 \supset & \int_{0}^{1} d \log \zeta \left[ -  \tfrac{1}{24} K^{\alpha \beta aa1} M_2{}^{\lambda}{}_{\beta a1a} R_{\alpha \lambda} -  \tfrac{1}{24} K^{\alpha \beta aa1} M_2{}_{\beta}{}^{\lambda}{}_{a1a} R_{\alpha \lambda} -  \tfrac{1}{24} K^{\alpha \beta aa1} M_2{}^{\lambda}{}_{\lambda a1a} R_{\alpha \beta } \right. \\\nonumber
			& \left. + \tfrac{1}{12} K^{\alpha \beta aa1} M_2{}^{\lambda \mu}{}_{a1a} R_{\lambda}{}^{\nu} g^{(2)}{}_{\alpha \beta \mu \nu} + \tfrac{1}{12} K^{\alpha \beta aa1} M_2{}^{\lambda \mu}{}_{a1a} R_{\mu}{}^{\alpha 11} g^{(2)}{}_{\alpha \beta \lambda \nu} -  \tfrac{1}{24} K^{\alpha \beta aa1} M_2{}_{\alpha }{}^{\lambda}{}_{a1a} R_{\beta \lambda} \right. \\\nonumber
			& \left. -  \tfrac{1}{24} K^{\alpha \beta aa1} M_2{}^{\lambda}{}_{\alpha a1a} R_{\beta \lambda} -  \tfrac{1}{24} K^{\alpha}{}_{\alpha }{}^{aa1} M_2{}^{\beta \lambda}{}_{a1a} R_{\beta \lambda} + \tfrac{1}{12} K^{\alpha \beta aa1} M_2{}^{\lambda \mu}{}_{a1a} R_{\alpha \mu \beta \lambda} \right.\\\nonumber
			& \left.  + \tfrac{1}{12} K^{\alpha \beta aa1} M_2{}^{\lambda \mu}{}_{a1a} R_{\alpha \lambda \beta \mu} -  \tfrac{1}{6} K^{\alpha \beta aa1} M_2{}^{\lambda \mu}{}_{a1a} R_{\lambda}{}^{\nu}{}_{\mu}{}^{\rho}  g^{(2)}{}_{\alpha \beta \nu \rho} + \tfrac{1}{48} K^{\alpha \beta aa1} M_2{}^{\lambda \mu}{}_{a1a} R g^{(2)}{}_{\alpha \beta \lambda \mu} \right] \, .
		\end{align}
	}
	Finally, the terms that depend neither on $\hcalR$ nor on $\hM$ are
	{\small
		\begin{align}\label{eq:result-int-no-Riem-M-terms}\nonumber
			{\rm tr} & \, \ha_2 \supset  \int_{0}^{1} d \log \zeta \left[ \tfrac{1}{45} K^{\alpha \beta a}{}_{a} R_{\alpha }{}^{\lambda} R_{\beta \lambda} -  \tfrac{1}{360} K^{\alpha }{}_{\alpha }{}^{a}{}_{a} R_{\beta \lambda} R^{\beta \lambda} -  \tfrac{1}{12} \calR_{\alpha \beta a1a} K^{\alpha \beta aa1} R - \tfrac{1}{36} K^{\alpha \beta a}{}_{a} R_{\alpha \beta } R \right. \\
			& \left. + \tfrac{1}{144} K^{\alpha }{}_{\alpha }{}^{a}{}_{a} R^2 -  \tfrac{1}{90} K^{\alpha \beta a}{}_{a} R_{\alpha }{}^{\lambda \mu \nu} R_{\beta \lambda \mu \nu}  -  \tfrac{1}{90} K^{\alpha \beta a}{}_{a} R^{\lambda \mu} R_{\alpha \lambda \beta \mu} + \tfrac{1}{360} K^{\alpha }{}_{\alpha }{}^{a}{}_{a} R_{\beta \lambda \mu \nu} R^{\beta \lambda \mu \nu} \right] \, .
		\end{align}
	}
	
	\subsection{Second part of the result}\label{subsect:result-non-integral}
	
	Now we focus on the part of the result that does not depend on the integration over the auxiliary parameter $\zeta$, i.e., Eq.\ \eqref{eq:non-integral-part-general}. As we have done in Sect.\ \ref{sect:non-min-op}, we split the final expression as a series in $\hV^\mu \equiv V^\mu{}_a{}^{a1}$.
	
	\subsubsection{Order $\hV^0$ terms}
	
	The contributions that do not depend on $\hV$ come from Eq.\ \eqref{eq:endomorphism-part-general}. They give rise to few scalar contractions when compared to the other ones, and they read
	{\small
		\begin{align}\label{eq:result-end-terms}
			{\rm tr} \ha_2\supset & \, - \tfrac{1}{6} K^{\alpha \beta aa1} R_{\alpha \beta } E_{a1a} + \tfrac{1}{12} K^{\alpha }{}_{\alpha }{}^{aa1} R E_{a1a} + \tfrac{1}{24} K^{\alpha \beta aa1} M_2{}^{\mu \nu}{}_{a1}{}^{a2} E_{a2a} g^{(2)}{}_{\alpha \beta \mu \nu} \\\nonumber
			& -  \tfrac{1}{2} \calR_{\alpha \beta a1a2} K^{\alpha \beta aa1} E^{a2}{}_{a} + \tfrac{1}{48} K^{\alpha \beta aa1} K^{\mu \nu a2a3}  E_{a1a2} E_{a3a} g^{(2)}{}_{\alpha \beta \mu \nu} \, .
		\end{align}
	}
	Thus, if one is interested in deriving the back-reaction of some matter fields on the pseudo-Riemannian structure, then, if the assumption of Subsect.\ \ref{subsect:tr-log-pric-part-simple} are met, it is sufficient to take into account this last equation, together with Eqs.\ \eqref{eq:result-int-no-Riem-terms}, \eqref{eq:result-int-no-FRiem-terms} and \eqref{eq:result-int-no-Riem-M-terms}.
	
	\subsubsection{Order $\hV^1$ terms}
	
	There are many terms which are linear in $\hV$, so we further divide them. Moreover, we choose to write down these terms so that no covariant derivative is acting on $\hV$. The contractions that involve the Ricci tensor and the Ricci scalar are
	{\small
		\begin{align}\label{eq:result-V1-Ricci}
			{\rm tr} \ha_2\supset & \, - \tfrac{1}{12} K^{\alpha \beta aa1} V^{\mu}{}_{a1a} \nabla_{\alpha }R_{\beta \mu} + \tfrac{1}{24} K^{\alpha \beta aa1} V_{\beta a1a} \nabla_{\alpha }R -  \tfrac{1}{12} K^{\alpha \beta aa1} V^{\mu}{}_{a1a} \nabla_{\beta}R_{\alpha \mu} \\\nonumber
			& + \tfrac{1}{24} K^{\alpha \beta aa1} V_{\alpha a1a} \nabla_{\beta}R + \tfrac{1}{24} K^{\alpha }{}_{\alpha }{}^{aa1} V^{\beta}{}_{a1a} \nabla_{\beta}R -  \tfrac{1}{12} K^{\alpha \beta aa1} V^{\mu}{}_{a1a} \nabla_{\mu}R_{\alpha \beta} \, ,
		\end{align}
	}
	while those that are linear in the generalized curvature $\hcalR_{\mu\nu}$ are
	{\small
		\begin{align}\label{eq:result-V1-FRiem}
			{\rm tr} \, \ha_2\supset & \, \tfrac{1}{3} K^{\alpha \beta aa1} V^{\mu a2}{}_{a} \hnabla_{\alpha }\calR_{\beta \mu a1a2} -  \tfrac{1}{3} K^{\alpha \beta aa1} V^{\mu a2}{}_{a} \hnabla_{\mu}\calR_{\alpha \beta a1a2} + \tfrac{1}{12} K^{\alpha \beta aa1} V_{\beta}{}^{a2}{}_{a} \hnabla^{\mu}\calR_{\alpha \mu a1a2} \\\nonumber
			&  + \tfrac{1}{12} K^{\alpha \beta aa1} V_{\alpha }{}^{a2}{}_{a} \hnabla^{\mu}\calR_{\beta \mu a1a2}  + \tfrac{1}{12} K^{\alpha }{}_{\alpha }{}^{aa1} V^{\beta a2}{}_{a} \hnabla^{\mu}\calR_{\beta \mu a1a2} \, .
		\end{align}
	}
	Finally, the remaining terms are given in the Appendix \ref{subect:app:final-result-V1}.

	\subsubsection{Order $\hV^2$ terms}
	
	There is a huge number of tensor structures which are quadratic in $\hV$. For this reason, in the main text we only present those involve the endomorphism $\hE$ and $\hM$. These can be written in the following simple form
		\begin{align}\label{eq:result-V2-terms-End-M2}
			{\rm tr} \, \ha_2  \supset & - \tfrac{1}{1920} K^{\alpha \beta aa1} K^{\mu \nu a2a4} K^{\rho \sigma a3a5} V^{\lambda}{}_{a1a2} V^{\tau}{}_{a4a3} E_{a5a} \, g^{(4)}{}_{\alpha \beta \mu \nu \rho \sigma \lambda \tau} \\\nonumber
			& - \tfrac{1}{1920} K^{\alpha \beta aa1} K^{\mu \nu a2a4} M_2{}^{\rho \sigma}{}_{a1}{}^{a3} V^{\lambda}{}_{a4a} V^{\tau}{}_{a3a2} \, g^{(4)}{}_{\alpha \beta \mu \nu \rho \sigma \lambda \tau}  \, .
		\end{align}
	All the other scalar contractions are listed in subsect.\ \ref{subect:app:final-result-V2} of App.\ \ref{sect:app:final-result-remaining-terms}.

	\subsubsection{Order $\hV^3$ terms}

	The number of independent tensor structures drastically decreases when we move on to consider contributions which are cubic in $\hV$. Indeed, they can be written as
		\begin{align}\label{eq:result-V3-terms}
			{\rm tr} \, \ha_2  \supset & - \tfrac{1}{5760} K^{\alpha \beta a}{}_{a1} K^{\mu \nu}{}_{a2}{}^{a5} K_{\rho}{}^{\sigma a3a4} V^{\lambda}{}_{a5a3} V^{\tau}{}_{a4a} \hnabla^{\rho}V^{\kappa a1a2} \, g^{(4)}{}_{\alpha \beta \mu \nu \sigma \lambda \tau \kappa} \\\nonumber
			& -  \tfrac{1}{2880} K^{\alpha \beta aa1} K^{\mu \nu}{}_{a2}{}^{a5} K_{\rho}{}^{\sigma a3}{}_{a4} V^{\lambda}{}_{a1a3} V^{\tau}{}_{a5a} \hnabla^{\rho}V^{\kappa a4a2} \, g^{(4)}{}_{\alpha \beta \mu \nu \sigma \lambda \tau \kappa}\\\nonumber
			& - \tfrac{1}{5760} K^{\alpha \beta a}{}_{a1} K^{\mu \nu}{}_{a2}{}^{a5} K^{\rho}{}_{\sigma}{}^{a3a4} V^{\lambda}{}_{a5a3} V^{\tau}{}_{a4a} \hnabla^{\sigma}V^{\kappa a1a2} \, g^{(4)}{}_{\alpha \beta \mu \nu \rho \lambda \tau \kappa} \\\nonumber
			& - \tfrac{1}{2880} K^{\alpha \beta aa1} K^{\mu \nu}{}_{a2}{}^{a5} K^{\rho}{}_{\sigma}{}^{a3}{}_{a4} V^{\lambda}{}_{a1a3} V^{\tau}{}_{a5a} \hnabla^{\sigma}V^{\kappa a4a2} \, g^{(4)}{}_{\alpha \beta \mu \nu \rho \lambda \tau \kappa}\\\nonumber
			& - \tfrac{1}{2880} K^{\alpha \beta aa1} K^{\mu \nu a2}{}_{a5} K^{\rho \sigma}{}_{a3}{}^{a4} V^{\lambda}{}_{a4a} V_{\tau a1a2} \hnabla^{\tau}V^{\kappa a5a3} \, g^{(4)}{}_{\alpha \beta \mu \nu \rho \sigma \lambda \kappa} \\\nonumber
			& -  \tfrac{1}{5760} K^{\alpha \beta aa1} K^{\mu \nu a2}{}_{a5} K^{\rho \sigma}{}_{a3}{}^{a4} V^{\lambda}{}_{a1a2} V_{\tau a4a} \hnabla^{\tau}V^{\kappa a5a3} \, g^{(4)}{}_{\alpha \beta \mu \nu \rho \sigma \lambda \kappa}\\\nonumber
			& + \tfrac{1}{5760} K^{\alpha \beta aa1} K^{\mu \nu a2}{}_{a5} K^{\rho \sigma}{}_{a3}{}^{a4} V^{\lambda}{}_{a1a2} V^{\tau}{}_{a4a} \hnabla^{\gamma}V^{\kappa a5a3} \, g^{(5)}{}_{\alpha \beta \mu \nu \rho \sigma \lambda \tau \kappa \gamma}  \, .
		\end{align}
	
	\subsubsection{Order $\hV^4$ terms}
	
	Finally, terms that are quartic in $\hV$ can be written in the simple fashion
	{\small
		\begin{align}\label{eq:result-V4-terms}
			{\rm tr} \, \ha_2  \supset \tfrac{1}{1290240} K^{\alpha \beta aa1} K^{\mu \nu a2a3} K^{\rho \sigma a4a5} K^{\lambda \kappa a6a7} V^{\tau}{}_{a1a2} V^{\gamma}{}_{a3a4} V^{\eta}{}_{a5a6} V^{\theta}{}_{a7a} \, g^{(6)}{}_{\alpha \beta \mu \nu \rho \sigma \lambda \kappa \tau \gamma \eta \theta} \, .
		\end{align}
	}
	
	\subsection{Limiting case}
	
	Let us take a look at our non-minimal second-order operator \eqref{eq:non-min-op}. If we switch off the non-minimal part $\hN$ the resulting operator is of Laplace type, and it has the derivative interaction which has been considered by Obukhov in \cite{Obukhov:1983mm}. In this limit the first part of our general result (that which is given by the integral over $\zeta$) automatically vanishes, see Subsect.\ \ref{subsect:result-integral}, while $K^{\mu\nu}{}_a{}^{a1} \rightarrow g^{\mu\nu} \delta_a{}^{a1}$, see Eq.\ \eqref{eq:def-K(n)}. Thus, the expansion performed in Sect.\ \ref{sect:non-min-op} must yield the same result given in \cite{Obukhov:1983mm} for the trace of the second Seeley-DeWitt coefficient. Indeed, we find complete agreement with \cite{Obukhov:1983mm} up to boundary terms, which is consistent since we have discarded surface terms from the onset.
	
	\subsection{Some words on other computational schemes}
	
	Recently, a new computational approach towards non-minimal differential operators was proposed in \cite{Barvinsky:2025jbw}, whose main concern is to derive the low-energy part of the heat kernel in an exact way. It relies on the introduction of spin-projectors, which are employed in a different manner with respect to previous studies \cite{Martini:2023apm}. Indeed, in flat space these projectors allow to express the untraced heat kernel as a sum over the spin-parity eigenspaces. When the background geometry is curved the projectors are replaced by quasi-projectors, and an appropriate subtraction procedure is introduced to deal with the non-locality of them. A systematic perturbative expansion in the background dimension is then carried out on the \emph{full} heat kernel, which is free of IR pathologies.
	
	When some simple operators are considered and a judicious choice of the endomorphism is made the approach of \cite{Barvinsky:2025jbw} yields an exact expression for the full untraced heat kernel. However, for more complicated differential operators this technique requires to deal with many integrations over proper time of some complicated nested commutators. The final expressions shall still be non-perturbative in the proper time, but they do not allow to implement the final result on a computer to yield an easy-to-use expression. Nevertheless, this result represents a cornerstone in the derivation of infrared properties of $1$-loop effective action stemming from non-minimal second order operators.
	
	On the other hand, the approach that we have followed in this paper was to focus on the traced heat kernel coefficients, and in particular on the second one. By doing this and employing the results of Sect.\ \ref{sect:funct-tr} on universal functional traces we were able to provide a model-independent expression for this coefficient. Furthermore, the open-source implementation in \texttt{Mathematica} of this Seeley-DeWitt coefficient significantly contributes to the applicability of our results. Finally, no subtraction procedure is needed in this case, since we are only interested in the UV divergent part of the effective action. Therefore, this paper and \cite{Barvinsky:2025jbw} provide two complementary results on the heat kernel of non-minimal second-order operators.

	\section{Applications}\label{sect:applications}
	
	In this section we shall present three applications of the results derived in the previous section. Though these examples will only involve bosonic quantum fields, our formulae can also be applied to fermionic higher spins. Concretely, our applications will involve a vector field, an antisymmetric $2$-tensor (Kalb-Ramond), and a vector-valued $2$-form (like the torsion). However, only in the first case we also take into account derivative interaction, thus fully using the results of Subsect.\ \ref{subsect:result-non-integral}. For simplicity none of our examples falls into the most-general case dealt with in Subsect.\ \ref{subsect:tr-log-pric-part-general}.
	
	\subsection{Vector field}\label{subsect:appl-vector}
	
	Let us consider a Lagrangian for a vector field $A_\mu$ which comprises a non-minimal kinetic term, as well as cubic and quartic self-interactions. Specifically, we parameterize it as
	\begin{align}\label{eq:lagrangian-vector}
		\mathcal{L} = - \tfrac{1}{2} \left( (\nabla_\mu A_\nu)^2 - \lambda (\nabla_\mu A^\mu)^2 \right) - \tfrac{b}{2} A^\mu A^\nu R_{\mu\nu} - \tfrac{c}{2} A^2 \, R -\tfrac{f}{3} A^\mu A^\nu \nabla_\mu A_\nu - \tfrac{g}{4} (A^\mu A_\mu)^2 \, .
	\end{align}
	For example, we may think of this Lagrangian to describe the torsion vector \cite{Sauro:2022chz,Sauro:2022hoh,Paci:2024ohq} or Weyl vectors \cite{Ghilencea:2018dqd} in a metric-affine geometry. Alternatively, such kind of dynamics may also arise by considering vector fields coupled to scalars as in \cite{Buchbinder:2024fwy}. The beta functions of higher-derivative gravity due to a non-minimal vector field's quantum fluctuations were also computed in \cite{Piva:2023noi}: the author did not consider the cubic interaction, but mixed loops of gravitons and vectors were taken into account. The resulting second-order differential operator which acts on contra-variant vectors $A^\nu$ takes the form
	\begin{equation}\label{eq:operator-vector}
		F^\mu{}_\nu (\nabla) = \square \delta^\mu{}_\nu - \lambda \nabla^\mu \nabla_\nu  - \tfrac{f}{3} \left( A_\nu \nabla^\mu - A^\mu \nabla_\nu \right) - b R^\mu{}_\nu  - c R \delta^\mu{}_\nu - g A^\alpha A_\alpha \delta^\mu{}_\nu - 2 g A^\mu A_\nu \, .
	\end{equation}
	When dealing with vectors we automatically have up to one non-minimal term, thus we can apply the simplified result of Subsect.\ \ref{subsect:tr-log-pric-part-simple}.
	
	In particular, the explicit expressions of the tensors that parameterized all the differential operators are
	\begin{subequations}\label{eqs:tensors-vector-case}
		\begin{align}
			& \calR_{\mu\nu}{}^\alpha{}_\beta = R^\alpha{}_{\beta\mu\nu} \, ;\\
			& K^{\alpha \beta \,\, \mu}{}_\nu = g^{\alpha\beta} \delta^\mu{}_\nu - \frac{\lambda}{\lambda-1} g^{\alpha\mu} \delta^\beta{}_\nu \, ;\\
			& M_2{}^{\alpha\beta \,\, \mu}{}_\nu = - \lambda g^{\alpha\mu} R^\beta{}_\nu - \frac{\lambda}{\lambda-1} R^{\alpha\mu} \delta^\beta{}_\nu \, ;\\
			& M_3{}^{\alpha \,\, \mu}{}_\nu = - \lambda \nabla^\mu R^\alpha{}_\nu - \frac{\lambda}{2} g^{\alpha\mu} \nabla_\nu R \, ;\\
			& E^\mu{}_\nu = - b R^\mu{}_\nu - c R \delta^\mu{}_\nu - g A^\alpha A_\alpha \delta^\mu{}_\nu - 2 g A^\mu A_\nu \, ;\\
			& V^{\alpha \,\, \mu}{}_\nu = \frac{f}{3} \left( A^\mu \delta^\alpha{}_\nu - g^{\alpha\mu} A_\nu \right) \, .
		\end{align}
	\end{subequations}
	In analogy with Eq.\ \eqref{eq:operator-vector}, we fix the ambiguity in ordering the covariant derivatives of $\hK$ by putting that which acts on $A^\nu$ to the right. The auxiliary integration parameter $\zeta$ is introduced by simply letting $\lambda \rightarrow \lambda \zeta$. Then, the final form of the trace of the second Seeley-DeWitt coefficient is found by inserting Eqs.\ \eqref{eqs:tensors-vector-case} into Eq.s \eqref{eq:result-int-no-FRiem-terms}, \eqref{eq:result-int-no-Riem-M-terms}, \eqref{eq:result-int-no-Riem-terms}, \eqref{eq:result-end-terms}, \eqref{eq:result-V1-FRiem}, \eqref{eq:result-V1-Ricci}, \eqref{eq:result-V1-terms-E-M}, \eqref{eq:result-V2-terms-End-M2}, \eqref{eq:result-V2-terms-FRiem}, \eqref{eq:result-V2-terms-Riem}, \eqref{eq:result-V2-terms-der2}, \eqref{eq:result-V3-terms} and \eqref{eq:result-V4-terms} and computing the $\zeta$ integrals. Furthermore, we parameterize the result in terms of $\gamma=\frac{\lambda}{1+\lambda}$. Accordingly, we have that
	{\small
	\begin{align}\label{eq:result-vector}
		& {\rm tr \, \ha_2} = \bigl(\tfrac{5}{1296} (1 + \gamma)^2 f^4 -  \tfrac{1}{9} (3 + 4 \gamma + \gamma^2) f^2 g + \tfrac{1}{4} (24 + 12 \gamma + 5 \gamma^2) g^2\bigr) A_{\alpha } A^{\alpha } A_{\beta} A^{\beta} \\\nonumber
		& + \tfrac{1}{432} \Bigl(\bigl(\gamma^2 (-5 + 4 b) - 4 \gamma (3 + 5 b) - 4 (5 + 6 b)\bigr) f^2 + 72 \bigl(\gamma^2 (-1 + b) + 12 b + \gamma (-4 + 6 b)\bigr) g\Bigr) A^{\alpha } A^{\beta} R_{\alpha \beta} \\\nonumber
		& + \tfrac{1}{360} \bigl(-8 + 15 \gamma (2 + \gamma) - 30 \gamma (4 + \gamma) b + 15 (12 + 6 \gamma + \gamma^2) b^2\bigr) R_{\alpha \beta} R^{\alpha \beta} \\\nonumber
		& + \left(- \tfrac{1}{216} (1 + \gamma) \bigl(-8 + 6 b + 36 c + \gamma (-5 + 5 b + 18 c)\bigr) f^2 + \tfrac{1}{6} \left(6 (-1 + b + 6 c) + \gamma^2 (-2 + 2 b + 9 c) \right.\right.\\\nonumber
		& \left.\left. + \gamma (-5 + 3 b + 18 c)\right) g\right) A_{\alpha } A^{\alpha } R + \tfrac{1}{144} \left(12 \gamma (-1 + b + 2 c) (-1 + 6 c) + 8 (-1 + 6 c) (-1 + 3 b + 6 c) \right.\\\nonumber
		& \left. + 3 \gamma^2 \bigl(1 + b^2 - 12 c + 24 c^2 + 2 b (-1 + 6 c)\bigr)\right) R^2 -  \tfrac{11}{180} R_{\alpha \beta \mu \nu} R^{\alpha \beta \mu \nu} \\\nonumber
		& + \tfrac{1}{36} \bigl(2 - 2 \gamma b + \gamma^2 (-2 + 2 b + 9 c)\bigr) f A^{\alpha } \nabla_{\alpha }R + \tfrac{1}{54} \bigl(- (1 + 3 \gamma + 2 \gamma^2) f^3 + 9 \gamma (4 + 5 \gamma) f g\bigr) A^{\alpha } A^{\beta} \nabla_{\beta}A_{\alpha } \\\nonumber
		& + \tfrac{1}{216} (4 + 4 \gamma + 7 \gamma^2) f^2 \nabla_{\alpha }A^{\alpha } \nabla_{\beta}A^{\beta} + \tfrac{1}{432} (-20 - 20 \gamma + \gamma^2) f^2 \nabla_{\beta}A_{\alpha } \nabla^{\beta}A^{\alpha }  \, .
	\end{align}}
	By setting $\gamma=0$ we recover the same result that can be found using the squaring procedure outlined in \cite{Obukhov:1983mm}.
	
	In \cite{Barvinsky:1985an} a different method is used for dealing with a special case of Eq.\ \eqref{eq:operator-vector}, i.e., when $c=f=g=0$ and $b=1$, which is tailored for gauge theories. In such a special case the dependence on the non-minimal parameter $\gamma$ identically drops out, which is a consequence of the Ward identity. In the presence of derivative- and self-interactions we observe that choosing $c=0$ and $b=1$ no longer suffices to annihilate the $\gamma$-dependence of the trace of $\ha_2$. Thus, for any such a vector theory the counter-terms stemming from graviton and mixed graviton-vector loops must be taken into account to address the gauge invariance of the effective action, or to prove the presence of a quantum anomaly.
	
	\subsection{Kalb-Ramond field}\label{subsect:appl-Kalb}

	Let us now consider a Kalb-Ramond field $B_{\mu\nu}=B_{[\mu\nu]}$ \cite{Kalb:1974yc}. In analogy with the previous example there exists a single divergence for this field, i.e., $\nabla_\mu B^{\mu\nu}$. Accordingly, the Lagrangian only depends on a single non-minimal parameter $b$, and it can be written as
	\begin{equation}\label{eq:lagrangian-kalb-ramond}
		\mathcal{L} = - \frac{1}{2} \left( \nabla_\alpha B_{\mu\nu} \nabla^\alpha B^{\mu\nu} + b \nabla_\mu B^{\mu\nu} \nabla_\alpha B^\alpha{}_\nu \right) + \frac{1}{2} B^{\mu\nu} E_{\mu\nu\alpha\beta} B^{\alpha\beta} \, ,
	\end{equation}
	where $E_{\mu\nu\alpha\beta}$ is an endomorphism. As a specific example one may consider the Lagrangian that arises by integrating some fermionic matter fields, and one may consider the quantum effective action associated to the Weyl anomaly in the presence of a Kalb-Ramond background as in \cite{Shapiro:2025pmn}. By letting $b \rightarrow b \, \zeta$ we obtain a non-minimal operator whose dependence on the auxiliary parameter $\zeta$ is linear, and which reads
	\begin{equation}\label{eq:op-kalb-ramond}
		F^{\mu\nu}{}_{\alpha\beta} = \square \, \mathds{1}^{\mu\nu}{}_{\alpha\beta} - b \zeta \nabla^{[\mu} \nabla_{[\beta} \delta^{\nu]}{}_{\alpha]} + E^{\mu\nu}{}_{\alpha\beta} \, ,
	\end{equation}
	where $\mathds{1}^{\mu\nu}{}_{\alpha\beta} = \frac{1}{2} \left( \delta^\mu{}_\alpha \delta^\nu{}_\beta - \delta^\nu{}_\alpha \delta^\mu{}_\beta \right)$ is the identity and the operator is acting on a quantum field bearing contra-variant indices. We first specialize to momentum space and invert the principal symbol, and then we calculate $\hM_2$, finding
	\begin{subequations}\label{eqs:tensors-kalb-ramond}
		\begin{align}
			& \calR_{\mu\nu}{}^{\alpha\beta}{}_{\rho\sigma} = \tfrac{1}{2} \delta^{\beta }{}_{\sigma } R^{\alpha }{}_{\rho \mu \nu } -  \tfrac{1}{2} \delta^{\beta }{}_{\rho } R^{\alpha }{}_{\sigma \mu \nu } -  \tfrac{1}{2} \delta^{\alpha }{}_{\sigma } R^{\beta }{}_{\rho \mu \nu } + \tfrac{1}{2} \delta^{\alpha }{}_{\rho } R^{\beta }{}_{\sigma \mu \nu } \, ; \\
			& D^{\alpha\beta \,\, \mu\nu}{}_{\rho\sigma} = - \tfrac{1}{2} \delta^{\mu }{}_{\sigma } \delta^{\nu }{}_{\rho } g^{\alpha \beta } + \tfrac{1}{2} \delta^{\mu }{}_{\rho } \delta^{\nu }{}_{\sigma } g^{\alpha \beta } - \tfrac{1}{4} b \zeta \delta^{\beta }{}_{\sigma } \delta^{\nu }{}_{\rho } g^{\alpha \mu } + \tfrac{1}{4} b \zeta \delta^{\beta }{}_{\rho } \delta^{\nu }{}_{\sigma } g^{\alpha\mu }\\\nonumber
			& \qquad\qquad\qquad + \tfrac{1}{4} b \zeta \delta^{\beta }{}_{\sigma } \delta^{\mu }{}_{\rho } g^{\alpha \nu } -  \tfrac{1}{4} b \zeta \delta^{\beta }{}_{\rho } \delta^{\mu}{}_{\sigma } g^{\alpha \nu } \, ;\\
			& K^{\alpha\beta \,\, \mu\nu}{}_{\rho\sigma} =  \tfrac{1}{2} \delta^{\mu }{}_{\rho } \delta^{\nu }{}_{\sigma } g^{\alpha \beta } - \tfrac{1}{2} \delta^{\mu }{}_{\sigma } \delta^{\nu }{}_{\rho } g^{\alpha \beta } + \frac{b \zeta}{4 + 2 b \zeta} \left( \delta^{\beta }{}_{\sigma } \delta^{\nu }{}_{\rho } g^{\alpha \mu } -  \delta^{\beta }{}_{\rho } \delta^{\nu }{}_{\sigma } g^{\alpha \mu } \right.\\\nonumber
			& \qquad\qquad\qquad \left. -  \delta^{\beta }{}_{\sigma } \delta^{\mu }{}_{\rho } g^{\alpha \nu } + \delta^{\beta }{}_{\rho } \delta^{\mu }{}_{\sigma } g^{\alpha \nu } \right) \, ;\\\nonumber
			& M_2{}^{\alpha\beta \,\, \mu\nu}{}_{\rho\sigma} = \frac{1}{8 + 4 b \zeta} \left[b \zeta \left(\delta^{\beta }{}_{\sigma } (2 \delta^{\nu }{}_{\rho } R^{\alpha \mu } - 2 \delta^{\mu }{}_{\rho } R^{\alpha \nu } + b \zeta g^{\alpha \nu } R^{\mu }{}_{\rho } -  b \zeta g^{\alpha \mu } R^{\nu }{}_{\rho } + 4 R^{\alpha \mu \nu }{}_{\rho } - 4 R^{\alpha \nu \mu }{}_{\rho }) \right.\right.\\\nonumber
			& \qquad\qquad\qquad \left.\left. + \delta^{\beta }{}_{\rho } (-2 \delta^{\nu }{}_{\sigma } R^{\alpha \mu } + 2 \delta^{\mu }{}_{\sigma } R^{\alpha \nu } -  b \zeta g^{\alpha \nu } R^{\mu }{}_{\sigma } + b \zeta g^{\alpha \mu } R^{\nu }{}_{\sigma } - 4 R^{\alpha \mu \nu }{}_{\sigma } + 4 R^{\alpha \nu \mu }{}_{\sigma }) \right.\right.\\\nonumber
			& \qquad\qquad\qquad \left.\left. + (2 + b \zeta) \left(\delta^{\nu }{}_{\sigma } g^{\alpha \mu } R^{\beta }{}_{\rho } -  \delta^{\mu }{}_{\sigma } g^{\alpha \nu } R^{\beta }{}_{\rho } -  \delta^{\nu }{}_{\rho } g^{\alpha \mu } R^{\beta }{}_{\sigma } + \delta^{\mu }{}_{\rho } g^{\alpha \nu } R^{\beta }{}_{\sigma } \right.\right.\right.\\
			& \qquad\qquad\qquad \left.\left.\left. + 2 g^{\alpha \nu } R^{\beta \mu }{}_{\rho \sigma } - 2 g^{\alpha \mu } R^{\beta \nu }{}_{\rho \sigma }\right)\right) \right] \, .
		\end{align}
	\end{subequations}
	As in the previous example, we have decided to sort the covariant derivatives such that the rightmost one generates divergences of $B^{\rho\sigma}$. Both $\hK$ and $\hM_2$ are singular for $b \zeta = -2$, which is exactly the value for which the Lagrangian \eqref{eq:lagrangian-kalb-ramond} propagates only the transverse modes. Due to the linear dependence on $\zeta$ in $D^{\alpha\beta \,\, \mu\nu}{}_{\rho\sigma}$ this example falls into the simplified category of non-minimal operators dealt with in Subsect.\ \ref{subsect:tr-log-pric-part-simple}. Thus, to find contribution to the trace of $\ha_2$ due to the non-minimal part of the operator we have to insert the tensors written in the previous equations \eqref{eqs:tensors-kalb-ramond} into Eqs.\ \eqref{eq:result-int-no-FRiem-terms}, \eqref{eq:result-int-no-Riem-M-terms} and \eqref{eq:result-int-no-Riem-terms}. Performing the $\zeta$ integral we find
	\begin{equation}\label{eq:integrand-result-KR}
		{\rm tr}\, \ha_2 \supset - \frac{b^2}{4 (2 + b)^2} R_{\alpha \beta} R^{\alpha \beta} + \frac{(-8 + b) b}{24 (2 + b)^2} R^2 + \frac{b^2}{8 (2 + b)^2} R_{\alpha \beta \mu \nu} R^{\alpha \beta \mu \nu} \, .
	\end{equation}
	Next, we consider the remaining part of the trace log of the operator \eqref{eq:op-kalb-ramond}, i.e., that which depends on the endomorphism. By using Eq.\ \eqref{eq:result-end-terms} we obtain the contribution to the trace of the second Seeley-DeWitt coefficient, and the complete form of the trace of this coefficient is given by
	\begin{align}\label{eq:final-result-KR}
		{\rm tr}\, \ha_2 = & \frac{b^2}{12 (2 + b)^2} E_{\alpha \mu \beta \nu} E^{\alpha \beta \mu \nu} + \frac{24 + 12 b + b^2}{12 (2 + b)^2} E^{\alpha\beta\mu\nu} E_{\mu\nu\alpha\beta} + \frac{b^2}{12 (2 + b)^2} E^{\alpha\beta}{}_{\alpha}{}^{\mu} E_\mu{}^\nu{}_{\beta\nu} \\\nonumber
		& -  \frac{b (8 + b)}{6 (2 + b)^2} E_{\alpha}{}^\mu{}_{\beta\mu} R^{\alpha\beta} - \left(\frac{1}{45} +  \frac{b^2}{4 (2 + b)^2}\right) R_{\alpha\beta} R^{\alpha\beta} + \frac{8 + 4 b + b^2}{12 (2 + b)^2} E^{\alpha\beta}{}_{\alpha\beta} R \\\nonumber
		& + \frac{16 - 8 b + 7 b^2}{72 (2 + b)^2} R^2 + \frac{b (4 + b)}{4 (2 + b)^2} E^{\alpha\beta\mu\nu} R_{\alpha\beta\mu\nu} + \left( \frac{b^2}{8 (2 + b)^2} - \frac{13}{90} \right) R_{\alpha\beta\mu\nu} R^{\alpha\beta\mu\nu}
	\end{align}
	The gauge theory of a Kalb-Ramond field is written in terms of the square of the field strength $\nabla_{[\mu} B_{\nu\rho]}$. The endomorphism that is obtained by choosing a Lagrangian quadratic in this field strength is
	\begin{equation}\label{eq:gauge-end-KR}
		E^{\mu\nu}{}_{\alpha\beta} = - \tfrac{1}{2} \delta_{\beta }{}^{\nu } R_{\alpha }{}^{\mu } + \tfrac{1}{2} \delta_{\beta }{}^{\mu } R_{\alpha }{}^{\nu } + \tfrac{1}{2} \delta_{\alpha }{}^{\nu } R_{\beta }{}^{\mu } -  \tfrac{1}{2} \delta_{\alpha }{}^{\mu } R_{\beta }{}^{\nu } + R_{\alpha \beta }{}^{\mu \nu } \, .
	\end{equation}
	In analogy with what is found for electrodynamics \cite{Barvinsky:1985an}, when this specific form of the endomorphism is plugged in Eq.\ \eqref{eq:final-result-KR} the dependence on the non-minimal parameter $b$ identically drops out, yielding
	\begin{equation}\label{eq:gauge-final-result-KR}
		{\rm tr}\, \ha_2 = - \frac{46}{45} R_{\alpha \beta} R^{\alpha \beta} + \frac{2}{9} R^2 + \frac{16}{45} R_{\alpha \beta \mu \nu} R^{\alpha \beta \mu \nu} \, .
	\end{equation}
	Thus, the gauge theory of a Kalb-Ramond field on top of a curved pseudo-Riemannian manifold is not anomalous.
	
	\subsection{Torsion field: A simple toy model}\label{subsect:appl-torsion}
	
	Let us now introduce the final example, i.e., a simplified model of dynamical torsion. The latter is a vector-valued two-form $T^\rho{}_{\mu\nu} = T^\rho{}_{[\mu\nu]}$, which is customarily decomposed into its algebraically irreducible components \cite{Shapiro:2001rz}. These are given by a vector, an antisymmetric $3$-tensor and a hook-antisymmetric tracefree tensor.
	
	Such an algebraic decomposition has been employed in \cite{Melichev:2025hcg} to derive logarithmically divergent contributions to the effective action due to torsion and graviton fluctuations. Decomposing the torsion in this way may provide a clear algorithm to fathom the possible realization of asymptotic safety in higher-derivative gravity, but it is by no means necessary. Indeed, a second-order operator that comprises all the nine kinetic terms for the torsion can also be written in terms of a principal part parameterized by five free parameters and a configuration-space metric with two unknown coefficients, see Eq.\ \eqref{eq:operator-from-hessian} and \cite{Barvinsky:1985an,Vilkovisky:1984st}. We argue that the latter approach is more straightforward if one wants to analyze the most general torsion theory. In this case, one may expect that the simplifying assumption of Subsect.\ \ref{subsect:tr-log-pric-part-simple} does not hold, thus using the result of Subsect.\ \ref{subsect:tr-log-pric-part-general} will be necessary.
	
	While a systematic derivation of the phase diagram in the most general metric-torsion theory would settle a common agreement regarding the physical properties of these theories, this is not the aim of the present discussion. Indeed, our aim is to provide a simple example for showing how to treat the presence of non-minimal kinetic terms in torsion theories, and to point out some phenomenological consequences of taking them into account. To this end, we consider the Lagrangian
	\begin{equation}\label{eq:lagrangian-torsion}
		\mathcal{L} = - \frac{1}{2} \left[ T_\rho{}^{\mu\nu} \square T^\rho{}_{\mu\nu} + c \nabla^\alpha T^\rho{}_{\alpha\beta} \nabla_\lambda T_\rho{}^{\lambda\beta} - T^\rho{}_{\mu\nu} E_\rho{}^{\mu\nu}{}_\lambda{}^{\alpha\beta} T^\lambda{}_{\alpha\beta} \right] \, .
	\end{equation}
	This comprises a single non-minimal term, which is parameterized by $c$, and we shall assume that $-2<c<0$. Furthermore, in part of the principal symbol which is proportional to the box operator the indices of the torsion are contracted in the simplest way, thus the configuration metric Eq.\ \eqref{eq:operator-from-hessian} is trivial. For the time being we do not assume any specific form of the endomorphism, but we make the following observation. Let us consider one of the possible ``field strengths" of the torsion, i.e., ${\cal{F}}^{\rho}{}_{\lambda\mu\nu}= \nabla_{[\lambda} T^{\rho}{}_{\mu\nu]}$. This type of gauge approach towards a vector-valued $2$-form has been used in flat-space in \cite{Curtright:1980yk}. Thus, a gauge-like origin of the previous Lagrangian Eq.\ \eqref{eq:lagrangian-torsion} is obtained by considering
	\begin{equation}\label{eq:lagrangian-gauge-torsion}
		\mathcal{L} = - \frac{3}{2} {\cal{F}}^{\rho}{}_{\lambda\mu\nu} {\cal{F}}_{\rho}{}^{\lambda\mu\nu}
	\end{equation}
	in a non-minimal gauge. Then, if we insist to start from the latter Lagrangian, the resulting endomorphism is
	\begin{align}\label{eq:end-gauge-torsion}
		E_\rho{}^{\mu\nu}{}_\lambda{}^{\alpha\beta} = \, & \tfrac{1}{2} g^{\beta \nu } g_{\lambda \rho } R^{\alpha \mu } -  \tfrac{1}{2} g^{\beta \mu } g_{\lambda \rho } R^{\alpha \nu } -  \tfrac{1}{2} g^{\alpha \nu } g_{\lambda \rho } R^{\beta \mu } + \tfrac{1}{2} g^{\alpha \mu } g_{\lambda \rho } R^{\beta \nu } -  g_{\lambda \rho } R^{\alpha \beta \mu \nu } \\\nonumber
		& -  \tfrac{1}{2} g^{\beta \nu } R^{\alpha \mu }{}_{\lambda \rho } + \tfrac{1}{2} g^{\beta \mu } R^{\alpha \nu }{}_{\lambda \rho } + \tfrac{1}{2} g^{\alpha \nu } R^{\beta \mu }{}_{\lambda \rho } -  \tfrac{1}{2} g^{\alpha \mu } R^{\beta \nu }{}_{\lambda \rho } \, .
	\end{align}
	Let us go back to the starting Lagrangian Eq.\ \eqref{eq:lagrangian-torsion}. The second-order differential operator which is associated to it reads as
	\begin{align}\label{eq:operator-torsion}
		F^\rho{}_{\mu\nu \,\lambda}{}^{\alpha\beta} = \square \, \mathds{1}^\rho{}_{\mu\nu \,\lambda}{}^{\alpha\beta} + c \, \delta^\rho{}_\lambda \delta^{[\alpha}{}_{[\mu} \nabla^{\beta]} \nabla_{\nu]} - E^\rho{}_{\mu\nu\,}{}_\lambda{}^{\alpha\beta} \, ,
	\end{align}
	where $\mathds{1}^\rho{}_{\mu\nu \,\lambda}{}^{\alpha\beta}= \delta^\rho{}_\lambda \delta_{[\mu}{}^{\alpha} \delta_{\nu]}{}^{\beta}$ is the identity in the space of vector-valued $2$-forms. We introduce the auxiliary parameter $\zeta$ by letting $c \rightarrow c \, \zeta$ as in the previous example, and we read off the principal symbol and its inverse in momentum space. The expressions of these tensors and of the generalized curvature are
	\begin{subequations}\label{eqs:tensors-torsion}
		\begin{align}
			& \calR_{\mu_1\mu_2\,\,}{}^\lambda{}_{\alpha\beta \,\,\rho}{}^{\mu\nu} =\tfrac{1}{2} \delta_{\beta }{}^{\nu } \delta^{\lambda }{}_{\rho } R_{\alpha }{}^{\mu \mu_1 \mu_2} - \tfrac{1}{2} \delta_{\beta }{}^{\mu } \delta^{\lambda }{}_{\rho } R_{\alpha }{}^{\nu \mu_1 \mu_2} -  \tfrac{1}{2} \delta_{\alpha }{}^{\nu } \delta^{\lambda }{}_{\rho } R_{\beta }{}^{\mu \mu_1 \mu_2} + \tfrac{1}{2} \delta_{\alpha }{}^{\mu } \delta^{\lambda }{}_{\rho } R_{\beta }{}^{\nu \mu_1 \mu_2}\\\nonumber
			& \qquad\qquad\qquad - \tfrac{1}{2} \delta_{\alpha }{}^{\nu } \delta_{\beta }{}^{\mu } R^{\lambda }{}_{\rho }{}^{\mu_1 \mu_2} + \tfrac{1}{2} \delta_{\alpha }{}^{\mu } \delta_{\beta }{}^{\nu } R^{\lambda }{}_{\rho }{}^{\mu_1 \mu_2} \, ;\\
			& D^{\mu_1\mu_2\,\,}{}^\lambda{}_{\alpha\beta \,\,\rho}{}^{\mu\nu} = \tfrac{1}{4} c \, \zeta \, \delta^{\lambda }{}_{\rho } \left(- \delta_{\alpha }{}^{\nu } \delta_{\beta }{}^{\mu_1} g^{\mu \mu_2} + \delta_{\alpha }{}^{\mu } \delta_{\beta }{}^{\mu_1} g^{\mu_2 \nu } + \delta_{\alpha }{}^{\mu_1} (\delta_{\beta }{}^{\nu } g^{\mu \mu_2} -  \delta_{\beta }{}^{\mu } g^{\mu_2 \nu })\right) \, ;\\
			& K^{\mu_1\mu_2\,\,}{}^\lambda{}_{\alpha\beta \,\,\rho}{}^{\mu\nu} = \frac{1}{4 + 2 c \zeta}\delta^{\lambda }{}_{\rho } \left[\delta_{\alpha }{}^{\nu } \left(c \zeta \delta_{\beta }{}^{\mu_1} g^{\mu \mu_2} -  (2 + c \zeta) \delta_{\beta }{}^{\mu } g^{\mu_1 \mu_2}\right) \right.\\\nonumber
			& \qquad\qquad\qquad \left. + c \zeta \delta_{\alpha }{}^{\mu_1} (- \delta_{\beta }{}^{\nu } g^{\mu \mu_2} + \delta_{\beta }{}^{\mu } g^{\mu_2 \nu }) + \delta_{\alpha }{}^{\mu } \left((2 + c \zeta) \delta_{\beta }{}^{\nu } g^{\mu_1 \mu_2} -  c \zeta \delta_{\beta }{}^{\mu_1} g^{\mu_2 \nu }\right)\right] \, .
		\end{align}
	\end{subequations}
	Once again, we have sorted the covariant derivatives to act with divergences on $T^\rho{}_{\mu\nu}$. Clearly, the last equation becomes singular as $c \, \zeta \rightarrow - 2$. Next, we promote momentum-space expressions to coordinate-based ones, and we compute the $\hM_2$ differential operator, whose tensor coefficients are given by the following equation
	\begin{align}\label{eq:M2-torsion}
		M_2{}^{\mu_1\mu_2\,\,}{}^\lambda{}_{\alpha\beta \,\,\rho}{}^{\mu\nu} = & \frac{c \zeta}{8 + 4 c \zeta} \left[2 \delta_{\beta }{}^{\mu } \delta^{\lambda }{}_{\rho } g^{\mu_2 \nu } R_{\alpha }{}^{\mu_1} -  c \zeta \delta_{\alpha }{}^{\mu_1} \delta^{\lambda }{}_{\rho } g^{\mu_2 \nu } R_{\beta }{}^{\mu } + 2 \delta_{\alpha }{}^{\nu } \delta^{\lambda }{}_{\rho } g^{\mu \mu_2} R_{\beta }{}^{\mu_1} \right.\\\nonumber
		& \left. - 2 \delta_{\alpha }{}^{\mu } \delta^{\lambda }{}_{\rho } g^{\mu_2 \nu } R_{\beta }{}^{\mu_1} + c \zeta \delta_{\alpha }{}^{\mu_1} \delta^{\lambda }{}_{\rho } g^{\mu \mu_2} R_{\beta }{}^{\nu } - 2 \delta_{\alpha }{}^{\mu_1} \delta_{\beta }{}^{\mu } \delta^{\lambda }{}_{\rho } R^{\mu_2 \nu } \right.\\\nonumber
		& \left. -  c \zeta \delta_{\alpha }{}^{\mu_1} \delta_{\beta }{}^{\mu } \delta^{\lambda }{}_{\rho } R^{\mu_2 \nu } + 4 \delta^{\lambda }{}_{\rho } g^{\mu_2 \nu } R_{\alpha }{}^{\mu }{}_{\beta }{}^{\mu_1} - 4 \delta^{\lambda }{}_{\rho } g^{\mu_2 \nu } R_{\alpha }{}^{\mu_1}{}_{\beta }{}^{\mu } \right.\\\nonumber
		& \left. + 4 \delta^{\lambda }{}_{\rho } g^{\mu \mu_2} R_{\alpha }{}^{\mu_1}{}_{\beta }{}^{\nu } - 4 \delta_{\beta }{}^{\mu } g^{\mu_2 \nu } R_{\alpha }{}^{\mu_1 \lambda }{}_{\rho } - 4 \delta^{\lambda }{}_{\rho } g^{\mu \mu_2} R_{\alpha }{}^{\nu }{}_{\beta }{}^{\mu_1} \right.\\\nonumber
		& \left. + c \zeta \delta_{\alpha }{}^{\mu_1} g^{\mu_2 \nu } R_{\beta }{}^{\mu \lambda }{}_{\rho } - 4 \delta_{\alpha }{}^{\nu } g^{\mu \mu_2} R_{\beta }{}^{\mu_1 \lambda }{}_{\rho } + 4 \delta_{\alpha }{}^{\mu } g^{\mu_2 \nu } R_{\beta }{}^{\mu_1 \lambda }{}_{\rho } \right.\\\nonumber
		& \left. + 4 \delta_{\alpha }{}^{\mu_1} \delta^{\lambda }{}_{\rho } R_{\beta }{}^{\mu_2 \mu \nu } + 2 c \zeta \delta_{\alpha }{}^{\mu_1} \delta^{\lambda }{}_{\rho } R_{\beta }{}^{\mu_2 \mu \nu } -  c \zeta \delta_{\alpha }{}^{\mu_1} g^{\mu \mu_2} R_{\beta }{}^{\nu \lambda }{}_{\rho } \right.\\\nonumber
		& \left. + \delta_{\beta }{}^{\nu } \left(\delta^{\lambda }{}_{\rho } \left(-2 g^{\mu \mu_2} R_{\alpha }{}^{\mu_1} + (2 + c \zeta) \delta_{\alpha }{}^{\mu_1} R^{\mu \mu_2}\right) + 4 g^{\mu \mu_2} R_{\alpha }{}^{\mu_1 \lambda }{}_{\rho } \right.\right.\\\nonumber
		& \left.\left. + 2 (2 + c \zeta) \delta_{\alpha }{}^{\mu_1} R^{\lambda }{}_{\rho }{}^{\mu \mu_2}\right) - 2 c \zeta \delta_{\alpha }{}^{\mu_1} \delta_{\beta }{}^{\mu_2} R^{\lambda }{}_{\rho }{}^{\mu \nu } + 4 \delta_{\alpha }{}^{\mu_1} \delta_{\beta }{}^{\mu } R^{\lambda }{}_{\rho }{}^{\mu_2 \nu } \right.\\\nonumber
		& \left. + 2 c \zeta \delta_{\alpha }{}^{\mu_1} \delta_{\beta }{}^{\mu } R^{\lambda }{}_{\rho }{}^{\mu_2 \nu } + \delta_{\beta }{}^{\mu_1} \left(- c \zeta g^{\mu_2 \nu } R_{\alpha }{}^{\mu \lambda }{}_{\rho } + \delta^{\lambda }{}_{\rho } \left(c \zeta g^{\mu_2 \nu } R_{\alpha }{}^{\mu } -  c \zeta g^{\mu \mu_2} R_{\alpha }{}^{\nu } \right.\right.\right.\\\nonumber
		& \left.\left.\left. -  (2 + c \zeta) (\delta_{\alpha }{}^{\nu } R^{\mu \mu_2} -  \delta_{\alpha }{}^{\mu } R^{\mu_2 \nu } + 2 R_{\alpha }{}^{\mu_2 \mu \nu })\right) + c \zeta g^{\mu \mu_2} R_{\alpha }{}^{\nu \lambda }{}_{\rho } - 4 \delta_{\alpha }{}^{\nu } R^{\lambda }{}_{\rho }{}^{\mu \mu_2} \right.\right.\\\nonumber
		& \left.\left. - 2 c \zeta \delta_{\alpha }{}^{\nu } R^{\lambda }{}_{\rho }{}^{\mu \mu_2} + 2 c \zeta \delta_{\alpha }{}^{\mu_2} R^{\lambda }{}_{\rho }{}^{\mu \nu } - 4 \delta_{\alpha }{}^{\mu } R^{\lambda }{}_{\rho }{}^{\mu_2 \nu } - 2 c \zeta \delta_{\alpha }{}^{\mu } R^{\lambda }{}_{\rho }{}^{\mu_2 \nu }\right)\right] \, .
	\end{align}
	
	Now we have all the necessary ingredients to compute the trace of the second Seeley-DeWitt coefficient of the non-minimal operator Eq.\ \eqref{eq:operator-torsion}. First, we focus on the trace log of the non-minimal part of the operator \eqref{eq:operator-torsion}. To this end, we insert Eqs.\ \eqref{eqs:tensors-torsion} and \eqref{eq:M2-torsion} in the general result, which given by the sum of Eqs.\  \eqref{eq:result-int-no-FRiem-terms}, \eqref{eq:result-int-no-Riem-M-terms} and \eqref{eq:result-int-no-Riem-terms}. By performing the $\zeta$ integral we obtain the following contribution to the trace of $\ha_2$
	\begin{equation}
		{\rm tr}\, \ha_2 \supset - \frac{c^2}{(2 + c)^2} R_{\alpha \beta} R^{\alpha \beta} + \frac{(-8 + c) c}{6 (2 + c)^2} R^2 -  \frac{(-1 + c) c^2}{4 (2 + c)^2} R_{\alpha \beta \mu \nu} R^{\alpha \beta \mu \nu} \, .
	\end{equation}	
	Notice the expected singular behavior in the $c\rightarrow -2$ limit. Next, we consider the part of ${\rm tr} \, \ha_2$ which depends on the endomorphism, i.e., Eq.\ \eqref{eq:result-end-terms}. In this case the contribution to the trace of the $\ha_2$ coefficient is
	\begin{align}
		{\rm tr}\, \ha_2 \supset & \frac{c^2}{12 (2 + c)^2} E^{\alpha\beta\mu\nu\rho\sigma} E_{\nu\beta\rho\alpha\mu\sigma} + \frac{24 + 12 c + c^2}{12 (2 + c)^2} E^{\alpha\beta\mu\nu\rho\sigma} E_{\nu\rho\sigma\alpha\beta\mu} \\\nonumber
		& + \frac{c^2}{12 (2 + c)^2} E^{\alpha\beta\mu\nu}{}_\beta{}^\rho E_{\nu\rho}{}^\sigma{}_{\alpha\mu\sigma} -  \frac{c (8 + c)}{6 (2 + c)^2} E^{\mu}{}_\alpha{}^\nu{}_{\mu\beta\nu} R^{\alpha\beta} -  \frac{2}{15} R_{\alpha\beta} R^{\alpha\beta} \\\nonumber
		& + \frac{8 + 4 c + c^2}{12 (2 + c)^2} E^{\alpha\beta\mu}{}_{\alpha\beta\mu} R + \frac{1}{3} R^2 + \frac{c (4 + c)}{2 (2 + c)^2} E^{\alpha\beta\mu\nu}{}_\beta{}^\rho R_{\alpha\nu\mu\rho} -  \frac{31}{30} R_{\alpha\beta\mu\nu} R^{\alpha\beta\mu\nu} \\\nonumber
		& -  \frac{c (3 + 2 c)}{3 (2 + c)^2} E^{\alpha\beta\mu}{}_\alpha{}^{\nu\rho} R_{\beta\mu\nu\rho} + \frac{c (24 + 11 c)}{6 (2 + c)^2} 	E^{\alpha\beta\mu}{}_\alpha{}^{\nu\rho} R_{\beta\nu\mu\rho} \, .
	\end{align}
	Thus, the final result is given by
	\begin{align}\label{eq:final-result-torsion}
		{\rm tr}\, \ha_2 = & \frac{c^2}{12 (2 + c)^2} E^{\alpha\beta\mu\nu\rho\sigma} E_{\nu\beta\rho\alpha\mu\sigma} + \frac{24 + 12 c + c^2}{12 (2 + c)^2} E^{\alpha\beta\mu\nu\rho\sigma} E_{\nu\rho\sigma\alpha\beta\mu} \\\nonumber
		& + \frac{c^2}{12 (2 + c)^2} E^{\alpha\beta\mu\nu}{}_\beta{}^\rho E_{\nu\rho}{}^\sigma{}_{\alpha\mu\sigma} -  \frac{c (8 + c)}{6 (2 + c)^2} E^{\mu}{}_\alpha{}^\nu{}_{\mu\beta\nu} R^{\alpha\beta}  \\\nonumber
		& + \frac{8 + 4 c + c^2}{12 (2 + c)^2} E^{\alpha\beta\mu}{}_{\alpha\beta\mu} R  + \frac{c (4 + c)}{2 (2 + c)^2} E^{\alpha\beta\mu\nu}{}_\beta{}^\rho R_{\alpha\nu\mu\rho}  \\\nonumber
		& -  \frac{c (3 + 2 c)}{3 (2 + c)^2} E^{\alpha\beta\mu}{}_\alpha{}^{\nu\rho} R_{\beta\mu\nu\rho} + \frac{c (24 + 11 c)}{6 (2 + c)^2} 	E^{\alpha\beta\mu}{}_\alpha{}^{\nu\rho} R_{\beta\nu\mu\rho} \\\nonumber
		& + \left( \frac{(-8 + c) c}{6 (2 + c)^2} + \frac{1}{3} \right)  R^2 - \left( \frac{c^2}{(2 + c)^2} + \frac{2}{15} \right) R_{\alpha\beta} R^{\alpha\beta} - \left(  \frac{(-1 + c) c^2}{4 (2 + c)^2} + \frac{31}{30} \right)  R_{\alpha\beta\mu\nu} R^{\alpha\beta\mu\nu} \, .
	\end{align}
	Finally, we specialize the final expression of the trace of the $\ha_2$ coefficient of the differential operator Eq.\ \eqref{eq:operator-torsion} to the special case of the endomorphism given by Eq.\ \eqref{eq:end-gauge-torsion}. This yields
	\begin{equation}
		{\rm tr}\, \ha_2 = - \frac{62}{15} R_{\alpha \beta} R^{\alpha \beta} + R^2 + \frac{236 + 118 c - 15 c^2}{60 (2 + c)} R_{\alpha \beta \mu \nu} R^{\alpha \beta \mu \nu} \, .
	\end{equation}
	Therefore, in strike contrast with the examples of electrodynamics \cite{Barvinsky:1985an} and of the Kalb-Ramond field the dependence on the non-minimal parameter $c$ does not disappear. Accordingly, we cannot treat the Lagrangian Eq.\ \eqref{eq:lagrangian-gauge-torsion} as describing a gauge theory of a vector-valued $2$-form as a consistent quantum field theory on a curved background. Consequently, taking into account the quantum fluctuations of the metric field becomes mandatory in order to understand if this theory generates a quantum anomaly. However, due to the first Bianchi identities, we cannot couple the generalized field strength ${\cal{F}}^\rho{}_{\lambda\mu\nu}$ to the Riemann tensor, which yields a significantly different behavior with respect to that of the vector theory treated in Subsect.\ \ref{subsect:appl-vector}.

	\section{Discussion and outlooks}\label{sect:outro}
	
	In this paper we have derived for the first time a general expression for the local part of the trace of the second Seeley-DeWitt coefficient of a non-minimal operator in $d=4$. To this end we have employed the trick of \cite{Barvinsky:1985an} for rewriting the trace log of a non-minimal operator as the integral over an auxiliary parameter of a sum of functional traces. However, we have followed \cite{Melichev:2025hcg} in splitting the calculation into two different parts, in such a way that only the first one involves the integration over the auxiliary parameter. As we have discussed at the end of Sect.\ \ref{sect:non-min-op} such a split significantly improves the computational speed for two reasons. On the one hand, the integrals that have to be computed usually are way less that in the method of \cite{Barvinsky:1985an}; on the other hand, we generate symmetrized metrics $g^{(n)}$ only up to $n=6$, compared to $n=8$ in \cite{Barvinsky:1985an}'s method. Thus, we have obtained a reduction of a factor $\tfrac{1}{13 \times 15}$ in the number of generated tensor structure at order $\hV^4$, and analogous improvements appear at all orders in $\hV$. All of the model-independent formulae for the local part of the trace of the second Seeley-DeWitt coefficient are also attached to this paper in a \texttt{Mathematica} notebook \cite{Sauro:2025nb}.
	
	At the end of Sect.\ \ref{sect:final-result} we have discussed the complete compatibility of our result with that of \cite{Obukhov:1983mm} as we switch off the non-minimal terms. Due to the very structure of the trick by \cite{Barvinsky:1985an}, the present method allows to access only the trace of the heat kernel coefficients, thus one has to rely on different approaches if he wants to derive the untraced coefficients \cite{Groh:2011dw}. Even though we have limited our analysis to the local part of the trace of the second heat kernel coefficient $\ha_2$, in principle the algorithm presented here can also be employed for deriving other traced coefficients as well as the boundary terms.
	
	In Sect.\ \ref{sect:applications} we have applied the general result to three simple cases. Our findings are compatible with known results in the literature \cite{Barvinsky:1985an,Melichev:2025hcg,Sangy:2025jxk}. An interesting future development of these applications is to derive the divergent part of the effective action for higher-spins, torsion and non-metricity fields. This would provide a comprehensive picture of the radiative effects of these quantum fields on the high-energy behavior of higher-derivative gravity. In particular, it would be possible to analyze which parts of the physical parameter space are compatible with asymptotic safety (see, e.g., \cite{Melichev:2025hcg} for a recent effort in this direction). Part of this program was carried out in \cite{Piva:2021nyj}, though the author did not consider non-minimal terms for higher-spins, while the radiative (in)stability of torsion and non-metricty theories was studied in \cite{Marzo:2021iok}.

	As we have mentioned in the Introduction \ref{sect:intro}, there exists another way of deriving the trace log of non-minimal operators within the heat kernel method, which relies on spin-parity decompositions \cite{Groh:2011dw}. Such a program was carried out on a maximally symmetric background in \cite{Martini:2023apm} integrating out torsion fluctuations, employing the spin-parity decomposition developed in \cite{Martini:2023rnv}. In general, if the (bosonic) quantum field that we are integrating out has spin $s$, we will be required to know the precise form of the traced heat kernel coefficients of minimal operators up to order $2+2 \, s$. Therefore, while pursuing this route might be preferable when dealing with the torsion or non-metricity tensors, we argue that it will be less convenient for fields of spin $s\geq4$.

	\section{Acknowledgements}
	
	The author whishes to thank Oleg Melichev for useful comments and interesting discussions related to the topic of this paper.
	
	The work of the author is supported by a Della Riccia foundation grant. The author is also thankful to the Theoretisch-Physikalisches Institut of the Friedrich-Schiller-Universität of Jena for its hospitality and financial support.
	
	\appendix

	\section{Commutator identities}\label{sect:app:comm-ids}
	
	Let us exploit a Laplace transform to write down the inverse of the $n$th power of $\hsquare$, i.e.,
	\begin{equation}\label{eq:Lapl-tranf-box-minus-n}
		\frac{\hid}{\,\,\hsquare^n} = \frac{(-i)^n}{(n-1)!} \int_{0}^{\infty} ds \, s^{n-1} \,e^{i s \hsquare} \, .
	\end{equation}
	By using this integral representation we can trade the commutation of $\hsquare^{-1}$ with the commutation of $\hsquare$, thus keeping under control the mass-dimension of the operators that we are dealing with. Indeed, the following set of equations holds
	\begin{align}\label{eq:box^{-1}-hM}\nonumber
		\frac{\hid}{\hsquare} \hM = & - i \int_0^\infty ds \, e^{i s \hsquare} \, \hM = - i \int_0^\infty ds \left( \hM + \sum_{k\geq1} \frac{(is)^k}{k!} \hsquare^k \hM \right) \\\nonumber
		= & - i \int_0^\infty ds \left[ \hM + \sum_{k\geq1} \frac{(is)^k}{k!} \hM \hsquare^k + \sum_{k\geq1} \frac{(is)^k}{k!} k [\hsquare,\hM] \hsquare^{k-1} \right.\\\nonumber
		& \qquad\qquad\quad \left. + \sum_{k\geq2} \frac{(is)^k}{k!} \frac{k(k-1)}{2} [\hsquare,[\hsquare,\hM]] \hsquare^{k-2} + \dots \right] \\
		= & \hM \frac{\hid}{\hsquare} - [\hsquare,\hM] \frac{\id}{\,\,\hsquare^2} + [ \hsquare , [ \hsquare , \hM ]] \frac{\id}{\,\,\hsquare^3} + \dots \, ,
	\end{align}
	where the ellipsis indicates that we are discarding higher-dimensional terms. In doing so we are tacitly assuming that $\hM$ has dimension one, whence triple commutators would at most give rise to boundary terms.
	
	Therefore, the commutator of the inverse of $\hsquare$ with $\hM$ reads
	\begin{equation}\label{eq:comm-box^(-1)-M}
		\left[\frac{\hid}{\hsquare} ,\hM\right] = - [\hsquare,\hM] \frac{\id}{\,\,\hsquare^2} + [ \hsquare , [ \hsquare , \hM ]] \frac{\id}{\,\,\hsquare^3} + \dots \, .
	\end{equation}
	Finally, using the fact that
	\begin{equation}
		[A \, B , C] = A [B,C] + [B,C] A \, ,
	\end{equation}
	we find that
	\begin{equation}\label{eq:comm-box^(-2)-M}
		\left[\frac{1}{\,\,\hsquare^2},\hM \right] = - 2 [\hsquare,\hM] \frac{\id}{\,\,\hsquare^3} + 3 [ \hsquare , [ \hsquare , \hM ]] \frac{\id}{\,\,\hsquare^4} + \dots \, .
	\end{equation}

	\section{Coincidence limits}\label{sect:app:coinc-limits}
	
	In this section we present all the coincidence limits up to dimension-four operators of the Synge world function, the reduced Van-Vleck determinant and the Seeley-DeWitt coefficients. As we have done in the main text, we discard surface terms. The Seeley-DeWitt coefficients are those of the minimal operator
	\begin{equation}
		\hF(\hnabla) = \hsquare + \hE \, .
	\end{equation}
	
	\subsection{Covariant derivatives of the Synge world function}\label{subsect:app:coinc-limits-world-function}
	
	We start by considering the Synge world function. All these results are derived by covariant differentiating Eq. \eqref{eq:world-function}, sorting the covariant derivatives and taking the coincidence limit. First of all, we have
	\begin{equation}\label{eq:coinc-limit-sigma}
		\lfloor \sigma \rfloor = 0 \, ,
	\end{equation}
	which is consistent with the geometrical interpretation of Sect.\ \ref{sect:schwinger-dewitt}. Since $\sigma$ has dimension of length squared, dimensional analysis tells us that 
	\begin{equation}\label{eq:coinc-limit-n1sigma}
			\lfloor \nabla_\mu \sigma \rfloor = 0 \, .
	\end{equation}
	On the other hand, when acting with two covariant derivatives we find
	\begin{equation}\label{eq:coinc-limit-n2sigma}
		\lfloor \nabla_\mu \nabla_\nu \sigma \rfloor = g_{\mu\nu} \, .
	\end{equation}
	Again, since the Synge world function only depends on the background pseudo-Riemannian structure, we have that
	\begin{equation}\label{eq:coinc-limit-n3sigma}
		\lfloor \nabla_\mu \nabla_\nu\nabla_\rho \sigma \rfloor = 0 \, .
	\end{equation}
	Curvature invariants pop up as soon as we start applying four derivatives. In this case we obtain
	\begin{align}\label{eq:coinc-limit-n4sigma}
		\lfloor \nabla_{\mu_1} \dots \nabla_{\mu_4} \sigma \rfloor = - \tfrac{2}{3} R_{(\mu_3 |\mu_1|\mu_4)\mu_2} \, .
	\end{align}
	By acting with one more derivative we expect to find expressions proportional to the covariant derivatives of the Riemann tensor. Indeed, the precise form is
	\begin{align}\label{eq:coinc-limit-n5sigma}
		\lfloor \nabla_{\mu_1} \dots \nabla_{\mu_5} \sigma \rfloor = - \tfrac{1}{2} \left[ \nabla_{\mu_1} R_{(\mu_4 |\mu_2|\mu_5)\mu_3} + \nabla_{\mu_2} R_{(\mu_4|\mu_1|\mu_5)\mu_3} + \nabla_{\mu_3} R_{(\mu_4|\mu_1|\mu_5)\mu_2} \right] \, .
	\end{align}
	Finally, the most involved result from a computational viewpoint is that whose mass dimension is equal to four. By resorting to computer algebra we find
	\begin{align}\label{eq:coinc-limit-n6sigma}
		& \lfloor \nabla_{\mu_1} \nabla_{\mu_2} \nabla_{\mu_3} \nabla_{\mu_4} \nabla_{\mu_5} \nabla_{\mu_6} \sigma \rfloor =  \tfrac{1}{45} R_{\mu_1}{}^{\alpha }{}_{\mu_5 \mu_6} R_{\mu_2 \alpha \mu_3 \mu_4} -  \tfrac{1}{45} R_{\mu_1 \mu_5 \mu_6}{}^{\alpha } R_{\mu_2 \alpha \mu_3 \mu_4}\\\nonumber
		& -  \tfrac{1}{15} R_{\mu_1 \mu_6 \mu_5}{}^{\alpha } R_{\mu_2 \alpha \mu_3 \mu_4} -  \tfrac{1}{45} R_{\mu_1}{}^{\alpha }{}_{\mu_4 \mu_6} R_{\mu_2 \alpha \mu_3 \mu_5} -  \tfrac{4}{15} R_{\mu_1 \mu_4 \mu_6}{}^{\alpha } R_{\mu_2 \alpha \mu_3 \mu_5} -  \tfrac{1}{45} R_{\mu_1 \mu_6 \mu_4}{}^{\alpha } R_{\mu_2 \alpha \mu_3 \mu_5}\\\nonumber
		& - \tfrac{1}{45} R_{\mu_1}{}^{\alpha }{}_{\mu_4 \mu_5} R_{\mu_2 \alpha \mu_3 \mu_6} -  \tfrac{1}{15} R_{\mu_1 \mu_4 \mu_5}{}^{\alpha } R_{\mu_2 \alpha \mu_3 \mu_6} -  \tfrac{1}{45} R_{\mu_1 \mu_5 \mu_4}{}^{\alpha } R_{\mu_2 \alpha \mu_3 \mu_6} -  \tfrac{1}{45} R_{\mu_1}{}^{\alpha }{}_{\mu_3 \mu_6} R_{\mu_2 \alpha \mu_4 \mu_5}\\\nonumber
		& -  \tfrac{4}{15} R_{\mu_1 \mu_3 \mu_6}{}^{\alpha } R_{\mu_2 \alpha \mu_4 \mu_5} -  \tfrac{1}{45} R_{\mu_1 \mu_6 \mu_3}{}^{\alpha } R_{\mu_2 \alpha \mu_4 \mu_5} -  \tfrac{1}{45} R_{\mu_1}{}^{\alpha }{}_{\mu_3 \mu_5} R_{\mu_2 \alpha \mu_4 \mu_6} -  \tfrac{1}{15} R_{\mu_1 \mu_3 \mu_5}{}^{\alpha } R_{\mu_2 \alpha \mu_4 \mu_6}\\\nonumber
		& -  \tfrac{1}{45} R_{\mu_1 \mu_5 \mu_3}{}^{\alpha } R_{\mu_2 \alpha \mu_4 \mu_6} + \tfrac{1}{45} R_{\mu_1}{}^{\alpha }{}_{\mu_3 \mu_4} R_{\mu_2 \alpha \mu_5 \mu_6} + \tfrac{4}{15} R_{\mu_1 \mu_3 \mu_4}{}^{\alpha } R_{\mu_2 \alpha \mu_5 \mu_6} + \tfrac{1}{45} R_{\mu_1 \mu_4 \mu_3}{}^{\alpha } R_{\mu_2 \alpha \mu_5 \mu_6}\\\nonumber
		& + \tfrac{1}{15} R_{\mu_1}{}^{\alpha }{}_{\mu_5 \mu_6} R_{\mu_2 \mu_3 \mu_4 \alpha } + \tfrac{2}{15} R_{\mu_1 \mu_5 \mu_6}{}^{\alpha } R_{\mu_2 \mu_3 \mu_4 \alpha } -  \tfrac{1}{5} R_{\mu_1 \mu_6 \mu_5}{}^{\alpha } R_{\mu_2 \mu_3 \mu_4 \alpha } - \tfrac{1}{15} R_{\mu_1}{}^{\alpha }{}_{\mu_4 \mu_6} R_{\mu_2 \mu_3 \mu_5 \alpha }\\\nonumber
		& + \tfrac{2}{15} R_{\mu_1 \mu_6 \mu_4}{}^{\alpha } R_{\mu_2 \mu_3 \mu_5 \alpha } -  \tfrac{1}{15} R_{\mu_1}{}^{\alpha }{}_{\mu_4 \mu_5} R_{\mu_2 \mu_3 \mu_6 \alpha } -  \tfrac{1}{5} R_{\mu_1 \mu_4 \mu_5}{}^{\alpha } R_{\mu_2 \mu_3 \mu_6 \alpha } + \tfrac{2}{15} R_{\mu_1 \mu_5 \mu_4}{}^{\alpha } R_{\mu_2 \mu_3 \mu_6 \alpha }\\\nonumber
		& + \tfrac{1}{45} R_{\mu_1}{}^{\alpha }{}_{\mu_5 \mu_6} R_{\mu_2\mu_4 \mu_3 \alpha } -  \tfrac{1}{45} R_{\mu_1 \mu_5 \mu_6}{}^{\alpha } R_{\mu_2 \mu_4 \mu_3 \alpha } - \tfrac{4}{15} R_{\mu_1 \mu_6 \mu_5}{}^{\alpha } R_{\mu_2 \mu_4 \mu_3 \alpha } -  \tfrac{1}{15} R_{\mu_1}{}^{\alpha }{}_{\mu_3 \mu_6} R_{\mu_2 \mu_4 \mu_5 \alpha }\\\nonumber
		& + \tfrac{2}{15} R_{\mu_1 \mu_6 \mu_3}{}^{\alpha } R_{\mu_2 \mu_4 \mu_5 \alpha } - \tfrac{1}{15} R_{\mu_1}{}^{\alpha }{}_{\mu_3 \mu_5} R_{\mu_2 \mu_4 \mu_6 \alpha } -  \tfrac{1}{5} R_{\mu_1 \mu_3 \mu_5}{}^{\alpha } R_{\mu_2 \mu_4 \mu_6 \alpha } + \tfrac{2}{15} R_{\mu_1 \mu_5 \mu_3}{}^{\alpha } R_{\mu_2 \mu_4 \mu_6 \alpha }\\\nonumber
		& - \tfrac{1}{45} R_{\mu_1}{}^{\alpha }{}_{\mu_4 \mu_6} R_{\mu_2 \mu_5 \mu_3 \alpha } -  \tfrac{1}{15} R_{\mu_1 \mu_4 \mu_6}{}^{\alpha } R_{\mu_2 \mu_5 \mu_3 \alpha } -  \tfrac{1}{45} R_{\mu_1 \mu_6 \mu_4}{}^{\alpha } R_{\mu_2 \mu_5 \mu_3 \alpha } -  \tfrac{1}{45} R_{\mu_1}{}^{\alpha }{}_{\mu_3 \mu_6} R_{\mu_2 \mu_5 \mu_4 \alpha }\\\nonumber
		& -  \tfrac{1}{15} R_{\mu_1 \mu_3 \mu_6}{}^{\alpha } R_{\mu_2 \mu_5 \mu_4 \alpha } -  \tfrac{1}{45} R_{\mu_1 \mu_6 \mu_3}{}^{\alpha } R_{\mu_2 \mu_5 \mu_4 \alpha } - \tfrac{1}{45} R_{\mu_1}{}^{\alpha }{}_{\mu_3 \mu_4} R_{\mu_2 \mu_5 \mu_6 \alpha } + \tfrac{1}{3} R_{\mu_1 \mu_3 \mu_4}{}^{\alpha } R_{\mu_2 \mu_5 \mu_6 \alpha }\\\nonumber
		& -  \tfrac{1}{45} R_{\mu_1 \mu_4 \mu_3}{}^{\alpha } R_{\mu_2 \mu_5 \mu_6 \alpha } - \tfrac{1}{45} R_{\mu_1}{}^{\alpha }{}_{\mu_4 \mu_5} R_{\mu_2 \mu_6 \mu_3 \alpha } -  \tfrac{4}{15} R_{\mu_1 \mu_4 \mu_5}{}^{\alpha } R_{\mu_2 \mu_6 \mu_3 \alpha } -  \tfrac{1}{45} R_{\mu_1 \mu_5 \mu_4}{}^{\alpha } R_{\mu_2 \mu_6 \mu_3 \alpha }\\\nonumber
		& - \tfrac{1}{45} R_{\mu_1}{}^{\alpha }{}_{\mu_3 \mu_5} R_{\mu_2 \mu_6 \mu_4 \alpha } -  \tfrac{4}{15} R_{\mu_1 \mu_3 \mu_5}{}^{\alpha } R_{\mu_2 \mu_6 \mu_4 \alpha } -  \tfrac{1}{45} R_{\mu_1 \mu_5 \mu_3}{}^{\alpha } R_{\mu_2 \mu_6 \mu_4 \alpha } - \tfrac{1}{15} R_{\mu_1}{}^{\alpha }{}_{\mu_3 \mu_4} R_{\mu_2 \mu_6 \mu_5 \alpha }\\\nonumber
		& + \tfrac{2}{15} R_{\mu_1 \mu_4 \mu_3}{}^{\alpha } R_{\mu_2 \mu_6 \mu_5 \alpha } -  \tfrac{4}{45} R_{\mu_1}{}^{\alpha }{}_{\mu_2 \mu_6} R_{\mu_3 \alpha \mu_4 \mu_5} -  \tfrac{2}{15} R_{\mu_1 \mu_2 \mu_6}{}^{\alpha } R_{\mu_3 \alpha \mu_4 \mu_5} + \tfrac{2}{45} R_{\mu_1 \mu_6 \mu_2}{}^{\alpha } R_{\mu_3 \alpha \mu_4 \mu_5}\\\nonumber
		& -  \tfrac{4}{45} R_{\mu_1}{}^{\alpha }{}_{\mu_2 \mu_5} R_{\mu_3 \alpha \mu_4 \mu_6} -  \tfrac{2}{15} R_{\mu_1 \mu_2 \mu_5}{}^{\alpha } R_{\mu_3 \alpha \mu_4 \mu_6} + \tfrac{2}{45} R_{\mu_1 \mu_5 \mu_2}{}^{\alpha } R_{\mu_3 \alpha \mu_4 \mu_6} + \tfrac{4}{45} R_{\mu_1}{}^{\alpha }{}_{\mu_2 \mu_4} R_{\mu_3 \alpha \mu_5 \mu_6}\\\nonumber
		& + \tfrac{2}{15} R_{\mu_1 \mu_2 \mu_4}{}^{\alpha } R_{\mu_3 \alpha \mu_5 \mu_6} - \tfrac{2}{45} R_{\mu_1 \mu_4 \mu_2}{}^{\alpha } R_{\mu_3 \alpha \mu_5 \mu_6} -  \tfrac{1}{15} R_{\mu_1}{}^{\alpha }{}_{\mu_2 \mu_6} R_{\mu_3 \mu_4 \mu_5 \alpha } -  \tfrac{1}{15} R_{\mu_1 \mu_6 \mu_2}{}^{\alpha } R_{\mu_3 \mu_4 \mu_5 \alpha }\\\nonumber
		& - \tfrac{1}{15} R_{\mu_1}{}^{\alpha }{}_{\mu_2 \mu_5} R_{\mu_3 \mu_4 \mu_6 \alpha } -  \tfrac{1}{15} R_{\mu_1 \mu_5 \mu_2}{}^{\alpha } R_{\mu_3 \mu_4 \mu_6 \alpha } -  \tfrac{4}{45} R_{\mu_1}{}^{\alpha }{}_{\mu_2 \mu_6} R_{\mu_3 \mu_5 \mu_4 \alpha } -  \tfrac{2}{15} R_{\mu_1 \mu_2 \mu_6}{}^{\alpha } R_{\mu_3 \mu_5 \mu_4 \alpha }\\\nonumber
		& + \tfrac{2}{45} R_{\mu_1 \mu_6 \mu_2}{}^{\alpha } R_{\mu_3 \mu_5 \mu_4 \alpha } -  \tfrac{4}{45} R_{\mu_1}{}^{\alpha }{}_{\mu_2 \mu_4} R_{\mu_3 \mu_5 \mu_6 \alpha } + \tfrac{4}{15} R_{\mu_1 \mu_2 \mu_4}{}^{\alpha } R_{\mu_3 \mu_5 \mu_6 \alpha } + \tfrac{2}{45} R_{\mu_1 \mu_4 \mu_2}{}^{\alpha } R_{\mu_3 \mu_5 \mu_6 \alpha }\\\nonumber
		& -  \tfrac{4}{45} R_{\mu_1}{}^{\alpha }{}_{\mu_2 \mu_5} R_{\mu_3 \mu_6 \mu_4 \alpha } -  \tfrac{2}{15} R_{\mu_1 \mu_2 \mu_5}{}^{\alpha } R_{\mu_3 \mu_6 \mu_4 \alpha } + \tfrac{2}{45} R_{\mu_1 \mu_5 \mu_2}{}^{\alpha } R_{\mu_3 \mu_6 \mu_4 \alpha } -  \tfrac{1}{15} R_{\mu_1}{}^{\alpha }{}_{\mu_2 \mu_4} R_{\mu_3 \mu_6 \mu_5 \alpha }\\\nonumber
		& -  \tfrac{1}{15} R_{\mu_1 \mu_4 \mu_2}{}^{\alpha } R_{\mu_3 \mu_6 \mu_5 \alpha } + \tfrac{4}{45} R_{\mu_1}{}^{\alpha }{}_{\mu_2 \mu_3} R_{\mu_4 \alpha \mu_5 \mu_6} + \tfrac{2}{15} R_{\mu_1 \mu_2 \mu_3}{}^{\alpha } R_{\mu_4 \alpha \mu_5 \mu_6} -  \tfrac{2}{45} R_{\mu_1 \mu_3 \mu_2}{}^{\alpha } R_{\mu_4 \alpha \mu_5 \mu_6}\\\nonumber
		& - \tfrac{4}{45} R_{\mu_1}{}^{\alpha }{}_{\mu_2 \mu_3} R_{\mu_4 \mu_5 \mu_6 \alpha } + \tfrac{4}{15} R_{\mu_1 \mu_2 \mu_3}{}^{\alpha } R_{\mu_4 \mu_5 \mu_6 \alpha } + \tfrac{2}{45} R_{\mu_1 \mu_3 \mu_2}{}^{\alpha } R_{\mu_4 \mu_5 \mu_6 \alpha } - \tfrac{1}{15} R_{\mu_1}{}^{\alpha }{}_{\mu_2 \mu_3} R_{\mu_4 \mu_6 \mu_5 \alpha }\\\nonumber
		& -  \tfrac{1}{15} R_{\mu_1 \mu_3 \mu_2}{}^{\alpha } R_{\mu_4 \mu_6 \mu_5 \alpha }
	\end{align}
	
	\subsection{Covariant derivatives of the VanVleck-Morette determinant}\label{subsect:app:coinc-limits-VanVleck}
	
	Now we turn to the coincidence limits of covariant derivatives acting on the VanVleck determinant. In this case, it is customary to derive such relations acting on the square root of $\Delta$, whose defining equation can be read off from Eq.\ \eqref{eq:van-vleck-world-function}. By looking at Eq.\ \eqref{eq:van-vleck} we realize that it has mass dimension zero, while from its geometrical interpretation we expect its coincidence limit to be given by unity, i.e.,
	\begin{equation}\label{eq:coinc-limit-Delta}
		\lfloor \Delta^{1/2} \rfloor = 1 \, .
	\end{equation}
	Thus, dimensional analysis tells us that
	\begin{equation}\label{eq:coinc-limit-n1Delta}
		\lfloor \nabla_\mu \Delta^{1/2} \rfloor = 0 \, .
	\end{equation}
	When acting with at least two covariant derivatives we find coincidence limits that depend on the curvature invariants. Indeed, the first non-trivial result is 
	\begin{align}\label{eq:coinc-limit-n2Delta}
		\lfloor \nabla_\mu \nabla_\nu \Delta^{1/2} \rfloor = \tfrac{1}{6} R_{\mu\nu} \, .
	\end{align}
	By applying three derivatives and employing the differential Bianchi identities we find  
	\begin{align}\label{eq:coinc-limit-n3Delta}
		\lfloor \nabla_\mu \nabla_\nu \nabla_\rho \Delta^{1/2} \rfloor = \tfrac{1}{12} \left[ \nabla_\rho R_{\mu\nu} + \nabla_\mu R_{\nu\rho} + \nabla_\nu R_{\mu\rho} \right] \, .
	\end{align}
	Finally, for dimension-four terms we make use of computer algebra, and we obtain
	\begin{align}\label{eq:coinc-limit-n4Delta}
		& \lfloor \nabla_{\mu_1} \nabla_{\mu_2} \nabla_{\mu_3} \nabla_{\mu_4} \Delta^{1/2} \rfloor  =  \tfrac{1}{36} R_{\mu_1 \mu_4} R_{\mu_2 \mu_3} + \tfrac{1}{36} R_{\mu_1 \mu_3} R_{\mu_2 \mu_4} \\\nonumber
		& + \tfrac{1}{36} R_{\mu_1 \mu_2} R_{\mu_3 \mu_4} + \tfrac{1}{360} R_{\mu_4}{}^{\alpha } R_{\mu_1 \alpha \mu_2 \mu_3} + \tfrac{1}{360} R_{\mu_3}{}^{\alpha } R_{\mu_1 \alpha \mu_2 \mu_4} + \tfrac{7}{120} R_{\mu_4}{}^{\alpha } R_{\mu_1 \mu_2 \mu_3 \alpha }\\\nonumber
		& + \tfrac{7}{120} R_{\mu_3}{}^{\alpha } R_{\mu_1 \mu_2 \mu_4 \alpha } + \tfrac{1}{360} R_{\mu_4}{}^{\alpha } R_{\mu_1 \mu_3 \mu_2 \alpha} + \tfrac{1}{180} R_{\mu_2}{}^{\alpha } R_{\mu_1 \mu_3 \mu_4 \alpha } + \tfrac{1}{360} R_{\mu_3}{}^{\alpha } R_{\mu_1 \mu_4 \mu_2 \alpha }\\\nonumber
		& + \tfrac{1}{18} R_{\mu_2}{}^{\alpha } R_{\mu_1 \mu_4 \mu_3 \alpha } + \tfrac{1}{90} R_{\mu_1}{}^{\alpha }{}_{\mu_4}{}^{\beta} R_{\mu_2 \alpha \mu_3 \beta} -  \tfrac{1}{20} R_{\mu_1 \mu_4}{}^{\alpha \beta} R_{\mu_2 \alpha \mu_3 \beta} + \tfrac{1}{90} R_{\mu_1}{}^{\alpha }{}_{\mu_3}{}^{\beta} R_{\mu_2 \alpha \mu_4 \beta}\\\nonumber
		& + \tfrac{1}{90} R_{\mu_1}{}^{\alpha }{}_{\mu_4}{}^{\beta} R_{\mu_2 \beta \mu_3 \alpha } + \tfrac{1}{90} R_{\mu_1}{}^{\alpha }{}_{\mu_3}{}^{\beta} R_{\mu_2 \beta \mu_4 \alpha } + \tfrac{1}{40} R_{\mu_1 \mu_4}{}^{\alpha \beta} R_{\mu_2 \mu_3 \alpha \beta} + \tfrac{1}{180} R_{\mu_1}{}^{\alpha } R_{\mu_2 \mu_3 \mu_4 \alpha }\\\nonumber
		& - \tfrac{1}{20} R_{\mu_1}{}^{\alpha }{}_{\mu_3}{}^{\beta} R_{\mu_2 \mu_4 \alpha \beta} + \tfrac{1}{40} R_{\mu_1 \mu_3}{}^{\alpha \beta} R_{\mu_2 \mu_4 \alpha \beta} + \tfrac{1}{18} R_{\mu_1}{}^{\alpha } R_{\mu_2 \mu_4 \mu_3 \alpha } + \tfrac{1}{90} R_{\mu_1}{}^{\alpha }{}_{\mu_2}{}^{\beta} R_{\mu_3 \alpha \mu_4 \beta}\\\nonumber
		& + \tfrac{1}{90} R_{\mu_1}{}^{\alpha }{}_{\mu_2}{}^{\beta} R_{\mu_3 \beta \mu_4 \alpha } \, .
	\end{align}

	\subsection{Covariant derivatives of the Seeley-DeWitt coefficient}\label{subsect:app:coinc-limits-SDW-coeff}

	After having derived the coincidence limits of the Synge world function and VanVleck-Morette determinant up to dimension-four terms, we focus on the Seeley-DeWitt coefficients. The zeroth coefficient $\ha_0$ is interpreted as a parallel displacement operator \cite{Barvinsky:1985an}. As such, its covariant Taylor expansion only depends on the pseudo-Riemannian structure and on the generalized curvature $\hcalR_{\mu\nu}$, and it cannot depend on the specific differential operator that we are dealing with. This geometrical interpretation and the initial conditions of the Cauchy system Eq.\ \eqref{eq:schroedinger-cauchy-problem} yield 
	\begin{align}\label{eq:coinc-limit-a0}
		\lfloor  \ha_0 \rfloor = \id \, .
	\end{align}
	Accordingly, dimensional analysis directly tells us that
	\begin{align}\label{eq:coinc-limit-n1a0}
		\lfloor \hnabla_\mu \ha_0 \rfloor = 0 \, .
	\end{align}
	By applying one more derivative and using Eqs.\ \eqref{eq:coinc-limit-n1sigma} and \eqref{eq:coinc-limit-n2sigma} we find the first non-trivial result
	\begin{align}\label{eq:coinc-limit-n2a0}
		\lfloor \hnabla_\mu \hnabla_\nu \ha_0 \rfloor = \tfrac{1}{2} \hcalR_{\mu\nu} \, .
	\end{align}
	Acting with three derivatives on $\ha_0$ and taking the coincidence limit we find
	\begin{align}\label{eq:coinc-limit-n3a0}
		\lfloor \hnabla_\mu \hnabla_\nu \hnabla_\rho \ha_0 \rfloor = \tfrac{2}{3} \hnabla_{(\mu} \hcalR_{\nu)\rho} \, .
	\end{align}
	Finally, by considering the mass-dimension four terms we have
	\begin{align}\label{eq:coinc-limit-n4a0}
		\lfloor \hnabla_\mu \hnabla_\nu \hnabla_\rho \hnabla_\sigma \ha_0\rfloor =& \tfrac{1}{8} \left[ \hcalR_{\mu\nu} \hcalR_{\rho\sigma} + \hcalR_{\mu\rho} \hcalR_{\nu\sigma} + \hcalR_{\mu\sigma} \hcalR_{\nu\rho} + \hcalR_{\rho\sigma} \hcalR_{\mu\nu} + \hcalR_{\nu\sigma} \hcalR_{\mu\rho} + \hcalR_{\nu\rho} \hcalR_{\mu\sigma} \right] \\\nonumber
		& + \tfrac{1}{6} \left[ R^\lambda{}_{(\sigma\rho)\nu} \hcalR_{\mu\lambda} + R^\lambda{}_{(\sigma\rho)\mu} \hcalR_{\nu\lambda} + R^\lambda{}_{(\sigma\nu)\mu} \hcalR_{\rho\lambda} + R^\lambda{}_{(\rho\mu)\nu} \hcalR_{\sigma\lambda} \right] \, .
	\end{align}
	
	Next, we take into account the coincidence limits of $\ha_1$ and its derivatives up to dimension-four operators. By taking the coincidence limit of Eq.\ \eqref{eq:rec-rel-min} with $n=1$ and using Eq.\ \eqref{eq:coinc-limit-n2Delta} we find
	\begin{equation}\label{eq:coinc-limit-a1}
		\lfloor \ha_1 \rfloor = \tfrac{1}{6} \,R\, \id + \hE \, .
	\end{equation}
	Dimension-three terms arise by computing the covariant derivative of $\ha_1$ and taking the coincidence limit, and they read
	\begin{align}\label{eq:coinc-limit-n1a1}
		\lfloor \hnabla_\mu \ha_1 \rfloor = \tfrac{1}{12} \nabla_\mu R + \tfrac{1}{2} \hnabla_\mu \hat{E} + \tfrac{1}{6} \hnabla^\lambda \hcalR_{\mu\lambda}  \, .
	\end{align}
	Eventually, dimension-four terms arise when we act with one more derivative, and they are given by
	\begin{align}\label{eq:coinc-limit-n2a1}
		\lfloor \hnabla_\mu \hnabla_\nu \ha_1 \rfloor = & \tfrac{1}{90} \left[ R^{\alpha\beta} R_{\alpha\mu\beta\nu} + R_{\alpha\beta\gamma\mu} R^{\alpha\beta\gamma}{}_\nu - 2 R_{\mu\alpha} R^\alpha{}_\nu \right] \id \\\nonumber
		& + \tfrac{1}{12} R \hcalR_{\mu\nu} + \tfrac{1}{6} \hat{E} \hcalR_{\mu\nu} + \tfrac{1}{3} \hcalR_{\mu\nu} \hat{E} + \tfrac{1}{6} \hcalR_{(\mu}{}^\alpha \hcalR_{\nu)\alpha} \, ,
	\end{align}
	where we have discarded boundary terms.
	
	Lastly, we present the coincidence limit of the $\ha_2$ for the minimal operator with endomorphism $\hE$
	\begin{align}\label{eq:coinc-limit-a2}
		\lfloor \ha_2 \rfloor = \tfrac{1}{180} \left( R^{\mu\nu\rho\sigma} R_{\mu\nu\rho\sigma} - R^{\mu\nu} R_{\mu\nu} + \tfrac{5}{2} R \right) \id + \tfrac{1}{6} \hE\, R + \tfrac{1}{2} \hE^2 + \tfrac{1}{12} \hcalR^{\mu\nu} \hcalR_{\mu\nu} \, .
	\end{align}
	
	\section{Final result: The remaining terms}\label{sect:app:final-result-remaining-terms}
	
	\subsection{Order $V^1$ terms}\label{subect:app:final-result-V1}
	
	The terms that are linear in $\hV$ which were not given in the main text \ref{subsect:tr-log-Y-part} are those that comprise either the endomorphism, or $\hM$. Explicitly, they read
	{\small
		\begin{align}\label{eq:result-V1-terms-E-M}
			& {\rm tr} \, \ha_2 \supset \tfrac{1}{24} K^{\alpha \beta aa1} M_3{}^{\mu}{}_{a1}{}^{a3} V^{\nu}{}_{a3a} g^{(2)}{}_{\alpha \beta \mu \nu} + \tfrac{1}{24} K_{\alpha }{}^{\beta a}{}_{a1} V^{\mu}{}_{a3a} \hnabla^{\alpha }M_2{}^{\nu \rho a1a3} g^{(2)}{}_{\beta \mu \nu \rho}\\\nonumber
			& + \tfrac{1}{24} K^{\alpha }{}_{\beta}{}^{a}{}_{a1} V^{\mu}{}_{a3a} \hnabla^{\beta}M_2{}^{\nu \rho a1a3} g^{(2)}{}_{\alpha \mu \nu \rho} + \tfrac{1}{24} K^{\alpha \beta a}{}_{a1} V_{\mu a3a} \hnabla^{\mu}M_2{}^{\nu \rho a1a3} g^{(2)}{}_{\alpha \beta \nu \rho} \\\nonumber
			& -  \tfrac{1}{48} K^{\alpha \beta a}{}_{a1} K_{\mu}{}^{\nu}{}_{a3}{}^{a2} V^{\rho}{}_{a2a} \hnabla^{\mu}E^{a1a3} g^{(2)}{}_{\alpha \beta \nu \rho} + \tfrac{1}{48} K^{\alpha \beta}{}_{a}{}^{a1} K_{\mu}{}^{\nu a3}{}_{a2}  V^{\rho}{}_{a1a3} \hnabla^{\mu}E^{a2a} g^{(2)}{}_{\alpha \beta \nu \rho}\\\nonumber
			& -  \tfrac{1}{48} K^{\alpha \beta a}{}_{a1} K^{\mu}{}_{\nu a3}{}^{a2} V^{\rho}{}_{a2a} \hnabla^{\nu}E^{a1a3} g^{(2)}{}_{\alpha \beta \mu \rho} + \tfrac{1}{48} K^{\alpha \beta}{}_{a}{}^{a1} K^{\mu}{}_{\nu}{}^{a3}{}_{a2}  V^{\rho}{}_{a1a3} \hnabla^{\nu}E^{a2a} g^{(2)}{}_{\alpha \beta \mu \rho}\\\nonumber
			& + \tfrac{1}{48} K^{\alpha \beta a}{}_{a1} K^{\mu \nu}{}_{a3}{}^{a2} V_{\rho a2a} \hnabla^{\rho}E^{a1a3} g^{(2)}{}_{\alpha \beta \mu \nu} -  \tfrac{1}{48} K^{\alpha \beta a}{}_{a1} V^{\mu}{}_{a3a} \hnabla^{\sigma}M_2{}^{\nu \rho a1a3} g^{(3)}{}_{\alpha \beta \mu \nu \rho \sigma}
		\end{align}
	}

	\subsection{Order $V^2$ terms}\label{subect:app:final-result-V2}
	
	All the terms that are quadratic in $\hV$ and linear in the Riemann tensor (or its contractions) yield the following contribution to the trace of the second Seeley-DeWitt coefficient
	{\small
		\begin{align}\label{eq:result-V2-terms-Riem}
			& {\rm tr} \, \ha_2 \supset  \tfrac{1}{144} K^{\alpha \beta aa1} K^{\mu \nu a2a3} R_{\beta \nu \mu \rho} V_{\alpha a3a} V^{\rho}{}_{a1a2} + \tfrac{1}{144} K^{\alpha \beta aa1} K^{\mu \nu a2a3} R_{\beta \rho \mu \nu} V_{\alpha a3a} V^{\rho}{}_{a1a2}\\\nonumber
			& + \tfrac{1}{144} K^{\alpha \beta aa1} K^{\mu \nu a2a3} R_{\alpha \nu \mu \rho} V_{\beta a3a} V^{\rho}{}_{a1a2} + \tfrac{1}{144} K^{\alpha \beta aa1} K^{\mu \nu a2a3} R_{\alpha \rho \mu \nu} V_{\beta a3a} V^{\rho}{}_{a1a2}\\\nonumber
			& - \tfrac{1}{144} K^{\alpha \beta aa1} K^{\mu \nu a2a3} R_{\alpha \mu \beta \nu} V_{\rho a3a} V^{\rho}{}_{a1a2} -  \tfrac{1}{144} K^{\alpha \beta aa1} K^{\mu \nu a2a3} R_{\alpha \nu \beta \mu} V_{\rho a3a} V^{\rho}{}_{a1a2}\\\nonumber
			& + \tfrac{1}{72} K^{\alpha \beta aa1} K^{\mu \nu a2a3} R_{\beta \mu \nu \rho} V_{\alpha a1a2} V^{\rho}{}_{a3a} + \tfrac{1}{72} K^{\alpha \beta aa1} K^{\mu \nu a2a3} R_{\beta \nu \mu \rho} V_{\alpha a1a2} V^{\rho}{}_{a3a}\\\nonumber
			& + \tfrac{1}{72} K^{\alpha \beta aa1} K^{\mu \nu a2a3} R_{\alpha \mu \nu \rho} V_{\beta a1a2} V^{\rho}{}_{a3a} + \tfrac{1}{72} K^{\alpha \beta aa1} K^{\mu \nu a2a3} R_{\alpha \nu \mu \rho} V_{\beta a1a2} V^{\rho}{}_{a3a}\\\nonumber
			& + \tfrac{1}{72} K^{\alpha \beta aa1} K_{\beta}{}^{\mu a2a3} R_{\alpha \mu \nu \rho} V^{\nu}{}_{a1a2} V^{\rho}{}_{a3a} + \tfrac{1}{144} K^{\alpha \beta aa1} K^{\mu}{}_{\beta}{}^{a2a3} R_{\alpha \mu \nu \rho} V^{\nu}{}_{a1a2} V^{\rho}{}_{a3a}\\\nonumber
			& + \tfrac{1}{72} K^{\alpha \beta aa1} K_{\beta}{}^{\mu a2a3} R_{\alpha \nu \mu \rho} V^{\nu}{}_{a1a2} V^{\rho}{}_{a3a} + \tfrac{1}{144} K^{\alpha \beta aa1} K^{\mu}{}_{\beta}{}^{a2a3} R_{\alpha \nu \mu \rho} V^{\nu}{}_{a1a2} V^{\rho}{}_{a3a}\\\nonumber
			& + \tfrac{1}{144} K_{\alpha }{}^{\mu a2a3} K^{\alpha \beta aa1} R_{\beta \mu \nu \rho} V^{\nu}{}_{a1a2} V^{\rho}{}_{a3a} -  \tfrac{1}{144} K^{\alpha }{}_{\alpha }{}^{aa1} K^{\beta \mu a2a3} R_{\beta \mu \nu \rho} V^{\nu}{}_{a1a2} V^{\rho}{}_{a3a}\\\nonumber
			& + \tfrac{1}{144} K_{\alpha }{}^{\mu a2a3} K^{\alpha \beta aa1} R_{\beta \nu \mu \rho} V^{\nu}{}_{a1a2} V^{\rho}{}_{a3a} -  \tfrac{1}{144} K^{\alpha }{}_{\alpha }{}^{aa1} K^{\beta \mu a2a3} R_{\beta \nu \mu \rho} V^{\nu}{}_{a1a2} V^{\rho}{}_{a3a} \\\nonumber
			& + \tfrac{1}{144} K^{\alpha \beta aa1} K^{\mu \nu a3a4} R_{\alpha \beta} V^{\rho}{}_{a1a3} V^{\sigma}{}_{a4a} \, g^{(2)}{}_{\mu \nu \rho \sigma} + \tfrac{1}{288} K^{\alpha \beta aa1} K^{\mu \nu a3a4}  R_{\alpha \mu} V^{\rho}{}_{a1a3} V^{\sigma}{}_{a4a} \, g^{(2)}{}_{\beta \nu \rho \sigma} \\\nonumber
			& + \tfrac{1}{144} K^{\alpha \beta aa1} K^{\mu \nu a3a4} R_{\alpha \nu} V^{\rho}{}_{a1a3} V^{\sigma}{}_{a4a} \, g^{(2)}{}_{\beta \mu \rho \sigma} + \tfrac{1}{144} K^{\alpha \beta aa1} K^{\mu \nu a3a4} R_{\alpha \rho} V^{\rho}{}_{a1a3} V^{\sigma}{}_{a4a} \, g^{(2)}{}_{\beta \mu \nu \sigma}\\\nonumber
			& + \tfrac{1}{144} K^{\alpha \beta aa1} K^{\mu \nu a3a4} R_{\alpha \sigma} V^{\rho}{}_{a1a3} V^{\sigma}{}_{a4a} \, g^{(2)}{}_{\beta \mu \nu \rho} + \tfrac{1}{288} K^{\alpha \beta aa1} K^{\mu \nu a3a4} R_{\beta \nu} V^{\rho}{}_{a1a3} V^{\sigma}{}_{a4a} \, g^{(2)}{}_{\alpha \mu \rho \sigma}\\\nonumber
			& + \tfrac{1}{144} K^{\alpha \beta aa1} K^{\mu \nu a3a4} R_{\beta \rho} V^{\rho}{}_{a1a3} V^{\sigma}{}_{a4a} \, g^{(2)}{}_{\alpha \mu \nu \sigma} + \tfrac{1}{144} K^{\alpha \beta aa1} K^{\mu \nu a3a4} R_{\beta \sigma} V^{\rho}{}_{a1a3} V^{\sigma}{}_{a4a} \, g^{(2)}{}_{\alpha \mu \nu \rho}\\\nonumber
			& + \tfrac{1}{288} K^{\alpha \beta aa1} K^{\mu \nu a3a4} R_{\rho \sigma} V^{\rho}{}_{a1a3} V^{\sigma}{}_{a4a} \, g^{(2)}{}_{\alpha \beta \mu \nu} + \tfrac{1}{96} K^{\alpha \beta aa1} K^{\mu \nu a3a4} R_{\alpha }{}^{\lambda}{}_{\beta}{}^{\tau} V^{\rho}{}_{a1a3} V^{\sigma}{}_{a4a} \, g^{(3)}{}_{\mu \nu \rho \sigma \lambda \tau} \\\nonumber
			&  - \tfrac{1}{192} K^{\alpha \beta aa1} K^{\mu \nu a3a4} R_{\alpha}{}^{\lambda} V^{\rho}{}_{a1a3} V^{\sigma}{}_{a4a} \, g^{(3)}{}_{\beta \mu \nu \rho \sigma \lambda} -  \tfrac{1}{192} K^{\alpha \beta aa1} K^{\mu \nu a3a4} R_{\beta}{}^{\lambda} V^{\rho}{}_{a1a3} V^{\sigma}{}_{a4a} \, g^{(3)}{}_{\alpha \mu \nu \rho \sigma \lambda} \\\nonumber
			& - \tfrac{1}{192} K^{\alpha \beta aa1} K^{\mu \nu a3a4} R_{\rho}{}^{\lambda} V^{\rho}{}_{a1a3} V^{\sigma}{}_{a4a} \, g^{(3)}{}_{\alpha \beta \mu \nu \sigma \lambda} -  \tfrac{1}{576} K^{\alpha \beta aa1} K^{\mu \nu a3a4} R V^{\rho}{}_{a1a3} V^{\sigma}{}_{a4a} \, g^{(3)}{}_{\alpha \beta \mu \nu \rho \sigma} \, .
		\end{align}
	}
	On the other hand, the terms quadratic in $\hV$ and linear in the generalized curvature $\hcalR_{\mu\nu}$ are
	{\small
		\begin{align}\label{eq:result-V2-terms-FRiem}\nonumber
			& {\rm tr} \, \ha_2 \supset - \tfrac{1}{96} \calR_{\rho \sigma a4a3} K^{\alpha \beta aa1} K^{\mu \nu a2a4}  V^{\rho}{}_{a1a2} V^{\sigma a3}{}_{a} \, g^{(2)}{}_{\alpha \beta \mu \nu} -  \tfrac{1}{96} \calR_{\nu \sigma a4a3} K^{\alpha \beta aa1} K^{\mu \nu a2a4} V^{\rho}{}_{a1a2} V^{\sigma a3}{}_{a} \, g^{(2)}{}_{\alpha \beta \mu \rho}  \\
			& -  \tfrac{1}{96} \calR_{\nu \rho a4a3} K^{\alpha \beta aa1} K^{\mu \nu a2a4} V^{\rho}{}_{a1a2} V^{\sigma a3}{}_{a} \, g^{(2)}{}_{\alpha \beta \mu \sigma} -  \tfrac{1}{96} \calR_{\mu \sigma a4a3} K^{\alpha \beta aa1} K^{\mu \nu a2a4} V^{\rho}{}_{a1a2} V^{\sigma a3}{}_{a} \, g^{(2)}{}_{\alpha \beta \nu \rho} \\\nonumber
			& - \tfrac{1}{96} \calR_{\mu \rho a4a3} K^{\alpha \beta aa1} K^{\mu \nu a2a4} V^{\rho}{}_{a1a2} V^{\sigma a3}{}_{a} \, g^{(2)}{}_{\alpha \beta \nu \sigma}  + \tfrac{1}{96} \calR_{\mu \nu a4a3} K^{\alpha \beta aa1} K^{\mu \nu a2a4} V^{\rho}{}_{a1a2} V^{\sigma a3}{}_{a} \, g^{(2)}{}_{\alpha \beta \rho \sigma} \\\nonumber
			& -  \tfrac{1}{96} \calR_{\beta \sigma a4a3} K^{\alpha \beta aa1} K^{\mu \nu a2a4}  V^{\rho}{}_{a1a2} V^{\sigma a3}{}_{a} \, g^{(2)}{}_{\alpha \mu \nu \rho} + \tfrac{1}{96} \calR_{\beta \rho a4a3} K^{\alpha \beta aa1} K^{\mu \nu a2a4} V^{\rho}{}_{a1a2} V^{\sigma a3}{}_{a} \, g^{(2)}{}_{\alpha \mu \nu \sigma} \\\nonumber
			& + \tfrac{1}{96} \calR_{\beta \nu a4a3} K^{\alpha \beta aa1} K^{\mu \nu a2a4} V^{\rho}{}_{a1a2} V^{\sigma a3}{}_{a} \, g^{(2)}{}_{\alpha \mu \rho \sigma} + \tfrac{1}{96} \calR_{\beta \mu a4a3} K^{\alpha \beta aa1} K^{\mu \nu a2a4} V^{\rho}{}_{a1a2} V^{\sigma a3}{}_{a} \, g^{(2)}{}_{\alpha \nu \rho \sigma} \\\nonumber
			& -  \tfrac{1}{96} \calR_{\alpha \sigma a4a3} K^{\alpha \beta aa1} K^{\mu \nu a2a4} V^{\rho}{}_{a1a2} V^{\sigma a3}{}_{a} \, g^{(2)}{}_{\beta \mu \nu \rho} + \tfrac{1}{96} \calR_{\alpha \rho a4a3} K^{\alpha \beta aa1} K^{\mu \nu a2a4} V^{\rho}{}_{a1a2} V^{\sigma a3}{}_{a} \, g^{(2)}{}_{\beta \mu \nu \sigma} \\\nonumber
			& + \tfrac{1}{96} \calR_{\alpha \nu a4a3} K^{\alpha \beta aa1} K^{\mu \nu a2a4} V^{\rho}{}_{a1a2} V^{\sigma a3}{}_{a} \, g^{(2)}{}_{\beta \mu \rho \sigma} + \tfrac{1}{96} \calR_{\alpha \mu a4a3} K^{\alpha \beta aa1} K^{\mu \nu a2a4} V^{\rho}{}_{a1a2} V^{\sigma a3}{}_{a} \, g^{(2)}{}_{\beta \nu \rho \sigma}\\\nonumber
			& + \tfrac{1}{96} \calR_{\alpha \beta a4a3} K^{\alpha \beta aa1} K^{\mu \nu a2a4} V^{\rho}{}_{a1a2} V^{\sigma a3}{}_{a} \, g^{(2)}{}_{\mu \nu \rho \sigma} + \tfrac{1}{96} \calR_{\mu}{}^{\lambda}{}_{a4a3} K^{\alpha \beta aa1} K^{\mu \nu a2a4}  V^{\rho}{}_{a1a2} V^{\sigma a3}{}_{a} \, g^{(3)}{}_{\alpha \beta \nu \rho \sigma \lambda} \\\nonumber
			& + \tfrac{1}{96} \calR_{\sigma}{}^{\lambda}{}_{a3a} K^{\alpha \beta aa1} K^{\mu \nu a2a4}  V^{\rho}{}_{a1a2} V^{\sigma}{}_{a4}{}^{a3} \, g^{(3)}{}_{\alpha \beta \mu \nu \rho \lambda} + \tfrac{1}{96} \calR_{\nu}{}^{\lambda}{}_{a4a3} K^{\alpha \beta aa1} K^{\mu \nu a2a4} V^{\rho}{}_{a1a2} V^{\sigma a3}{}_{a} \, g^{(3)}{}_{\alpha \beta \mu \rho \sigma \lambda}  \, .
		\end{align}
	}
	Finally, the contributions proportional to two covariant derivatives acting on $\hV$ are given by
	{\small
		\begin{align}\label{eq:result-V2-terms-der2}\nonumber
			& {\rm tr} \, \ha_2  \supset \tfrac{1}{48} K^{\alpha \beta}{}_{a}{}^{a1} K_{\mu \nu}{}^{a2}{}_{a3} V^{\rho}{}_{a1a2} \hnabla^{\mu}\hnabla^{\nu}V^{\sigma a3a} \, g^{(2)}{}_{\alpha \beta \rho \sigma} + \tfrac{1}{48} K^{\alpha \beta}{}_{a}{}^{a1} K_{\mu}{}^{\nu a2}{}_{a3} V_{\rho a1a2} \hnabla^{\rho}\hnabla^{\mu}V^{\sigma a3a} \, g^{(2)}{}_{\alpha \beta \nu \sigma} \\
			& + \tfrac{1}{48} K^{\alpha \beta}{}_{a}{}^{a1} K^{\mu}{}_{\nu}{}^{a2}{}_{a3} V_{\rho a1a2} \hnabla^{\rho}\hnabla^{\nu}V^{\sigma a3a} \, g^{(2)}{}_{\alpha \beta \mu \sigma}	-  \tfrac{1}{96} K^{\alpha \beta}{}_{a}{}^{a1} K_{\mu}{}^{\nu a2}{}_{a3} V^{\rho}{}_{a1a2} \hnabla^{\mu}\hnabla^{\lambda}V^{\sigma a3a} \, g^{(3)}{}_{\alpha \beta \nu \rho \sigma \lambda} \\\nonumber
			& -  \tfrac{1}{96} K^{\alpha \beta}{}_{a}{}^{a1} K^{\mu}{}_{\nu}{}^{a2}{}_{a3} V^{\rho}{}_{a1a2} \hnabla^{\nu}\hnabla^{\lambda}V^{\sigma a3a} \, g^{(3)}{}_{\alpha \beta \mu \rho \sigma \lambda} -  \tfrac{1}{96} K^{\alpha \beta a}{}_{a1} K^{\mu \nu}{}_{a2}{}^{a3} V_{\rho a3a} \hnabla^{\rho}\hnabla^{\lambda}V^{\sigma a1a2} \, g^{(3)}{}_{\alpha \beta \mu \nu \sigma \lambda}\\\nonumber
			& - \tfrac{1}{192} K^{\alpha \beta a}{}_{a1} K^{\mu \nu}{}_{a2}{}^{a3} V^{\rho}{}_{a3a} \hnabla^{\lambda}\hnabla_{\lambda}V^{\sigma a1a2} \, g^{(3)}{}_{\alpha \beta \mu \nu \rho \sigma}  + \tfrac{1}{320} K^{\alpha \beta a}{}_{a1} K^{\mu \nu}{}_{a2}{}^{a3} V^{\rho}{}_{a3a} \hnabla^{\tau}\hnabla^{\lambda}V^{\sigma a1a2} \, g^{(4)}{}_{\alpha \beta \mu \nu \rho \sigma \lambda \tau} \, .
		\end{align}
	}
	
	\subsection{Integrand: The general case}\label{subsect:app-final-result-integrand-general-case}
	
	Here we present the full result for the modification of the local part of ${\rm tr}\, \ha_2$ due to the principal part of the operator in the most general case, whose derivation has been outlined in Subsect.\ \ref{subsect:tr-log-pric-part-general}. Since many independent scalar contractions are present, we split this result into multiple pieces. Furthermore, we denote the $\zeta$ derivative of the non-minimal symbol $\frac{d\hN}{d\zeta}$ simply by $d \hN$ for notational convenience.
	
	We start by taking into account the terms that involve the generalized curvature $\hcalR_{\mu\nu}$. The contribution which is quadratic in the latter is given by
{\small
	\begin{align}\label{eq:result-integand-general-FRiem^2}\nonumber
		{\rm tr}\, \ha_2 \supset \int_0^1 d\zeta & \left[ \tfrac{1}{24} dN^{\alpha \beta aa1} \calR_{\beta}{}^{\nu a2}{}_{a} \calR_{\mu \nu a3a2} K_{\alpha }{}^{\mu}{}_{a1}{}^{a3} + \tfrac{1}{24} dN^{\alpha \beta aa1} \calR_{\beta}{}^{\nu}{}_{a3}{}^{a2} \calR_{\mu \nu a2a} K_{\alpha }{}^{\mu}{}_{a1}{}^{a3} \right.\\\nonumber
		& \left. + \tfrac{1}{24} dN^{\alpha \beta aa1} \calR_{\alpha }{}^{\nu a2}{}_{a} \calR_{\mu \nu a3a2} K_{\beta}{}^{\mu}{}_{a1}{}^{a3} + \tfrac{1}{24} dN^{\alpha \beta aa1} \calR_{\alpha }{}^{\nu}{}_{a3}{}^{a2} \calR_{\mu \nu a2a} K_{\beta}{}^{\mu}{}_{a1}{}^{a3} \right.\\\nonumber
		& \left. + \tfrac{1}{24} dN^{\alpha }{}_{\alpha }{}^{aa1} \calR_{\beta}{}^{\nu a2}{}_{a} \calR_{\mu \nu a3a2} K^{\beta \mu}{}_{a1}{}^{a3} + \tfrac{1}{24} dN^{\alpha }{}_{\alpha }{}^{aa1} \calR_{\beta}{}^{\nu}{}_{a3}{}^{a2} \calR_{\mu \nu a2a} K^{\beta \mu}{}_{a1}{}^{a3} \right.\\\nonumber
		& \left. + \tfrac{1}{24} dN^{\alpha \beta aa1} \calR_{\beta}{}^{\nu a2}{}_{a} \calR_{\mu \nu a3a2} K^{\mu}{}_{\alpha a1}{}^{a3} + \tfrac{1}{24} dN^{\alpha \beta aa1} \calR_{\beta}{}^{\nu}{}_{a3}{}^{a2} \calR_{\mu \nu a2a} K^{\mu}{}_{\alpha a1}{}^{a3} \right.\\\nonumber
		& \left. + \tfrac{1}{24} dN^{\alpha \beta aa1} \calR_{\alpha }{}^{\nu a2}{}_{a} \calR_{\mu \nu a3a2} K^{\mu}{}_{\beta a1}{}^{a3} + \tfrac{1}{24} dN^{\alpha \beta aa1} \calR_{\alpha }{}^{\nu}{}_{a3}{}^{a2} \calR_{\mu \nu a2a} K^{\mu}{}_{\beta a1}{}^{a3} \right.\\\nonumber
		& \left. + \tfrac{1}{24} dN^{\alpha \beta aa1} \calR_{\alpha }{}^{\nu a2}{}_{a} \calR_{\beta \nu a3a2} K^{\mu}{}_{\mu a1}{}^{a3} + \tfrac{1}{24} dN^{\alpha \beta aa1} \calR_{\alpha }{}^{\nu}{}_{a3}{}^{a2} \calR_{\beta \nu a2a} K^{\mu}{}_{\mu a1}{}^{a3} \right.\\\nonumber
		& \left. - \tfrac{1}{8} dN^{\alpha \beta aa1} \calR_{\alpha \nu}{}^{a2}{}_{a} \calR_{\beta \mu a3a2} K^{\mu \nu}{}_{a1}{}^{a3} -  \tfrac{1}{8} dN^{\alpha \beta aa1} \calR_{\alpha \nu a3}{}^{a2} \calR_{\beta \mu a2a} K^{\mu \nu}{}_{a1}{}^{a3} \right.\\\nonumber
		& \left. - \tfrac{1}{8} dN^{\alpha \beta aa1} \calR_{\alpha \mu}{}^{a2}{}_{a} \calR_{\beta \nu a3a2} K^{\mu \nu}{}_{a1}{}^{a3} -  \tfrac{1}{8} dN^{\alpha \beta aa1} \calR_{\alpha \mu a3}{}^{a2} \calR_{\beta \nu a2a} K^{\mu \nu}{}_{a1}{}^{a3} \right.\\\nonumber
		& \left. -  \tfrac{1}{8} dN^{\alpha \beta aa1} \calR_{\alpha \beta}{}^{a2}{}_{a} \calR_{\mu \nu a3a2} K^{\mu \nu}{}_{a1}{}^{a3} -  \tfrac{1}{8} dN^{\alpha \beta aa1} \calR_{\alpha \beta a3}{}^{a2} \calR_{\mu \nu a2a} K^{\mu \nu}{}_{a1}{}^{a3} \right.\\
		& \left. - \tfrac{1}{48} dN^{\alpha \beta aa1} \calR_{\rho \sigma a2a} \calR^{\rho \sigma}{}_{a3}{}^{a2} K^{\mu \nu}{}_{a1}{}^{a3} g^{(2)}{}_{\alpha \beta \mu \nu} \right] \, .
\end{align}}

Next, we consider the terms that are linear in both the generalized curvature and the Riemann tensor (or its contractions). These are
{\small
	\begin{align}\label{eq:result-integand-general-FRiem-Riem}
		& {\rm tr}\, \ha_2 \supset \int_0^1 d \zeta \left[ - \tfrac{1}{12} dN^{\alpha \beta aa1} \calR_{\mu \nu a2a} K^{\mu \nu}{}_{a1}{}^{a2} R_{\alpha \beta} -  \tfrac{1}{12} dN^{\alpha \beta aa1} \calR_{\beta \nu a2a} K^{\mu \nu}{}_{a1}{}^{a2} R_{\alpha \mu} \right.\\\nonumber
		& \left. -  \tfrac{1}{12} dN^{\alpha \beta aa1} \calR_{\beta \mu a2a} K^{\mu \nu}{}_{a1}{}^{a2} R_{\alpha \nu} -  \tfrac{1}{12} dN^{\alpha \beta aa1} \calR_{\alpha \nu a2a} K^{\mu \nu}{}_{a1}{}^{a2} R_{\beta \mu} -  \tfrac{1}{12} dN^{\alpha \beta aa1} \calR_{\alpha \mu a2a} K^{\mu \nu}{}_{a1}{}^{a2} R_{\beta \nu} \right.\\\nonumber
		& \left. - \tfrac{1}{12} dN^{\alpha \beta aa1} \calR_{\alpha \beta a2a} K^{\mu \nu}{}_{a1}{}^{a2} R_{\mu \nu} + \tfrac{1}{24} dN^{\alpha \beta aa1} \calR_{\beta \mu a2a} K_{\alpha }{}^{\mu}{}_{a1}{}^{a2} R + \tfrac{1}{24} dN^{\alpha \beta aa1} \calR_{\alpha \mu a2a} K_{\beta}{}^{\mu}{}_{a1}{}^{a2} R \right.\\\nonumber
		& \left. + \tfrac{1}{24} dN^{\alpha }{}_{\alpha }{}^{aa1} \calR_{\beta \mu a2a} K^{\beta \mu}{}_{a1}{}^{a2} R + \tfrac{1}{24} dN^{\alpha \beta aa1} \calR_{\beta \mu a2a} K^{\mu}{}_{\alpha a1}{}^{a2} R + \tfrac{1}{24} dN^{\alpha \beta aa1} \calR_{\alpha \mu a2a} K^{\mu}{}_{\beta a1}{}^{a2} R \right.\\\nonumber
		& \left. + \tfrac{1}{24} dN^{\alpha \beta aa1} \calR_{\alpha \beta a2a} K^{\mu}{}_{\mu a1}{}^{a2} R + \tfrac{1}{6} dN^{\alpha \beta aa1} \calR_{\nu}{}^{\rho}{}_{a2a} K^{\mu \nu}{}_{a1}{}^{a2} R_{\alpha \beta \mu \rho} -  \tfrac{1}{12} dN^{\alpha \beta aa1} \calR_{\mu}{}^{\rho}{}_{a2a} K^{\mu \nu}{}_{a1}{}^{a2} R_{\alpha \beta \nu \rho} \right.\\\nonumber
		& \left. - \tfrac{1}{12} dN^{\alpha \beta aa1} \calR_{\nu}{}^{\rho}{}_{a2a} K^{\mu \nu}{}_{a1}{}^{a2} R_{\alpha \mu \beta \rho} - \tfrac{1}{12} dN^{\alpha \beta aa1} \calR_{\beta}{}^{\rho}{}_{a2a} K^{\mu \nu}{}_{a1}{}^{a2} R_{\alpha \mu \nu \rho} \right.\\\nonumber
		& \left. - \tfrac{1}{12} dN^{\alpha \beta aa1} \calR_{\mu}{}^{\rho}{}_{a2a} K^{\mu \nu}{}_{a1}{}^{a2} R_{\alpha \nu \beta \rho} - \tfrac{1}{12} dN^{\alpha \beta aa1} \calR_{\beta}{}^{\rho}{}_{a2a} K^{\mu \nu}{}_{a1}{}^{a2} R_{\alpha \nu \mu \rho} \right.\\\nonumber
		& \left. - \tfrac{1}{12} dN^{\alpha \beta aa1} \calR_{\alpha }{}^{\rho}{}_{a2a} K^{\mu \nu}{}_{a1}{}^{a2} R_{\beta \mu \nu \rho} -  \tfrac{1}{12} dN^{\alpha \beta aa1} \calR_{\alpha }{}^{\rho}{}_{a2a} K^{\mu \nu}{}_{a1}{}^{a2} R_{\beta \nu \mu \rho}
		\, \right] \, .
\end{align}}
	Finally, we present the part of the modification to the trace of $\ha_2$ which depends on the dimension-two tensor $\hM_2$ and $\hcalR_{\mu\nu}$, which read
	{\small
		\begin{align}\label{eq:result-integand-general-FRiem-M2}\nonumber
			& {\rm tr}\, \ha_2 \supset \int_0^1 d \zeta \left[ \tfrac{1}{48} dN^{\alpha \beta aa1} \calR_{\rho \sigma a2a} K^{\mu \nu}{}_{a1}{}^{a3} M_2{}^{\rho \sigma}{}_{a3}{}^{a2} g^{(2)}{}_{\alpha \beta \mu \nu} + \tfrac{1}{48} dN^{\alpha \beta aa1} \calR_{\nu \sigma a2a} K^{\mu \nu}{}_{a1}{}^{a3} M_2{}^{\rho \sigma}{}_{a3}{}^{a2} g^{(2)}{}_{\alpha \beta \mu \rho} \right.\\\nonumber
			& \left. + \tfrac{1}{48} dN^{\alpha \beta aa1} \calR_{\nu \rho a2a} K^{\mu \nu}{}_{a1}{}^{a3} M_2{}^{\rho \sigma}{}_{a3}{}^{a2} g^{(2)}{}_{\alpha \beta \mu \sigma} + \tfrac{1}{48} dN^{\alpha \beta aa1} \calR_{\mu \sigma a2a} K^{\mu \nu}{}_{a1}{}^{a3} M_2{}^{\rho \sigma}{}_{a3}{}^{a2} g^{(2)}{}_{\alpha \beta \nu \rho} \right.\\\nonumber
			& \left. + \tfrac{1}{48} dN^{\alpha \beta aa1} \calR_{\mu \rho a2a} K^{\mu \nu}{}_{a1}{}^{a3} M_2{}^{\rho \sigma}{}_{a3}{}^{a2} g^{(2)}{}_{\alpha \beta \nu \sigma} + \tfrac{1}{48} dN^{\alpha \beta aa1} \calR_{\mu \nu a2a} K^{\mu \nu}{}_{a1}{}^{a3} M_2{}^{\rho \sigma}{}_{a3}{}^{a2} g^{(2)}{}_{\alpha \beta \rho \sigma} \right.\\\nonumber
			& \left. + \tfrac{1}{48} dN^{\alpha \beta aa1} \calR_{\beta \sigma a2a} K^{\mu \nu}{}_{a1}{}^{a3} M_2{}^{\rho \sigma}{}_{a3}{}^{a2} g^{(2)}{}_{\alpha \mu \nu \rho} + \tfrac{1}{48} dN^{\alpha \beta aa1} \calR_{\beta \rho a2a} K^{\mu \nu}{}_{a1}{}^{a3} M_2{}^{\rho \sigma}{}_{a3}{}^{a2} g^{(2)}{}_{\alpha \mu \nu \sigma} \right.\\\nonumber
			& \left. + \tfrac{1}{48} dN^{\alpha \beta aa1} \calR_{\beta \nu a2a} K^{\mu \nu}{}_{a1}{}^{a3} M_2{}^{\rho \sigma}{}_{a3}{}^{a2} g^{(2)}{}_{\alpha \mu \rho \sigma} + \tfrac{1}{48} dN^{\alpha \beta aa1} \calR_{\beta \mu a2a} K^{\mu \nu}{}_{a1}{}^{a3} M_2{}^{\rho \sigma}{}_{a3}{}^{a2} g^{(2)}{}_{\alpha \nu \rho \sigma} \right.\\\nonumber
			& \left. + \tfrac{1}{48} dN^{\alpha \beta aa1} \calR_{\alpha \sigma a2a} K^{\mu \nu}{}_{a1}{}^{a3} M_2{}^{\rho \sigma}{}_{a3}{}^{a2} g^{(2)}{}_{\beta \mu \nu \rho} + \tfrac{1}{48} dN^{\alpha \beta aa1} \calR_{\alpha \rho a2a} K^{\mu \nu}{}_{a1}{}^{a3} M_2{}^{\rho \sigma}{}_{a3}{}^{a2} g^{(2)}{}_{\beta \mu \nu \sigma} \right.\\\nonumber
			& \left. + \tfrac{1}{48} dN^{\alpha \beta aa1} \calR_{\alpha \nu a2a} K^{\mu \nu}{}_{a1}{}^{a3} M_2{}^{\rho \sigma}{}_{a3}{}^{a2} g^{(2)}{}_{\beta \mu \rho \sigma} + \tfrac{1}{48} dN^{\alpha \beta aa1} \calR_{\alpha \mu a2a} K^{\mu \nu}{}_{a1}{}^{a3} M_2{}^{\rho \sigma}{}_{a3}{}^{a2} g^{(2)}{}_{\beta \nu \rho \sigma} \right.\\\nonumber
			& \left. + \tfrac{1}{48} dN^{\alpha \beta aa1} \calR_{\alpha \beta a2a} K^{\mu \nu}{}_{a1}{}^{a3} M_2{}^{\rho \sigma}{}_{a3}{}^{a2} g^{(2)}{}_{\mu \nu \rho \sigma} + \tfrac{1}{48} dN^{\alpha \beta aa1} \calR_{\sigma}{}^{\lambda}{}_{a2a} K^{\mu \nu}{}_{a1}{}^{a3} M_2{}^{\rho \sigma}{}_{a3}{}^{a2} g^{(3)}{}_{\alpha \beta \mu \nu \rho \lambda} \right.\\
			& \left. + \tfrac{1}{48} dN^{\alpha \beta aa1} \calR_{\rho}{}^{\lambda}{}_{a2a} K^{\mu \nu}{}_{a1}{}^{a3} M_2{}^{\rho \sigma}{}_{a3}{}^{a2} g^{(3)}{}_{\alpha \beta \mu \nu \sigma \lambda} \right] \, .
		\end{align}
	}
The part of the result that depends quadratically on the dimension-two tensor $\hM_2$ has the following simple form
\begin{equation}\label{eq:result-integand-general-M2^2}
	{\rm tr}\, \ha_2 \supset - \tfrac{1}{1920} \int_0^1 d \zeta dN^{\alpha \beta aa1} K^{\mu \nu}{}_{a1}{}^{a3} M_2{}^{\rho \sigma}{}_{a3}{}^{a2} M_2{}^{\lambda \tau}{}_{a2a} g^{(4)}{}_{\alpha \beta \mu \nu \rho \sigma \lambda \tau} \, .
\end{equation}
On the other hand, the number of terms that involve both $\hM_2$ and the Riemann tensor (or its traces) is huge. Thus, we further split them into two parts, the first of which features only the Ricci tensor and scalar, i.e.,
{\small
	\begin{align}\label{eq:result-integand-general-Ricci-M2}\nonumber
		& {\rm tr}\, \ha_2 \supset \int_0^1 d \zeta \, \left[ \tfrac{1}{144} dN^{\alpha \beta aa1} K^{\mu \nu}{}_{a1}{}^{a2} M_2{}^{\rho \sigma}{}_{a2a} R_{\alpha \beta} g^{(2)}{}_{\mu \nu \rho \sigma} + \tfrac{1}{144} dN^{\alpha \beta aa1} K^{\mu \nu}{}_{a1}{}^{a2} M_2{}^{\rho \sigma}{}_{a2a} R_{\alpha \mu} g^{(2)}{}_{\beta \nu \rho \sigma} \right.\\\nonumber
		& \left. + \tfrac{1}{144} dN^{\alpha \beta aa1} K^{\mu \nu}{}_{a1}{}^{a2} M_2{}^{\rho \sigma}{}_{a2a} R_{\alpha \nu} g^{(2)}{}_{\beta \mu \rho \sigma}  + \tfrac{1}{144} dN^{\alpha \beta aa1} K^{\mu \nu}{}_{a1}{}^{a2} M_2{}^{\rho \sigma}{}_{a2a} R_{\alpha \rho} g^{(2)}{}_{\beta \mu \nu \sigma} \right.\\\nonumber
		& \left. + \tfrac{1}{144} dN^{\alpha \beta aa1} K^{\mu \nu}{}_{a1}{}^{a2} M_2{}^{\rho \sigma}{}_{a2a} R_{\alpha \sigma} g^{(2)}{}_{\beta \mu \nu \rho} + \tfrac{1}{144} dN^{\alpha \beta aa1} K^{\mu \nu}{}_{a1}{}^{a2} M_2{}^{\rho \sigma}{}_{a2a} R_{\beta \mu} g^{(2)}{}_{\alpha \nu \rho \sigma} \right.\\\nonumber
		& \left. + \tfrac{1}{144} dN^{\alpha \beta aa1} K^{\mu \nu}{}_{a1}{}^{a2} M_2{}^{\rho \sigma}{}_{a2a} R_{\beta \nu} g^{(2)}{}_{\alpha \mu \rho \sigma} + \tfrac{1}{144} dN^{\alpha \beta aa1} K^{\mu \nu}{}_{a1}{}^{a2} M_2{}^{\rho \sigma}{}_{a2a} R_{\beta \rho} g^{(2)}{}_{\alpha \mu \nu \sigma} \right.\\\nonumber
		& \left. + \tfrac{1}{144} dN^{\alpha \beta aa1} K^{\mu \nu}{}_{a1}{}^{a2} M_2{}^{\rho \sigma}{}_{a2a} R_{\beta \sigma} g^{(2)}{}_{\alpha \mu \nu \rho} + \tfrac{1}{144} dN^{\alpha \beta aa1} K^{\mu \nu}{}_{a1}{}^{a2} M_2{}^{\rho \sigma}{}_{a2a} R_{\mu \nu} g^{(2)}{}_{\alpha \beta \rho \sigma} \right.\\\nonumber
		& \left. + \tfrac{1}{144} dN^{\alpha \beta aa1} K^{\mu \nu}{}_{a1}{}^{a2} M_2{}^{\rho \sigma}{}_{a2a} R_{\mu \rho} g^{(2)}{}_{\alpha \beta \nu \sigma} + \tfrac{1}{144} dN^{\alpha \beta aa1} K^{\mu \nu}{}_{a1}{}^{a2} M_2{}^{\rho \sigma}{}_{a2a} R_{\mu \sigma} g^{(2)}{}_{\alpha \beta \nu \rho} \right.\\\nonumber
		& \left. + \tfrac{1}{144} dN^{\alpha \beta aa1} K^{\mu \nu}{}_{a1}{}^{a2} M_2{}^{\rho \sigma}{}_{a2a} R_{\nu \rho} g^{(2)}{}_{\alpha \beta \mu \sigma} + \tfrac{1}{144} dN^{\alpha \beta aa1} K^{\mu \nu}{}_{a1}{}^{a2} M_2{}^{\rho \sigma}{}_{a2a} R_{\nu \sigma} g^{(2)}{}_{\alpha \beta \mu \rho} \right.\\\nonumber
		& \left. + \tfrac{1}{144} dN^{\alpha \beta aa1} K^{\mu \nu}{}_{a1}{}^{a2} M_2{}^{\rho \sigma}{}_{a2a} R_{\rho \sigma} g^{(2)}{}_{\alpha \beta \mu \nu} -  \tfrac{1}{96} dN^{\alpha \beta aa1} K^{\mu \nu}{}_{a1}{}^{a2} M_2{}^{\rho \sigma}{}_{a2a} R_{\rho}{}^{\lambda} g^{(3)}{}_{\alpha \beta \mu \nu \sigma \lambda} \right.\\
		& \left. -  \tfrac{1}{96} dN^{\alpha \beta aa1} K^{\mu \nu}{}_{a1}{}^{a2} M_2{}^{\rho \sigma}{}_{a2a} R_{\sigma}{}^{\lambda} g^{(3)}{}_{\alpha \beta \mu \nu \rho \lambda} -  \tfrac{1}{288} dN^{\alpha \beta aa1} K^{\mu \nu}{}_{a1}{}^{a2} M_2{}^{\rho \sigma}{}_{a2a} R g^{(3)}{}_{\alpha \beta \mu \nu \rho \sigma} \right] \, ,
	\end{align}
}
while those that involve the Riemann tenor explicitly are given by
{\small
	\begin{align}\label{eq:result-integand-general-Riemann-M2}\nonumber
		{\rm tr}\, \ha_2 \supset \int_0^1 d \zeta \, & \left[ - \tfrac{1}{72} dN^{\alpha \beta aa1} K^{\mu \nu}{}_{a1}{}^{a2} M_2{}^{\rho}{}_{\rho a2a} R_{\alpha \mu \beta \nu} - \tfrac{1}{72} dN^{\alpha \beta aa1} K^{\mu \nu}{}_{a1}{}^{a2} M_2{}_{\nu}{}^{\rho}{}_{a2a} R_{\alpha \mu \beta \rho} \right.\\\nonumber
		& \left. - \tfrac{1}{72} dN^{\alpha \beta aa1} K^{\mu \nu}{}_{a1}{}^{a2} M_2{}^{\rho}{}_{\nu a2a} R_{\alpha \mu \beta \rho} - \tfrac{1}{72} dN^{\alpha \beta aa1} K^{\mu \nu}{}_{a1}{}^{a2} M_2{}^{\rho}{}_{\rho a2a} R_{\alpha \nu \beta \mu} \right.\\\nonumber
		& \left. - \tfrac{1}{72} dN^{\alpha \beta aa1} K^{\mu \nu}{}_{a1}{}^{a2} M_2{}_{\mu}{}^{\rho}{}_{a2a} R_{\alpha \nu \beta \rho} - \tfrac{1}{72} dN^{\alpha \beta aa1} K^{\mu}{}_{\mu a1}{}^{a2} M_2{}^{\nu \rho}{}_{a2a} R_{\alpha \nu \beta \rho} \right.\\\nonumber
		& \left. - \tfrac{1}{72} dN^{\alpha \beta aa1} K^{\mu \nu}{}_{a1}{}^{a2} M_2{}^{\rho}{}_{\mu a2a} R_{\alpha \nu \beta \rho} - \tfrac{1}{72} dN^{\alpha \beta aa1} K^{\mu \nu}{}_{a1}{}^{a2} M_2{}_{\beta}{}^{\rho}{}_{a2a} R_{\alpha \nu \mu \rho} \right.\\\nonumber
		& \left. - \tfrac{1}{72} dN^{\alpha \beta aa1} K_{\beta}{}^{\mu}{}_{a1}{}^{a2} M_2{}^{\nu \rho}{}_{a2a} R_{\alpha \nu \mu \rho} - \tfrac{1}{72} dN^{\alpha \beta aa1} K^{\mu}{}_{\beta a1}{}^{a2} M_2{}^{\nu \rho}{}_{a2a} R_{\alpha \nu \mu \rho} \right.\\\nonumber
		& \left. - \tfrac{1}{72} dN^{\alpha \beta aa1} K^{\mu \nu}{}_{a1}{}^{a2} M_2{}^{\rho}{}_{\beta a2a} R_{\alpha \nu \mu \rho} - \tfrac{1}{72} dN^{\alpha \beta aa1} K^{\mu \nu}{}_{a1}{}^{a2} M_2{}_{\nu}{}^{\rho}{}_{a2a} R_{\alpha \rho \beta \mu} \right.\\\nonumber
		& \left. - \tfrac{1}{72} dN^{\alpha \beta aa1} K^{\mu \nu}{}_{a1}{}^{a2} M_2{}^{\rho}{}_{\nu a2a} R_{\alpha \rho \beta \mu} - \tfrac{1}{72} dN^{\alpha \beta aa1} K^{\mu \nu}{}_{a1}{}^{a2} M_2{}_{\mu}{}^{\rho}{}_{a2a} R_{\alpha \rho \beta \nu} \right.\\\nonumber
		& \left. - \tfrac{1}{72} dN^{\alpha \beta aa1} K^{\mu}{}_{\mu a1}{}^{a2} M_2{}^{\nu \rho}{}_{a2a} R_{\alpha \rho \beta \nu} - \tfrac{1}{72} dN^{\alpha \beta aa1} K^{\mu \nu}{}_{a1}{}^{a2} M_2{}^{\rho}{}_{\mu a2a} R_{\alpha \rho \beta \nu} \right.\\\nonumber
		& \left. - \tfrac{1}{72} dN^{\alpha \beta aa1} K^{\mu \nu}{}_{a1}{}^{a2} M_2{}_{\beta}{}^{\rho}{}_{a2a} R_{\alpha \rho \mu \nu} - \tfrac{1}{72} dN^{\alpha \beta aa1} K_{\beta}{}^{\mu}{}_{a1}{}^{a2} M_2{}^{\nu \rho}{}_{a2a} R_{\alpha \rho \mu \nu} \right.\\\nonumber
		& \left. - \tfrac{1}{72} dN^{\alpha \beta aa1} K^{\mu}{}_{\beta a1}{}^{a2} M_2{}^{\nu \rho}{}_{a2a} R_{\alpha \rho \mu \nu} - \tfrac{1}{72} dN^{\alpha \beta aa1} K^{\mu \nu}{}_{a1}{}^{a2} M_2{}^{\rho}{}_{\beta a2a} R_{\alpha \rho \mu \nu} \right.\\\nonumber
		& \left. - \tfrac{1}{72} dN^{\alpha \beta aa1} K^{\mu \nu}{}_{a1}{}^{a2} M_2{}_{\alpha }{}^{\rho}{}_{a2a} R_{\beta \nu \mu \rho} - \tfrac{1}{72} dN^{\alpha \beta aa1} K_{\alpha }{}^{\mu}{}_{a1}{}^{a2} M_2{}^{\nu \rho}{}_{a2a} R_{\beta \nu \mu \rho} \right.\\\nonumber
		& \left. - \tfrac{1}{72} dN^{\alpha }{}_{\alpha }{}^{aa1} K^{\beta \mu}{}_{a1}{}^{a2} M_2{}^{\nu \rho}{}_{a2a} R_{\beta \nu \mu \rho} - \tfrac{1}{72} dN^{\alpha \beta aa1} K^{\mu}{}_{\alpha a1}{}^{a2} M_2{}^{\nu \rho}{}_{a2a} R_{\beta \nu \mu \rho} \right.\\\nonumber
		& \left. -  \tfrac{1}{72} dN^{\alpha \beta aa1} K^{\mu \nu}{}_{a1}{}^{a2} M_2{}^{\rho}{}_{\alpha a2a} R_{\beta \nu \mu \rho} -  \tfrac{1}{72} dN^{\alpha \beta aa1} K^{\mu \nu}{}_{a1}{}^{a2} M_2{}_{\alpha }{}^{\rho}{}_{a2a} R_{\beta \rho \mu \nu} \right.\\\nonumber
		& \left. - \tfrac{1}{72} dN^{\alpha \beta aa1} K_{\alpha }{}^{\mu}{}_{a1}{}^{a2} M_2{}^{\nu \rho}{}_{a2a} R_{\beta \rho \mu \nu} - \tfrac{1}{72} dN^{\alpha }{}_{\alpha }{}^{aa1} K^{\beta \mu}{}_{a1}{}^{a2} M_2{}^{\nu \rho}{}_{a2a} R_{\beta \rho \mu \nu} \right.\\\nonumber
		& \left. - \tfrac{1}{72} dN^{\alpha \beta aa1} K^{\mu}{}_{\alpha a1}{}^{a2} M_2{}^{\nu \rho}{}_{a2a} R_{\beta \rho \mu \nu} -  \tfrac{1}{72} dN^{\alpha \beta aa1} K^{\mu \nu}{}_{a1}{}^{a2} M_2{}^{\rho}{}_{\alpha a2a} R_{\beta \rho \mu \nu} \right.\\
		& \left. + \tfrac{1}{48} dN^{\alpha \beta aa1} K^{\mu \nu}{}_{a1}{}^{a2} M_2{}^{\rho \sigma}{}_{a2a} R_{\rho}{}^{\lambda}{}_{\sigma}{}^{\tau} g^{(3)}{}_{\alpha \beta \mu \nu \lambda \tau}\right] \, .
\end{align}}
We are left only with those terms that depend quadratically on the pseudo-Riemannian curvatures. Since many tensor structures arise, we further split them into. Thus, we first write down those terms that do not explicitly depend on the Riemann tensor
{\small
	\begin{align}\label{eq:result-integand-general-Ricci^2}\nonumber
		& {\rm tr}\, \ha_2 \supset \int_0^1 d \zeta \left[ - \tfrac{1}{36} dN^{\alpha \beta aa1} K^{\mu \nu}{}_{a1a} R_{\alpha \nu} R_{\beta \mu} -  \tfrac{1}{36} dN^{\alpha \beta aa1} K^{\mu \nu}{}_{a1a} R_{\alpha \mu} R_{\beta \nu} \right.\\\nonumber
		& \left. -  \tfrac{1}{90} dN^{\alpha \beta aa1} K^{\mu}{}_{\mu a1a} R_{\alpha }{}^{\nu} R_{\beta \nu} -  \tfrac{1}{36} dN^{\alpha \beta aa1} K^{\mu \nu}{}_{a1a} R_{\alpha \beta} R_{\mu \nu} - \tfrac{1}{90} dN^{\alpha \beta aa1} K_{\beta}{}^{\mu}{}_{a1a} R_{\alpha }{}^{\nu} R_{\mu \nu} \right.\\\nonumber
		& \left. -  \tfrac{1}{90} dN^{\alpha \beta aa1} K^{\mu}{}_{\beta a1a} R_{\alpha }{}^{\nu} R_{\mu \nu} -  \tfrac{1}{90} dN^{\alpha \beta aa1} K_{\alpha }{}^{\mu}{}_{a1a} R_{\beta}{}^{\nu} R_{\mu \nu} -  \tfrac{1}{90} dN^{\alpha }{}_{\alpha }{}^{aa1} K^{\beta \mu}{}_{a1a} R_{\beta}{}^{\nu} R_{\mu \nu} \right.\\\nonumber
		& \left. -  \tfrac{1}{90} dN^{\alpha \beta aa1} K^{\mu}{}_{\alpha a1a} R_{\beta}{}^{\nu} R_{\mu \nu} + \tfrac{1}{720} dN^{\alpha \beta aa1} K^{\mu \nu}{}_{a1a}  R_{\rho \sigma} R^{\rho \sigma} g^{(2)}{}_{\alpha \beta \mu \nu} + \tfrac{1}{72} dN^{\alpha \beta aa1} K^{\mu}{}_{\mu a1a} R_{\alpha \beta} R \right.\\\nonumber
		& \left. + \tfrac{1}{72} dN^{\alpha \beta aa1} K_{\beta}{}^{\mu}{}_{a1a} R_{\alpha \mu} R + \tfrac{1}{72} dN^{\alpha \beta aa1} K^{\mu}{}_{\beta a1a} R_{\alpha \mu} R + \tfrac{1}{72} dN^{\alpha \beta aa1} K_{\alpha }{}^{\mu}{}_{a1a} R_{\beta \mu} R \right.\\
		& \left. + \tfrac{1}{72} dN^{\alpha }{}_{\alpha }{}^{aa1} K^{\beta \mu}{}_{a1a} R_{\beta \mu} R + \tfrac{1}{72} dN^{\alpha \beta aa1} K^{\mu}{}_{\alpha a1a} R_{\beta \mu} R -  \tfrac{1}{288} dN^{\alpha \beta aa1} K^{\mu \nu}{}_{a1a}  R^2 g^{(2)}{}_{\alpha \beta \mu \nu} \right] \, .
	\end{align}
}
On the other hand, the terms that only depend on the Riemann tensor are
{\small
	\begin{align}\label{eq:result-integand-general-Riemann^2}\nonumber
		& {\rm tr}\, \ha_2 \supset \int_0^1 d \zeta \left[- \tfrac{1}{40} dN^{\alpha \beta aa1} K^{\mu \nu}{}_{a1a} R_{\alpha \nu}{}^{\rho \sigma} R_{\beta \mu \rho \sigma} - \tfrac{1}{40} dN^{\alpha \beta aa1} K^{\mu \nu}{}_{a1a} R_{\alpha \mu}{}^{\rho \sigma} R_{\beta \nu \rho \sigma} \right.\\\nonumber
		& \left. + \tfrac{1}{180} dN^{\alpha \beta aa1} K^{\mu}{}_{\mu a1a} R_{\alpha }{}^{\nu \rho \sigma} R_{\beta \nu \rho \sigma} + \tfrac{1}{20} dN^{\alpha \beta aa1} K^{\mu \nu}{}_{a1a} R_{\alpha }{}^{\rho}{}_{\mu}{}^{\sigma} R_{\beta \nu \rho \sigma} + \tfrac{1}{20} dN^{\alpha \beta aa1} K^{\mu \nu}{}_{a1a} R_{\alpha \nu}{}^{\rho \sigma} R_{\beta \rho \mu \sigma} \right.\\\nonumber
		& \left. - \tfrac{1}{90} dN^{\alpha \beta aa1} K^{\mu \nu}{}_{a1a} R_{\alpha }{}^{\rho}{}_{\nu}{}^{\sigma} R_{\beta \rho \mu \sigma} -  \tfrac{1}{90} dN^{\alpha \beta aa1} K^{\mu \nu}{}_{a1a} R_{\alpha }{}^{\rho}{}_{\mu}{}^{\sigma} R_{\beta \rho \nu \sigma} -  \tfrac{1}{90} dN^{\alpha \beta aa1} K^{\mu \nu}{}_{a1a} R_{\alpha }{}^{\rho}{}_{\nu}{}^{\sigma} R_{\beta \sigma \mu \rho} \right.\\\nonumber
		& \left. -  \tfrac{1}{90} dN^{\alpha \beta aa1} K^{\mu \nu}{}_{a1a} R_{\alpha }{}^{\rho}{}_{\mu}{}^{\sigma} R_{\beta \sigma \nu \rho} + \tfrac{1}{180} dN^{\alpha \beta aa1} K_{\beta}{}^{\mu}{}_{a1a} R_{\alpha }{}^{\nu \rho \sigma} R_{\mu \nu \rho \sigma} + \tfrac{1}{180} dN^{\alpha \beta aa1} K^{\mu}{}_{\beta a1a} R_{\alpha }{}^{\nu \rho \sigma} R_{\mu \nu \rho \sigma} \right.\\\nonumber
		& \left. + \tfrac{1}{180} dN^{\alpha \beta aa1} K_{\alpha }{}^{\mu}{}_{a1a} R_{\beta}{}^{\nu \rho \sigma} R_{\mu \nu \rho \sigma} + \tfrac{1}{180} dN^{\alpha }{}_{\alpha }{}^{aa1} K^{\beta \mu}{}_{a1a} R_{\beta}{}^{\nu \rho \sigma} R_{\mu \nu \rho \sigma} + \tfrac{1}{180} dN^{\alpha \beta aa1} K^{\mu}{}_{\alpha a1a} R_{\beta}{}^{\nu \rho \sigma} R_{\mu \nu \rho \sigma} \right.\\
		& \left. - \tfrac{1}{90} dN^{\alpha \beta aa1} K^{\mu \nu}{}_{a1a} R_{\alpha }{}^{\rho}{}_{\beta}{}^{\sigma} R_{\mu \rho \nu \sigma} -  \tfrac{1}{90} dN^{\alpha \beta aa1} K^{\mu \nu}{}_{a1a} R_{\alpha }{}^{\rho}{}_{\beta}{}^{\sigma} R_{\mu \sigma \nu \rho} -  \tfrac{1}{720} dN^{\alpha \beta aa1} K^{\mu \nu}{}_{a1a}  R_{\rho \sigma \lambda \tau} R^{\rho \sigma \lambda \tau} g^{(2)}{}_{\alpha \beta \mu \nu} \right] \, .
	\end{align}
}
Finally, the terms that display both the Riemann tensor and its contractions are
{\small
	\begin{align}\label{eq:result-integand-general-Riemann-Ricci}\nonumber
		& {\rm tr}\, \ha_2 \supset \int_0^1 d \zeta \left[ - \tfrac{7}{120} dN^{\alpha \beta aa1} K^{\mu \nu}{}_{a1a} R_{\nu}{}^{\rho} R_{\alpha \beta \mu \rho} - \tfrac{7}{120} dN^{\alpha \beta aa1} K^{\mu \nu}{}_{a1a} R_{\mu}{}^{\rho} R_{\alpha \beta \nu \rho} \right.\\\nonumber
		& \left. - \tfrac{1}{36} dN^{\alpha \beta aa1} K^{\mu \nu}{}_{a1a} R R_{\alpha \mu \beta \nu} -  \tfrac{1}{360} dN^{\alpha \beta aa1} K^{\mu \nu}{}_{a1a} R_{\nu}{}^{\rho} R_{\alpha \mu \beta \rho} - \tfrac{1}{180} dN^{\alpha \beta aa1} K^{\mu \nu}{}_{a1a} R_{\beta}{}^{\rho} R_{\alpha \mu \nu \rho} \right.\\\nonumber
		& \left. -  \tfrac{1}{36} dN^{\alpha \beta aa1} K^{\mu \nu}{}_{a1a} R R_{\alpha \nu \beta \mu} - \tfrac{1}{360} dN^{\alpha \beta aa1} K^{\mu \nu}{}_{a1a} R_{\mu}{}^{\rho} R_{\alpha \nu \beta \rho} + \tfrac{1}{180} dN^{\alpha \beta aa1} K^{\mu}{}_{\mu a1a} R^{\nu \rho} R_{\alpha \nu \beta \rho} \right.\\\nonumber
		& \left. -  \tfrac{1}{18} dN^{\alpha \beta aa1} K^{\mu \nu}{}_{a1a} R_{\beta}{}^{\rho} R_{\alpha \nu \mu \rho} + \tfrac{1}{180} dN^{\alpha \beta aa1} K_{\beta}{}^{\mu}{}_{a1a} R^{\nu \rho} R_{\alpha \nu \mu \rho} + \tfrac{1}{180} dN^{\alpha \beta aa1} K^{\mu}{}_{\beta a1a} R^{\nu \rho} R_{\alpha \nu \mu \rho} \right.\\\nonumber
		& \left. -  \tfrac{1}{360} dN^{\alpha \beta aa1} K^{\mu \nu}{}_{a1a} R_{\nu}{}^{\rho} R_{\alpha \rho \beta \mu} -  \tfrac{1}{360} dN^{\alpha \beta aa1} K^{\mu \nu}{}_{a1a} R_{\mu}{}^{\rho} R_{\alpha \rho \beta \nu} - \tfrac{1}{180} dN^{\alpha \beta aa1} K^{\mu \nu}{}_{a1a} R_{\alpha }{}^{\rho} R_{\beta \mu \nu \rho} \right.\\\nonumber
		& \left. -  \tfrac{1}{18} dN^{\alpha \beta aa1} K^{\mu \nu}{}_{a1a} R_{\alpha }{}^{\rho} R_{\beta \nu \mu \rho} + \tfrac{1}{180} dN^{\alpha \beta aa1} K_{\alpha }{}^{\mu}{}_{a1a} R^{\nu \rho} R_{\beta \nu \mu \rho} + \tfrac{1}{180} dN^{\alpha }{}_{\alpha }{}^{aa1} K^{\beta \mu}{}_{a1a} R^{\nu \rho} R_{\beta \nu \mu \rho} \right.\\
		& \left. + \tfrac{1}{180} dN^{\alpha \beta aa1} K^{\mu}{}_{\alpha a1a} R^{\nu \rho} R_{\beta \nu \mu \rho} \right] \, .
	\end{align}
}

	{\small
	\begin{align}\label{eq:result-integand-general-Riem^2}\nonumber
		& {\rm tr}\, \ha_2 \supset \int_0^1 d \zeta \left[ - \tfrac{1}{36} dN^{\alpha \beta aa1} K^{\mu \nu}{}_{a1a} R_{\alpha \nu} R_{\beta \mu} -  \tfrac{1}{36} dN^{\alpha \beta aa1} K^{\mu \nu}{}_{a1a} R_{\alpha \mu} R_{\beta \nu} \right.\\\nonumber
		& \left. -  \tfrac{1}{90} dN^{\alpha \beta aa1} K^{\mu}{}_{\mu a1a} R_{\alpha }{}^{\nu} R_{\beta \nu} -  \tfrac{1}{36} dN^{\alpha \beta aa1} K^{\mu \nu}{}_{a1a} R_{\alpha \beta} R_{\mu \nu} -  \tfrac{1}{90} dN^{\alpha \beta aa1} K_{\beta}{}^{\mu}{}_{a1a} R_{\alpha }{}^{\nu} R_{\mu \nu} \right.\\\nonumber
		& \left. -  \tfrac{1}{90} dN^{\alpha \beta aa1} K^{\mu}{}_{\beta a1a} R_{\alpha }{}^{\nu} R_{\mu \nu} -  \tfrac{1}{90} dN^{\alpha \beta aa1} K_{\alpha }{}^{\mu}{}_{a1a} R_{\beta}{}^{\nu} R_{\mu \nu} -  \tfrac{1}{90} dN^{\alpha }{}_{\alpha }{}^{aa1} K^{\beta \mu}{}_{a1a} R_{\beta}{}^{\nu} R_{\mu \nu} \right.\\\nonumber
		& \left. -  \tfrac{1}{90} dN^{\alpha \beta aa1} K^{\mu}{}_{\alpha a1a} R_{\beta}{}^{\nu} R_{\mu \nu} + \tfrac{1}{720} dN^{\alpha \beta aa1} K^{\mu \nu}{}_{a1a}  R_{\rho \sigma} R^{\rho \sigma} g^{(2)}{}_{\alpha \beta \mu \nu} -  \tfrac{1}{288} dN^{\alpha \beta aa1} K^{\mu \nu}{}_{a1a} R^2 g^{(2)}{}_{\alpha \beta \mu \nu} \right.\\\nonumber
		& \left. -  \tfrac{7}{120} dN^{\alpha \beta aa1} K^{\mu \nu}{}_{a1a} R_{\nu}{}^{\rho} R_{\alpha \beta \mu \rho} -  \tfrac{7}{120} dN^{\alpha \beta aa1} K^{\mu \nu}{}_{a1a} R_{\mu}{}^{\rho} R_{\alpha \beta \nu \rho} - \tfrac{1}{360} dN^{\alpha \beta aa1} K^{\mu \nu}{}_{a1a} R_{\nu}{}^{\rho} R_{\alpha \mu \beta \rho}\right.\\\nonumber
		& \left. - \tfrac{1}{180} dN^{\alpha \beta aa1} K^{\mu \nu}{}_{a1a} R_{\beta}{}^{\rho} R_{\alpha \mu \nu \rho} -  \tfrac{1}{360} dN^{\alpha \beta aa1} K^{\mu \nu}{}_{a1a} R_{\mu}{}^{\rho} R_{\alpha \nu \beta \rho} + \tfrac{1}{180} dN^{\alpha \beta aa1} K^{\mu}{}_{\mu a1a} R^{\nu \rho} R_{\alpha \nu \beta \rho} \right.\\\nonumber
		& \left. - \tfrac{1}{18} dN^{\alpha \beta aa1} K^{\mu \nu}{}_{a1a} R_{\beta}{}^{\rho} R_{\alpha \nu \mu \rho} + \tfrac{1}{180} dN^{\alpha \beta aa1} K_{\beta}{}^{\mu}{}_{a1a} R^{\nu \rho} R_{\alpha \nu \mu \rho} + \tfrac{1}{180} dN^{\alpha \beta aa1} K^{\mu}{}_{\beta a1a} R^{\nu \rho} R_{\alpha \nu \mu \rho} \right.\\\nonumber
		& \left. - \tfrac{1}{360} dN^{\alpha \beta aa1} K^{\mu \nu}{}_{a1a} R_{\nu}{}^{\rho} R_{\alpha \rho \beta \mu} - \tfrac{1}{360} dN^{\alpha \beta aa1} K^{\mu \nu}{}_{a1a} R_{\mu}{}^{\rho} R_{\alpha \rho \beta \nu} - \tfrac{1}{180} dN^{\alpha \beta aa1} K^{\mu \nu}{}_{a1a} R_{\alpha }{}^{\rho} R_{\beta \mu \nu \rho} \right.\\\nonumber
		& \left. -  \tfrac{1}{40} dN^{\alpha \beta aa1} K^{\mu \nu}{}_{a1a} R_{\alpha \nu}{}^{\rho \sigma} R_{\beta \mu \rho \sigma} -  \tfrac{1}{18} dN^{\alpha \beta aa1} K^{\mu \nu}{}_{a1a} R_{\alpha }{}^{\rho} R_{\beta \nu \mu \rho} + \tfrac{1}{180} dN^{\alpha \beta aa1} K_{\alpha }{}^{\mu}{}_{a1a} R^{\nu \rho} R_{\beta \nu \mu \rho} \right.\\\nonumber
		& \left. + \tfrac{1}{180} dN^{\alpha }{}_{\alpha }{}^{aa1} K^{\beta \mu}{}_{a1a} R^{\nu \rho} R_{\beta \nu \mu \rho} + \tfrac{1}{180} dN^{\alpha \beta aa1} K^{\mu}{}_{\alpha a1a} R^{\nu \rho} R_{\beta \nu \mu \rho} -  \tfrac{1}{40} dN^{\alpha \beta aa1} K^{\mu \nu}{}_{a1a} R_{\alpha \mu}{}^{\rho \sigma} R_{\beta \nu \rho \sigma} \right.\\\nonumber
		& \left. + \tfrac{1}{180} dN^{\alpha \beta aa1} K^{\mu}{}_{\mu a1a} R_{\alpha }{}^{\nu \rho \sigma} R_{\beta \nu \rho \sigma} + \tfrac{1}{20} dN^{\alpha \beta aa1} K^{\mu \nu}{}_{a1a} R_{\alpha }{}^{\rho}{}_{\mu}{}^{\sigma} R_{\beta \nu \rho \sigma} + \tfrac{1}{20} dN^{\alpha \beta aa1} K^{\mu \nu}{}_{a1a} R_{\alpha \nu}{}^{\rho \sigma} R_{\beta \rho \mu \sigma} \right.\\\nonumber
		& \left. -  \tfrac{1}{90} dN^{\alpha \beta aa1} K^{\mu \nu}{}_{a1a} R_{\alpha }{}^{\rho}{}_{\nu}{}^{\sigma} R_{\beta \rho \mu \sigma} -  \tfrac{1}{90} dN^{\alpha \beta aa1} K^{\mu \nu}{}_{a1a} R_{\alpha }{}^{\rho}{}_{\mu}{}^{\sigma} R_{\beta \rho \nu \sigma} -  \tfrac{1}{90} dN^{\alpha \beta aa1} K^{\mu \nu}{}_{a1a} R_{\alpha }{}^{\rho}{}_{\nu}{}^{\sigma} R_{\beta \sigma \mu \rho} \right.\\\nonumber
		& \left. -  \tfrac{1}{90} dN^{\alpha \beta aa1} K^{\mu \nu}{}_{a1a} R_{\alpha }{}^{\rho}{}_{\mu}{}^{\sigma} R_{\beta \sigma \nu \rho} + \tfrac{1}{180} dN^{\alpha \beta aa1} K_{\beta}{}^{\mu}{}_{a1a} R_{\alpha }{}^{\nu \rho \sigma} R_{\mu \nu \rho \sigma} + \tfrac{1}{180} dN^{\alpha \beta aa1} K^{\mu}{}_{\beta a1a} R_{\alpha }{}^{\nu \rho \sigma} R_{\mu \nu \rho \sigma} \right.\\\nonumber
		& \left. + \tfrac{1}{180} dN^{\alpha \beta aa1} K_{\alpha }{}^{\mu}{}_{a1a} R_{\beta}{}^{\nu \rho \sigma} R_{\mu \nu \rho \sigma} + \tfrac{1}{180} dN^{\alpha }{}_{\alpha }{}^{aa1} K^{\beta \mu}{}_{a1a} R_{\beta}{}^{\nu \rho \sigma} R_{\mu \nu \rho \sigma} + \tfrac{1}{180} dN^{\alpha \beta aa1} K^{\mu}{}_{\alpha a1a} R_{\beta}{}^{\nu \rho \sigma} R_{\mu \nu \rho \sigma} \right.\\\nonumber
		& \left. - \tfrac{1}{90} dN^{\alpha \beta aa1} K^{\mu \nu}{}_{a1a} R_{\alpha }{}^{\rho}{}_{\beta}{}^{\sigma} R_{\mu \rho \nu \sigma} -  \tfrac{1}{90} dN^{\alpha \beta aa1} K^{\mu \nu}{}_{a1a} R_{\alpha }{}^{\rho}{}_{\beta}{}^{\sigma} R_{\mu \sigma \nu \rho} \right.\\
		& \left. -  \tfrac{1}{720} dN^{\alpha \beta aa1} K^{\mu \nu}{}_{a1a} R_{\rho \sigma \lambda \tau} R^{\rho \sigma \lambda \tau} g^{(2)}{}_{\alpha \beta \mu \nu} \right] \, .
	\end{align}
}

	
	\bibliographystyle{chetref}


	\end{document}